\documentclass{aa}  

\usepackage{graphicx}
\usepackage{txfonts}
\usepackage{hyperref}
\newcommand{\enzo}{{\small ENZO}}
\newcommand{\rh}{R_{100}}

\begin{document}

   \title{ On the nature  of LOFAR   RMs  and  new constraints on magnetic fields in cosmic filaments and on magnetogenesis scenarios }

   \subtitle{}

   \author{E.~Carretti\inst{1}\and
          F.~Vazza\inst{2}\inst{,1}\inst{,3}\and
S.~P.~O'Sullivan\inst{4}\and
V.~Vacca\inst{5}\and
A.~Bonafede\inst{2}\inst{,1}\and
G.~Heald\inst{6}\inst{,7}\and
C.~Horellou\inst{8} \and
S.~Mtchedlidze\inst{2}\inst{,9} \and
T.~Vernstrom\inst{6}\inst{,10}
}
   \institute{INAF - Istituto di Radioastronomia, Via Gobetti 101, 40129, Bologna, Italy\\
              \email{carretti@ira.inaf.it (EC)}
\and
Dipartimento di Fisica e Astronomia, Universit\'a di Bologna, via Gobetti 93/2, 40122 Bologna, Italy
\and
Hamburger Sternwarte, University of Hamburg, Gojenbergsweg 112, 21029 Hamburg, Germany
\and
Departamento de Física de la Tierra y Astrofísica \& IPARCOS-UCM, Universidad Complutense de Madrid, 28040 Madrid, Spain
\and
INAF - Osservatorio Astronomico di Cagliari, Via della Scienza 5, 09047 Selargius (CA), Italy
\and
ATNF, CSIRO Space \& Astronomy, P.O. Box 1130, Bentley, WA 6102, Australia 
\and
SKA Observatory, SKA-Low Science Operations Centre, 26 Dick Perry Avenue, Kensington WA 6151, Australia
\and 
Department of Space, Earth and Environment, Chalmers University of Technology, Onsala Space Observatory, 439 92, Onsala, Sweden 
\and
School of Natural Sciences and Medicine, Ilia State University, 3-5 Cholokashvili St., 0194 Tbilisi, Georgia
\and
ICRAR, The University of Western Australia, 35 Stirling Hw, 6009 Crawley, Australia
             }

   \date{Received XXX; Accepted YYY }

 
  \abstract{
 The measurement of magnetic fields  in cosmic web  filaments can be used to reveal the magnetogenesis of  the Universe. In previous work, we produced first estimates of the field strength and its redshift evolution using  the Faraday Rotation Measure (RM) catalogue of extragalactic background sources at low frequency obtained with LOFAR observations. Here we refine our analysis  by selecting  sources with low Galactic RM, which reduces its residual contamination. We also conduct  a comprehensive analysis of the different contributions to the extragalactic RMs along the line of sight, and confirm   that they are dominated  by the cosmic filaments component, with  only 21 percent originating in galaxy clusters and the circumgalactic medium (CGM)  of galaxies. We find  a possible hint of a shock at the  virial radius of massive galaxies.  We also  find that the fractional polarization of background sources  might be a valuable CGM tracer.  The newly selected RMs have a steeper evolution  with  redshift than previously found. The field strength in filaments ($B_f$)  and its evolution  are estimated  assuming $B_f$ evolves as a power-law $B_f=B_{f,0}\,(1+z)^\alpha$.  Our  analysis finds an average strength at $z=0$ of  $B_{f,0} =11$--15~nG, with an error of 4 nG, and a slope $\alpha=2.3$--$2.6 \pm  0.5$, which is steeper than what we previously found. The comoving field has a slope of $\beta=$ [0.3, 0.6$]\pm 0.5$ that is consistent with being invariant with redshift. Primordial magnetogenesis scenarios are favoured  by our data, together with a sub-dominant  astrophysical-origin  RM component increasing  with redshift. 
   }

\titlerunning{Magnetic fields in cosmic filaments}
   \maketitle
%

\section{Introduction}
\label{sec:intro}
The origin of the magnetic field of the Universe can be investigated through the evolution of several indirect proxies  with cosmic time  \citep[e.g.][]{2016RPPh...79g6901S,2021RPPh...84g4901V, 2021MNRAS.505.5038A,2021MNRAS.500.5350V}. The magnetic field in cosmic filaments is a sweet spot for such an investigation. There the  field is not yet as processed as it is in galaxy clusters, where the memory of the initial conditions has been erased \citep{2014ApJ...797..133C,2017CQGra..34w4001V},  keeping the information of the primeval conditions, whilst it is stronger than in voids, which makes the detection easier \citep{2010Sci...328...73N,2017CQGra..34w4001V, 2022ApJ...929..127M}. There are two major groups of magnetogenesis models: primordial scenarios, where the field was produced early, during cosmic Inflation or in phase transitions before recombination \citep[e.g.][]{1988PhRvD..37.2743T, 1994RPPh...57..325K, 2019JCAP...11..028P}, and astrophysical scenarios, where the magnetisation of the Universe was generated  late in galaxies and AGNs and then injected in the intergalactic medium (IGM) \citep[][]{1994RPPh...57..325K, 2006MNRAS.370..319B,  2009MNRAS.392.1008D}{}{}. 

The Faraday Rotation Measure (RM) is a measure  of the polarization angle rotation by birifrangence of the two circular polarization states, produced when a polarized radiation travels trough a magneto-ionic medium with a magnetic field and populated with free electrons (a plasma). The RM depends on the magnetic field component along the line-of-sight (LOS), weighted by the electron number density, and integrated along the LOS from the source to the observer. It is an effective  tool  to  study  magnetic fields and the ionised medium in the Galaxy \citep[][]{2016A&A...596A.103P, 2022ApJ...940...75D}{}{},  the circumgalactic medium (CGM) of galaxies \citep[][]{2023A&A...670L..23H, 2023arXiv230811391B}{}{}, the environment local to the source \citep[][]{2008MNRAS.391..521L}{}{}, and the cosmic web \citep[][]{2022MNRAS.515..256P, 2023MNRAS.518.2273C, 2023SciA....9E7233V}{}{}. It helps  not only to investigate magnetic fields, but also to reveal the warm-hot intergalactic medium (WHIM) \citep[][]{2021PASA...38...20A}{}{} that, having low temperatures of 10$^4$--10$^6$~K, is  hard to detect in low density environments at other wavelengths (e.g., X-ray).

The evolution with redshift of the extragalactic  RM was investigated in the past decades, mostly finding no evolution \citep{1971PThPS..49..181F, 1972A&A....19..104R, 1977A&A....61..771K, 1979PASJ...31..125S, 1982ApJ...263..518K, 1982MNRAS.201..365T, 1984ApJ...279...19W, 1995ApJ...445..624O, 2003AcASn..44S.155Y, 2008Natur.454..302B, 2018MNRAS.475.1736V,  2020PASA...37...29R}. The extragalactic RMs at GHz frequencies are dominated by the contribution local to the source  \citep{2022MNRAS.512..945C}. At these frequencies  
\citet{1971PThPS..49..181F}, \cite{2008ApJ...676...70K} and \citet{2016ApJ...829....5L} found  a hint of evolution using samples of up to a few hundreds sources, which was attributed to RM originating local to the source \citep{2008ApJ...676...70K}. However, this evolution was not confirmed using a much larger sample of $\approx 4000$ sources \citep{2012arXiv1209.1438H}, thus the results are inconclusive thus far.

Limits on the IGM magnetic field (i.e. the mean magnetic field strength of the Universe, averaged over all cosmic web structures such as voids, walls, and filaments) are set with several tracers. Rotation measures of  extragalactic sources  give an upper limit of 1.5 nG when measured at 144 MHz \citep[][]{2022MNRAS.515..256P}{}{} and of 15 nG when  measured at 1400 MHz \citep[][]{2019ApJ...878...92V}{}{}. The investigation  of the synchrotron  emission of the cosmic web  has led to upper limits of 10-200 nG \citep[][]{2017MNRAS.467.4914V, 2017MNRAS.468.4246B}{}{}.  The temporal smearing distribution of Fast Radio Bursts gives an upper  limit  of 10 nG \citep[][]{2023ApJ...946L..18P}{}{}.  Delays in the arrival time of GeV emission from TeV $\gamma$-ray blazars  yield  lower limits of    $4\times 10^{-14}$ to $7\times 10^{-16}$ G in voids, depending on the blazar duty cycle \citep[][]{2023ApJ...950L..16A}. The time delays of lensed blazars give a lower limit of $2\times 10^{-17}$ G \citep[][]{2023arXiv231101207E}{}{}.  Time delays of $\gamma$-rays from Gamma Ray Bursts give lower limits of a few $10^{-19}$ G \citep[e.g.][]{2023arXiv230605970H, 2023arXiv230607672V}{}{}. Using blazars, the filling factor of the IGM field has been constrained to $f \gtrsim 0.67$, which favours primordial over astrophysical models \citep*[][]{2024ApJ...963..135T}. 

The magnetic field in cosmic filaments  has been measured at 30--60 nG using  synchrotron emission stacking \citep[][]{2021MNRAS.505.4178V}{}{} and at 40--80 nG with extragalactic RMs at low frequencies \citep[][]{2022MNRAS.512..945C,  2023MNRAS.518.2273C}.  \citet[][]{2023SciA....9E7233V} found that  filaments are highly polarized, which implies a significant ordered  magnetic field component in filaments, consistent with the detection of an RM signal from  them  \citep[][]{2022MNRAS.512..945C}{}{}. The high polarization fraction  also  supports the theoretical paradigm of  filament  accretion  with shocks from infalling matter. \citet[][]{2023MNRAS.518.2273C}{}{} estimated the evolution of the magnetic field strength with redshift. Their results are  consistent with no evolution of the physical  field out to $z=3$. Upper limits  were  found by cross correlating the synchrotron emission at low frequency with X-ray emission \citep[][]{2023MNRAS.523.6320H}{}{}, comparing synchrotron emission with simulations \citep[][]{2018MNRAS.479..776V}{}{}, and cross-correlating RMs with cosmic web tracers \citep[][]{2021MNRAS.503.2913A}{}{}. Cosmic filaments have been recently detected in Lyman-$\alpha$ \citep[][]{2023NatAs...7.1390M} and X-ray diffuse emission \citep[][]{2024arXiv240117281D}. Magnetic fields have also been recently  detected in the cosmic web just off galaxy groups \citep{2024arXiv240720325A}, where they have a strength of 200-600 nG and then drop under their dataset sensitivity at projected separations from  groups beyond seven splashback radii.

In previous work we   estimated the magnetic field strength in cosmic filaments using extragalactic source RMs at low frequencies of $\approx$144 MHz \citep[][hereafter Paper I]{2022MNRAS.512..945C} and  its evolution with redshift  \citep[][hereafter Paper II]{2023MNRAS.518.2273C}. We used the RM catalogue of  \citet[][]{2023MNRAS.519.5723O}{}{} derived from the LOFAR Two-metre Sky Survey data \citep[LoTSS;][]{2022A&A...659A...1S}{}{}. We also  compared the RM--redshift evolution with that predicted by cosmological models, finding that primordial stochastic models are favoured. We  found that the sight-lines  of polarized sources tend to avoid the high density environments of galaxy clusters at this low frequency, and   that the RM of the extragalactic sources is dominated by that produced by the intervening cosmic filaments, with a minor contribution from the environment local to the source.  

In this work we refine  our analysis by improving our sample selection, specifically, only selecting sources with low Galactic RM. We also conduct a  comprehensive analysis of the different contributions to the extragalactic RMs we use and find again that, once the Galactic RM is subtracted off,  they are dominated by the IGM/cosmic web term. Finally, we estimate the strength and evolution with redshift of the magnetic field in cosmic filaments, using a Bayesian analysis, and compare the results with the predictions of magnetogenesis  models, including several new primordial stochastic ones,  which our previous work found favoured compared to other magnetogenesis scenarios.  

This paper is organised as follows. In Sect. \ref{sec:data} the RM sample at 144 MHz   we use is described, and the RM as a function  of the redshift  ($z$) is computed. Section \ref{sec:simul} describes the MHD cosmological simulations used in this work. In Sect. \ref{sec:tests} we investigate the origin of our RMs, specifically  whether there is a significant contribution from intervening galaxy clusters, galaxies and their CGM, or the radio sources themselves.  In Sect. \ref{sec:zevo} we fit the strength and redshift evolution of the magnetic field in filaments to the observed extragalactic RM as a function of redshift. We also compare the observed RMs with those computed for a number of magnetogenesis scenarios. Finally, in Sect. \ref{sec:disc} we discuss our results and draw our conclusions. 

Throughout the paper we assume a flat $\Lambda$CDM cosmological model  with $H_0 = 67.8$~km~s$^{-1}$~Mpc$^{-1}$, $\Omega_M = 0.308$, $\Omega_\Lambda =0.692$,  $\Omega_b = 0.0468$, and $\sigma_8=0.815$ \citep{2016A&A...594A..13P}. Errors  refer to 1-$\sigma$ uncertainties. 

\section{Faraday Rotation Measure data}
\label{sec:data}

%

%
\subsection{LoTSS RM catalogue at 144~MHz}
%
   \begin{figure*}
   \centering
    \includegraphics[width=0.49\hsize]{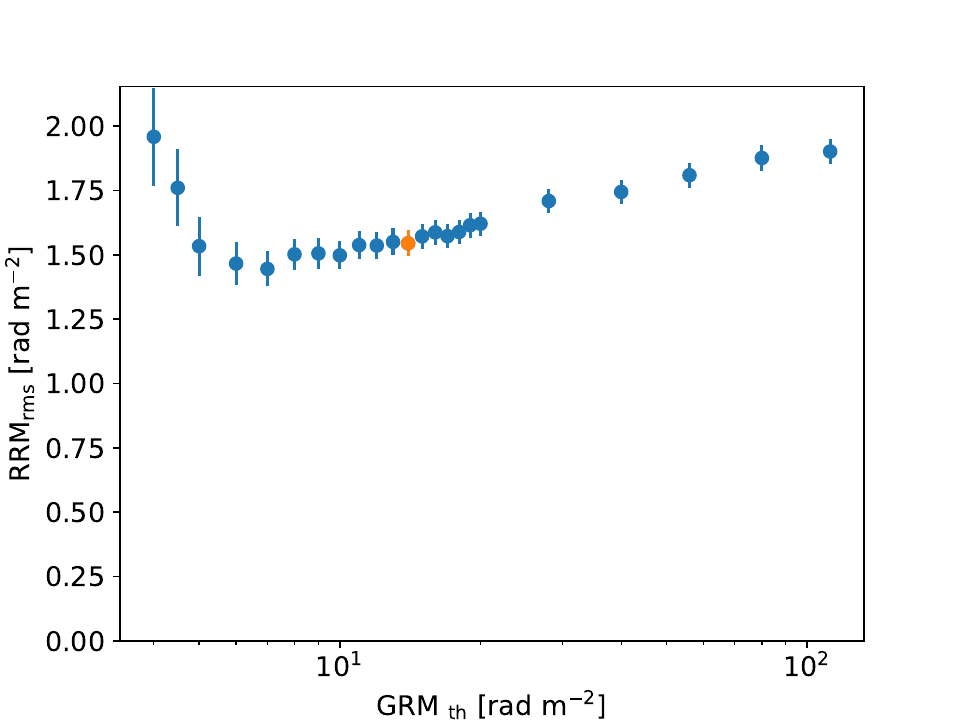}
    \includegraphics[width=0.49\hsize]{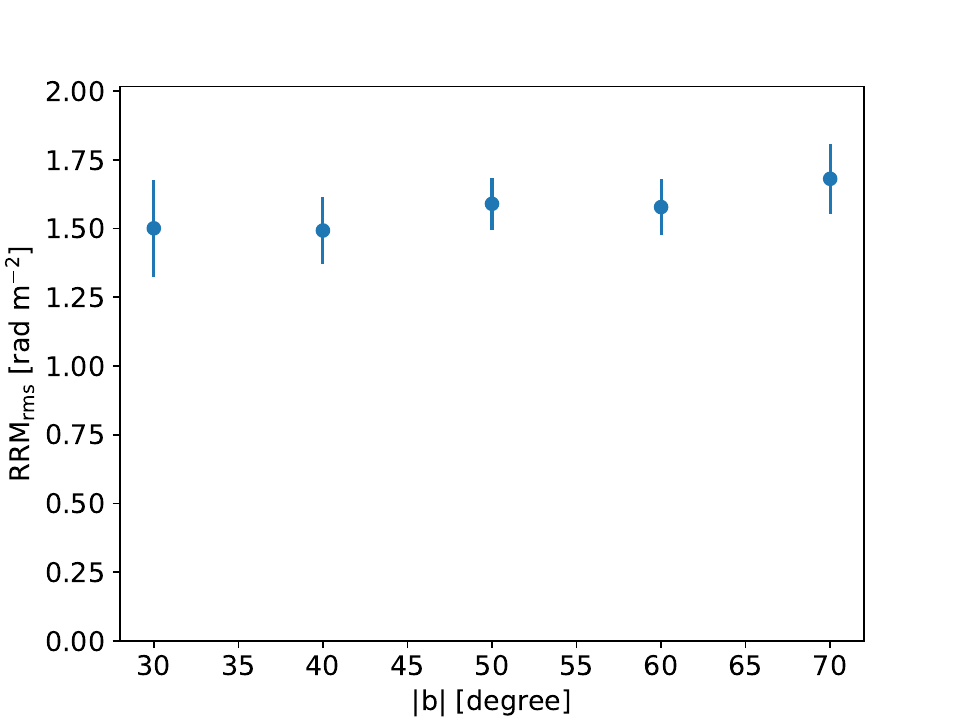}
   \caption{ { Left}: RRM rms of the spectroscopic redshift sample filtered by   different GRM limits. The orange circle highlights the case with a limit of 14~rad~m$^{-2}$.  { Right}:  RRM rms as a function of Galactic latitude $|b|$ for the case filtered with GRM$_{\rm th} =14$~rad~m$^{-2}$. }
   \label{fig:grm}
    \end{figure*}

   \begin{figure}
   \centering
    \includegraphics[width=0.99\hsize]{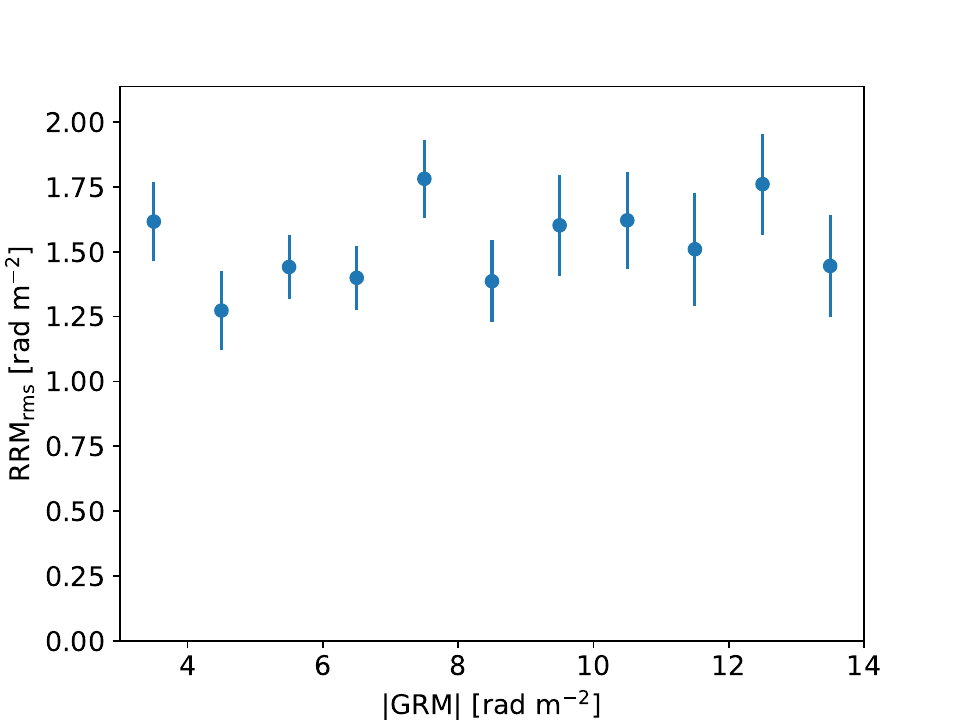}
   \caption{RRM rms  as a function of |GRM|    of the spectroscopic redshift sample filtered by $\rm |GRM| < 14$~rad~m$^{-2}$.}
              \label{fig:rrmvgrm}%
    \end{figure}

  \begin{figure}
   \centering
    \includegraphics[width=0.99\hsize]{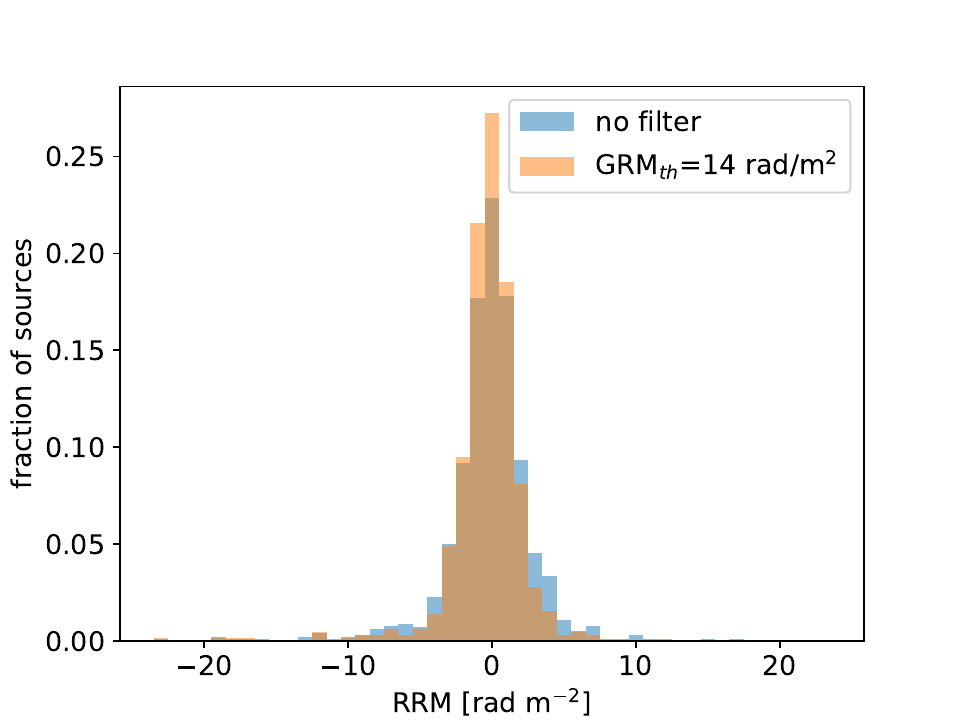}
   \caption{RRM distribution   of the spectroscopic redshift sample filtered by $\rm |GRM| < 14$~rad~m$^{-2}$. The case with no GRM filter is also shown for comparison.}
              \label{fig:rrmdist}%
    \end{figure}

   \begin{figure*}
   \centering
    \includegraphics[width=0.49\hsize]{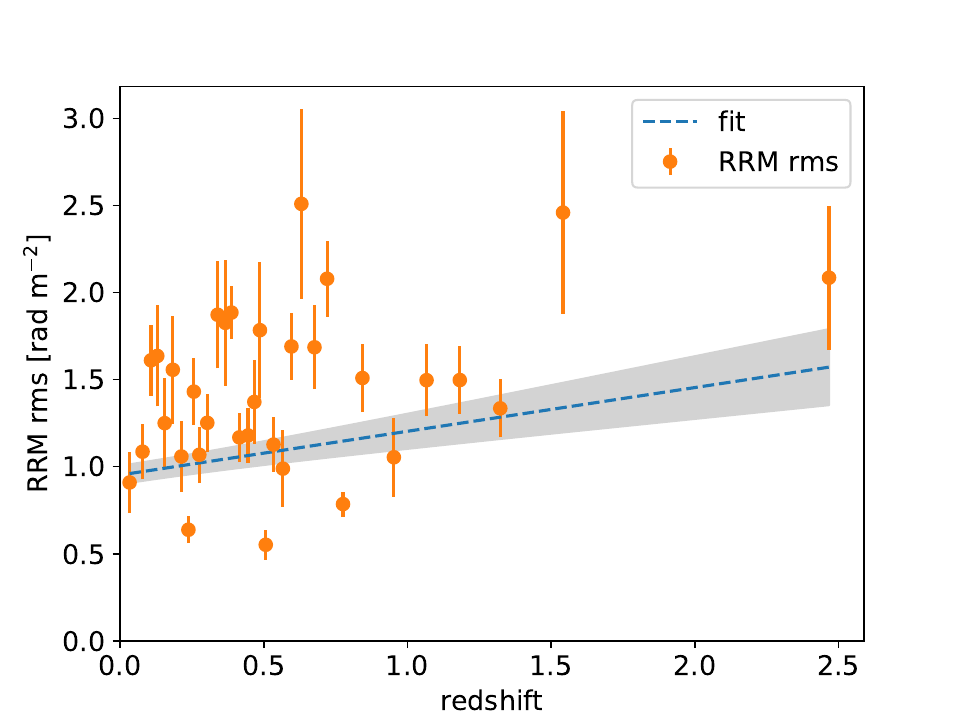}
    \includegraphics[width=0.49\hsize]{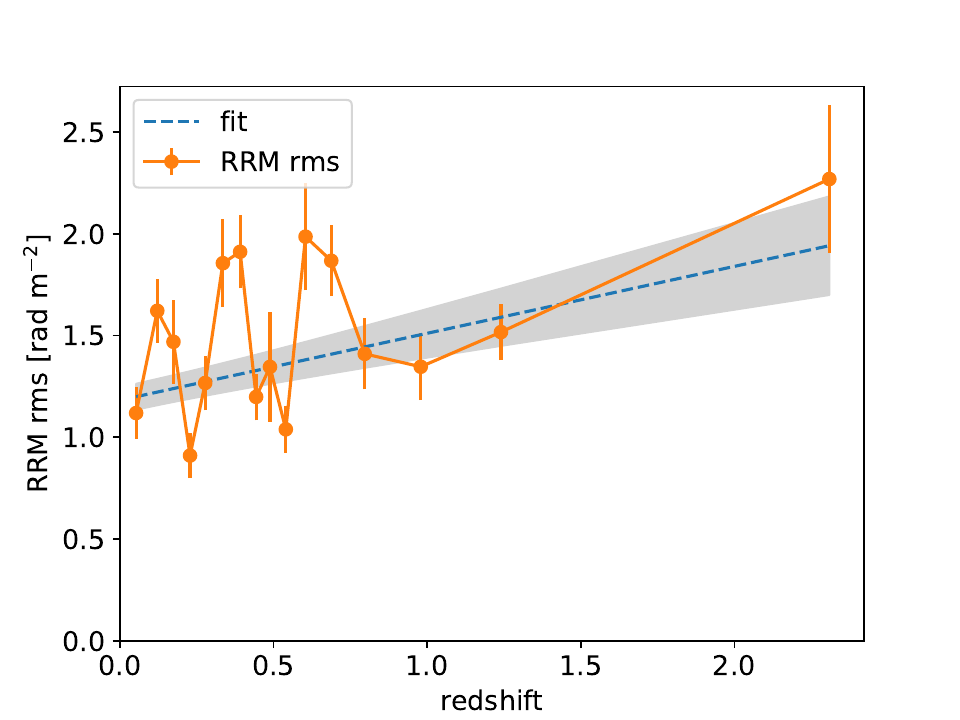}
   \caption{RRM rms in redshift bins of 20 (left) and 40 (right) sources each of the spectroscopic redshift sample filtered by $\rm |GRM| < 14$~rad~m$^{-2}$. The  linear fit (dashed line) and its uncertainty (shaded area) are also reported. }
              \label{fig:zdisp}%
    \end{figure*}

We use the RMs from the   catalogue  obtained from the LoTSS DR2 (LOFAR Two-metre Sky Survey Data Release 2) polarization data \citep[][]{ 2023MNRAS.519.5723O,2022A&A...659A...1S}. All details can be found in the paper describing the catalogue  \citep{2023MNRAS.519.5723O}.  Here we summarise the information relevant to this work. The observations were taken with the High Band Antenna (HBA) of the LOFAR telescope  in a 48~MHz broad  band centred at 144~MHz, with a frequency resolution of 97.6~kHz, and an angular resolution of 20~arcsec.      The catalogue covers  an area of 5720~deg$^2$ and consists of 2461 RM components, 1949 of which have an associated redshift, either photometric or spectroscopic. The RMs are  measured through RM-Synthesis \citep{1966MNRAS.133...67B, 2005A&A...441.1217B}. The catalogue excludes the  RM range of     [-1, 3]~rad~m$^{-2}$ to avoid  sources possibly contaminated by instrumental polarization leakage. 

As done in Paper I and II, we filtered the sample by only keeping sources with a spectroscopic redshift ($z$) and at Galactic latitudes $|b|>25^\circ$, the latter is  to avoid regions with large Galactic RM (GRM) contamination. This results in a sample of 1016  sources, the median and maximum redshift are 0.48 and 3.37.  

The measured RMs are a combination of the GRM (more generally,  the RM structures correlated on scales larger than $\approx 1^\circ$, which are dominated by  the Milky Way ISM and  its circumgalactic medium), the extragalactic RM (RM$_{\rm eg}$), and  the noise ($N$): 
\begin{equation}
    {\rm RM = GRM + RM_{eg}} + N. 
\end{equation}
The extragalactic term  consists of an RM of astrophysical origin, either local to the source, mostly produced  in the environment surrounding the source \citep{2008MNRAS.391..521L}, or objects intervening along the sight line such as galaxy clusters or galaxies, and an RM generated in the cosmic web, mostly in filaments because the contribution from voids is marginal  (see Paper II).
The extragalactic term that we use in this work  is estimated as the Residual RM: 
\begin{equation}
    \rm RRM = RM - GRM .
    \label{eq:rrm}
\end{equation}
At the low frequency of 144 MHz we work at, the astrophysical term is  small  (see Sect.~\ref{sec:tests}) and the RRM is dominated by the cosmic filament term.

The GRM contribution to each source is estimated from the GRM map of \citet[][]{2022A&A...657A..43H}{}{} as the median of  a disc of radius of $0.5^\circ$ centred at the source to avoid a known bias at the source position  (see Paper I and \citealt{2022A&A...657A..43H} for details). The radius of $0.5^\circ$  is set to match the average  separation of $\approx 1^\circ$ between the sources used by \citet[][]{2022A&A...657A..43H}{}{}  to derive their GRM map. The error is estimated by bootstrapping.\footnote{Bootstrapping consists of resampling the sample a quantity is estimated with by randomly selecting sample elements, and then the quantity is estimated. This is repeated for a number of resamplings. From the distribution of the results the standard deviation is finally estimated.}  \citet[][]{2024arXiv240313418B} have computed  RRMs for the same LoTSS sample, estimating GRMs either with the same GRM map as used here or from  \citet[][]{2020A&A...633A.150H} and a method similar to ours. They obtain RRMs similar to ours with either map, which supports the robustness of our GRM estimate method. 

The noise term consists of a  measurement  sensitivity term ($N_m$)  and a GRM error ($N_{\rm GRM}$). The two noise terms quadratically add up in an RRM rms computation and they are subtracted off throughout   the paper: 
\begin{equation}
    {\rm RRM} \,\, {\rm rms} = \left(\left<{\rm RRM}^2\right> - \left<N_m^2\right> - \left<N_{\rm GRM}^2\right>\right)^{1/2}.
    \label{eq:rmscorr}
\end{equation}

\subsection{RRM rms versus  GRM}
\label{sec:grmcuts}

The RRMs computed with Eq.~\ref{eq:rrm} can still be affected by GRM estimate errors that  are expected to be somewhat proportional to the GRM value.  We thus filtered the sample for different GRM thresholds (GRM$_{\rm th}$), keeping sources with $\rm |GRM| < GRM_{\rm th}$, and computed the RRM rms of each resulting sample. We excluded  2-$\sigma$  outliers from the rms computation. 

The result is shown in Fig.~\ref{fig:grm}, left panel, that reports the RRM rms as a function of GRM$_{\rm th}$. The unfiltered  sample, with a maximum $\rm |GRM|$ of  $ 105.0$~rad~m$^{-2}$, has an RRM rms of $1.90\pm 0.05$~rad~m$^{-2}$.  The minimum rms is at GRM$_{\rm th} = 7$~rad~m$^{-2}$ at  $1.44 \pm 0.07$~rad~m$^{-2}$. Lower GRM$_{\rm th}$ values do not give any reduction of the RRM rms and  we infer that   the  residual GRM contributions are  marginal, at the uncertainty level.  However, this case with  GRM$_{\rm th} = 7$~rad~m$^{-2}$  only includes   338 sources,   a mere 1/3 of the full sample.

A more valuable   option is the case   with GRM$_{\rm th} =14$~rad~m$^{-2}$. It  gives  an RRM rms of $1.54\pm 0.05$~rad~m$^{-2}$, which  differs  from the minimum at some 1-$\sigma$ confidence level (1.2-$\sigma)$.  It  is still statistically consistent with the minimum, but it contains a much larger  number of sources (653). The median GRM is 6.7~rad~m$^{-2
}$. The median and maximum redshift of this sample are  0.47 and 3.22. Larger GRM$_{\rm th}$ values return  RRM rms values that differ from the minimum  by more than 1.6-$\sigma$ and we consider them not to be an option. 

To check whether the  GRM$_{\rm th} =14$~rad~m$^{-2}$ case  still retains  a significant residual GRM contribution we computed the RRM rms in Galactic latitude $|b|$ bins, of bin-width  of $\Delta\,|b| =10^\circ$. If a contribution is still present, a decrease of the RRM rms with $|b|$ would be expected. Fig.~\ref{fig:grm}, right panel, shows a rather flat behaviour with no obvious trend with $|b|$. The Spearman's correlation rank of $\rm |RRM|$ as a function of $|b|$ is $\rho = -0.018$ with a $p$-value of 0.64, which  supports that RRM and $b$ are uncorrelated in this sample.

To further check whether there is significant residual GRM contamination in our RRMs, we compute the RRM rms in   |GRM| bins  out to 14~rad~m$^{-2}$ (Fig.~\ref{fig:rrmvgrm}). Data with |GRM| $< 3$~rad~m$^{-2}$ are excluded to avoid the gap in RM that could produce an artificial correlaton.  There  is  no  obvious correlation, which is confirmed by a Spearman's correlation rank of |RRM| versus |GRM| of $\rho=0.011$ with an high $p$-value of $p=0.77$.  This supports a marginal residual GRM contamination.

Fig. ~\ref{fig:rrmdist} shows the RRM distribution of the same filtered sample. Compared to  the RRM distribution  of the  unfiltered sample, also shown in Fig. \ref{fig:rrmdist}.  a reduction of the wings can be noticed. 

 These sanity checks are informative and necessary in order to assess  to which extent can any strategy to separate Galactic and extragalactic contributions to the RMs of a sample of sources. Simpler procedures to remove the Galactic contribution can be biased (see the analysis in Paper I).  This  would result into a distribution of putative extragalactic RRMs that would instead show statistical correlations with the Galactic RM, signalling an incorrect estimate of the truly extragalactic component. 

\subsection{Behaviour of extragalactic RM with redshift} 
\label{sec:rmrms}

For the best fitting analysis of this work we use the spectroscopic redshift sample filtered for $\rm |GRM| < 14$~rad~m$^{-2}$ discussed in Sect.~\ref{sec:grmcuts}. We 
computed the RRM rms in 20-source redshift bins (Fig.~\ref{fig:zdisp}, left panel). The RRM rms is computed as for Eq.~\ref{eq:rmscorr} and excluding 2-$\sigma$ outliers. Errors are estimated by bootstrapping. 

Compared  to the RRM rms of the full spectroscopic redshift sample (see Paper II), the  RRM rms(z)  of the GRM filtered sample computed here is steeper and lower, and at the $z=0$ end it gets closer to 0. A weighted linear fit gives a slope of $0.25\pm 0.08$ which is not flat at 3-$\sigma$ significance. 

We also show the plot of RRM rms as a function of redshift with 40-source bins (Fig.~\ref{fig:zdisp}, right panel), which gives smaller errors. We note the presence of wiggles. We believe it is unlikely that they are related to the residual GRM contamination. They were also present in Paper II, where we used the full, unfiltered spectroscopic sample, and here, where we use a sample with reduced GRM contamination, they are still there. We discuss more on the wiggles in Sect.~\ref{sec:galaxies}.   

\section{Cosmological MHD simulations}
\label{sec:simul}

Compared to our previous work  (Paper II), here we use a much improved set of cosmological magneto-hydrodynamical simulations produced using {\enzo}\footnote{enzo-project.org}, designed to investigate the dependence between RRM and different models of magnetogenesis with increased resolution and improved physical models. 

 We produced multiple resimulations of the same comoving volume of $(42.5\ \mathrm{Mpc})^3$ with a static grid of $1024^3$ cells, giving a constant spatial resolution of $41.5$ kpc/cell and a constant mass resolution of  $1.01 \times 10^{7}\ \mathrm{M_{\odot}}$ per dark matter particle. Also based on our previous works, this volume was chosen to provide a reasonable compromise between a good spatial resolution in the medium or low density regions which are mostly contributing to the observed RRM (after the excision of denser halo regions), as well as a large enough cosmic volume to provide us a fair sampling of voids and filaments along simulated lines of sight.
 All runs include equilibrium gas cooling, a  "sub-grid" dynamo amplification model at run-time, which allows the estimation of the maximum contribution of a dynamo in low density environments \citep[see][]{2008Sci...320..909R}, while the treatment of primordial magnetic fields and feedback from galaxy formation processes varies.
In detail: 

\begin{enumerate}
 \item \textit{``primordial uniform model''}:  a primordial uniform volume-filling comoving magnetic field strength of  $B_0=0.1\ \mathrm{nG}$ initialised at the beginning of the simulation ($z=40$);
  \item \textit{``primordial stochastic models''}:  three different  initially tangled  seed primordial magnetic fields, whose distribution of field scales follows a simple power law spectrum: $P_B(k) = P_{B0}k^{\alpha_s}$ characterised by a constant spectral index and an amplitude, commonly referred after smoothing the fields within a scale $\lambda=1 \rm ~Mpc$, using the same approach of \citet[][]{2021MNRAS.500.5350V}. Here we simulated the $\alpha_s=-1.0$, $=0.0$,  $=1.0$ and  $=2.0$ cases, i.e. from very "red" to very "blue" primordial spectra, using the normalisation parameters given in \citet{2021MNRAS.500.5350V} and based on observational constraints from the Cosmic Microwave Background obtained by \citet{2019JCAP...11..028P}. "Red" spectra with a lower $\alpha_s$ have most of their magnetic energy on the largest scales, mimicking the result of inflationary processes, while "blue" spectra with higher $\alpha_s$ have more energy at smaller scales, with our limiting case here of $\alpha_s=2.0$ which mimics the outcome of "causal" generation processes, in which the initial magnetic field is correlated only on scales smaller than the cosmological horizon. In detail, the adopted normalisation for each model is $B_{\rm 1~Mpc}=0.003\,\rm nG$ for $\alpha_s=2.0$, $B_{\rm 1~Mpc}=0.042 \, \rm nG$ for $\alpha_s=1.0$, $B_{\rm 1~Mpc}=0.35 \,\rm nG$ for $\alpha_s=0.0$ and $B_{\rm 1~Mpc}=1.87 \, \rm nG$ for the $\alpha_s=-1.0$. As will be discussed in Sec.~\ref{sec:synth_rrm}, unlike in all other cases we found that downscaling the amplitude of the last, $\alpha_s=-1.0$ model, to $B_{\rm 1~Mpc}=0.37 \, \rm nG$ can produce a reasonable match to LOFAR RRMs.

 \item \textit{``astrophysical models''}: two models in which magnetic fields were injected at run-time in the simulation, both by star forming particles and by simulated active galactic nuclei. In the first case, we used the star formation recipe by  \citet[][]{2003ApJ...590L...1K}, designed to reproduce the observed Kennicutt's law \citep[][]{1998ApJ...498..541K} and with free parameters calibrated to reasonably reproduce the integrated star formation history and the stellar mass function of galaxies at $z\leq 2$. 
 The feedback from star forming particles assumes a fixed fraction of energy/momentum/mass ejected per each formed star particles, $E_{SN}= \epsilon_{SF} m_* c^2$, with efficiency calibrated to $\epsilon_{SF}=10^{-8}$ as in previous work \citep[][]{2017CQGra..34w4001V}.
 We also consider that $90\%$ of the feedback energy is released in the thermal form (i.e. hot supernovae-driven winds), distributed to the 27 nearest cells around the star particle, and $10\%$ in the form of magnetic energy, assigned to magnetic dipoles by each feedback burst.

The feedback from active galactic nuclei is treated by assuming that, at each  timestep of the simulation, the highest density peaks in the simulation harbour a supermassive black hole, to which we assign a realistic mass based on  observed scaling relations \citep[e.g.][]{2019ApJ...884..169G}. We then  compute the instantaneous mass growth rate onto each supermassive black hole by following the standard Bondi–Hoyle formalism, in which we include (as typically for simulations at this resolution) an ad-hoc "boost" parameter meant to compensate for the lack of resolution around the Bondi radius. Depending on the temperature of the accreted gas, we use either "cold gas accretion" feedback (in which most of the energy is distributed in the form of thermal energy in the neighborhood of each simulated AGN) or "hot gas accretion" feedback (in which most of the energy is released in the form of bipolar kinetic jets).
In both cases, $10\% $ of the feedback energy is released in the form of magnetic energy, through pairs of magnetised loops wrapped around the direction of kinetic jets. This magnetic field is added to a negligible uniform initial seed field of $B_0 = 10^{-11}\ \mathrm{nG}$ (comoving), leading to "magnetic bubbles" correlated with halos in the simulated volume. The two variations studied in this work concerns two different set of parameters for the efficiency of feedback from the hot and cold gas accretion and are calibrated to well reproduce the radio luminosity functions of real radio galaxies in the local Universe, with the "astroph 2" model having an overall slightly more efficient feedback energetics, when integrated over the entire lifetime of the simulation. A more detailed descriptions of all parameters used in this model, as well as of the main comparison with simulated and observed galaxy properties is the subject of a forthcoming paper (Vazza et al., submitted). 

 \item \textit{``mixed model''}: we combined the astrophysical model giving the largest contribution to the magnetisation of the cosmic web of the two discussed  above ("astroph 2"), and the stochastic primordial model that  gives the best match to LOFAR RRMs (see Sect.~\ref{sec:synth_rrm}), which is that with  $\alpha_s=-1.0$ and downscaled normalisation  $B_{\rm 1~Mpc}=0.37 \, \rm nG$.

\end{enumerate}

An overview of the numerical parameters of the tested magnetic fields in given in Table\ref{tab:sim}.
The adopted cosmological parameters are as for Sect.~\ref{sec:intro}.  
The production of these new simulations was motivated in order to produce long lines-of-sight (LOS) with a finely sampled redshift evolution of gas and magnetic field quantities from $z=3$ to $z=0$, which was not available in existing simulations. 

To allow a comparison with the observed RM, we  generated 100 LOS through each simulated volume, with information of gas density and 3D magnetic field from $z=2.98$ to $z=0$. Each LOS is $\approx 6.502$  comoving $\rm ~Gpc$ long and was produced by replicating the simulated volume 153 times, using a set of 17 snapshots saved at nearly equally spaced redshifts, and by randomly varying the volume-to-volume crossing position. In total, for each simulated model we extracted 100 LOS, each containing physical field values for  a total of $156,672$ cells. Our analysis shows that the RRM integrated over such long LOS gives very stable trends, and already with a sample of 100 LOS the uncertainties on the RRM within each redshift bin are $\leq 0.1\%$ (see e.g. Fig.\ref{fig:rm_synth}).

\begin{table}
	\centering
	\caption{Parameters of the primordial magnetic fields tested in our suite of cosmological simulations.  The columns are: (1) model name;  (2) spectral index of the magnetic power spectrum ("-" for uniform field setups); (3) amplitude of the comoving magnetic field smoothed on a comoving $1$~Mpc scale.
    } 
	\label{tab:sim}
	\begin{tabular}{ccc}
	  \hline 
   (1)  &  (2)  &  (3)  \\
 model &   $\alpha_s$ & $B_{\rm 1Mpc}$ \\
  &   & $[\rm nG]$   \\
 \hline
primordial uniform  & -  & $0.1$   \\
primordial stochastic   & $-1.0$ & $ 0.37 $ \\
primordial stochastic    & $0.0 $    & $0.35$   \\
primordial stochastic    & $1.0$     & $0.042$   \\
primordial stochastic    & $2.0$     & $0.003$   \\
astroph 1    & -    & $10^{-11}$  \\
astroph 2   & -    & $10^{-11}$   \\
astroph 2 + stochastic &$-1.0$  & $0.37$   \\
\hline
	\end{tabular}
\end{table}

   \begin{figure*}
   \centering
    \includegraphics[width=0.49\hsize]{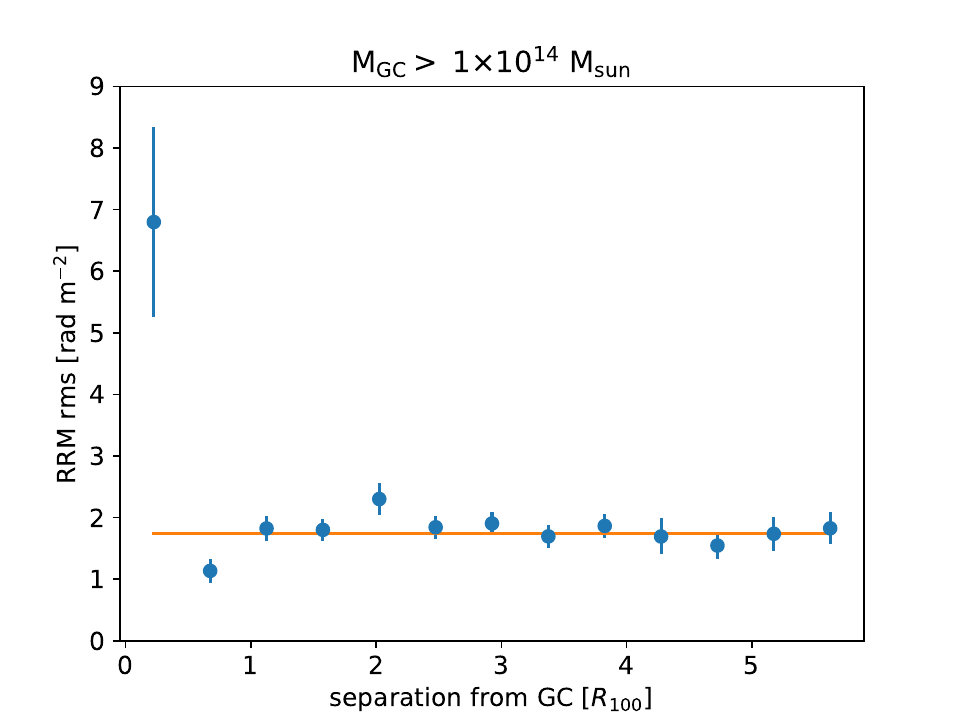}
    \includegraphics[width=0.49\hsize]{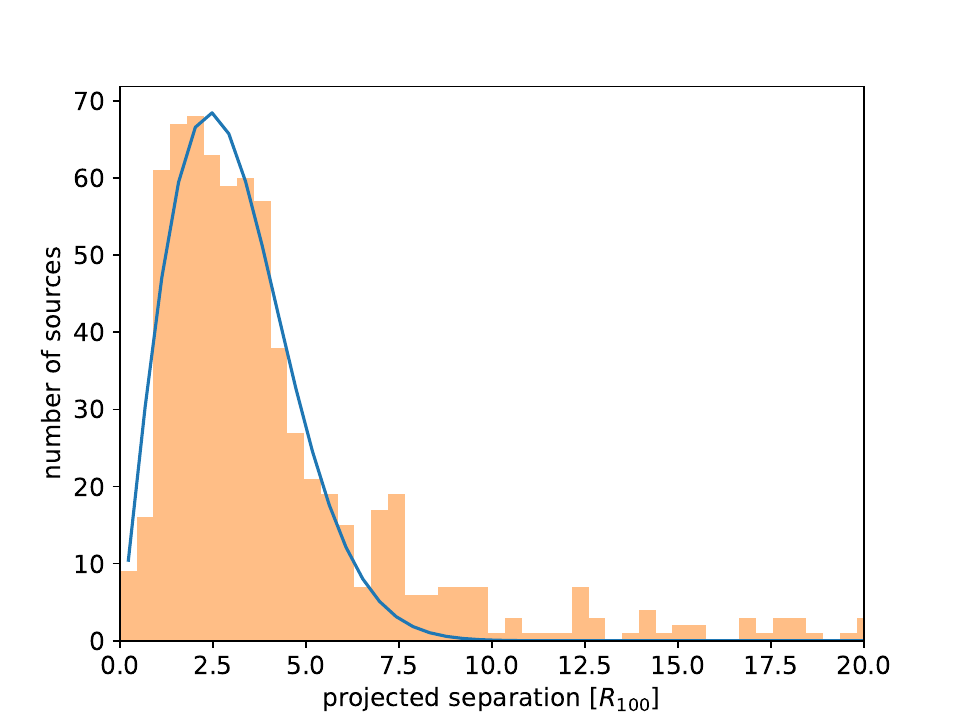}
    \includegraphics[width=0.49\hsize]{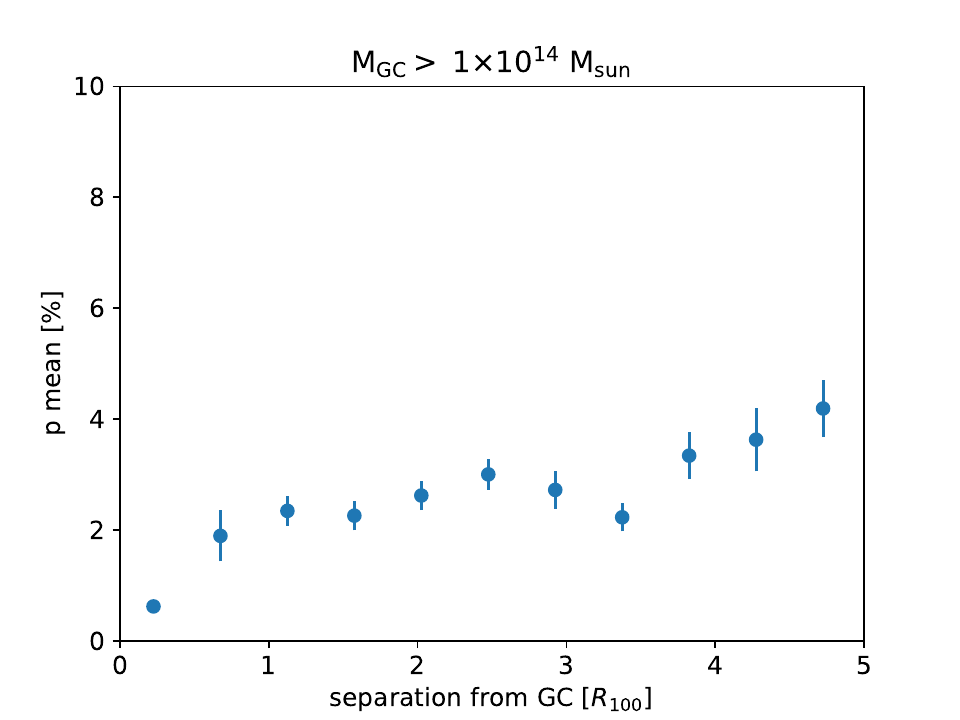}
    \includegraphics[width=0.49\hsize]{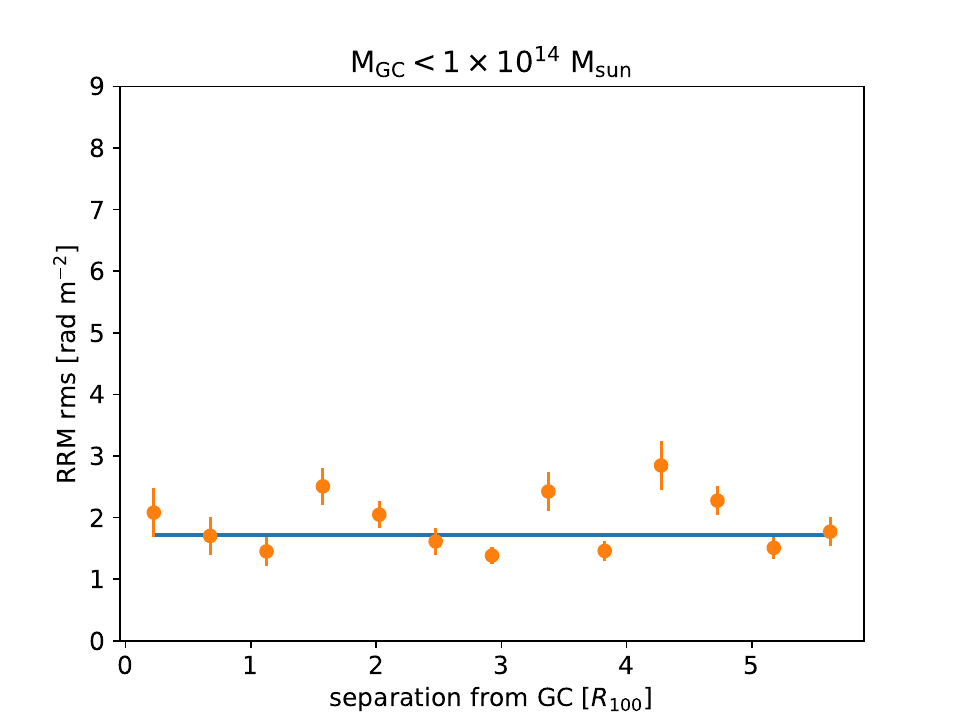}
   \caption{RRM rms  (top-left), distribution  (top-right), and mean fractional polarization (bottom-left) of the full spectroscopic sample at 144~MHz in   bins of the separation from the nearest galaxy cluster of mass $M > 1.0 \times 10^{14}\, \rm M_\odot$   along the LOS.  Bottom  right: Same as the for the top left panel except it is for  galaxy clusters of  mass   $M < 1.0 \times 10^{14}\, \rm M_\odot$.}
              \label{fig:clusters}%
    \end{figure*}
   \begin{figure}
   \centering
    \includegraphics[width=0.96\hsize]{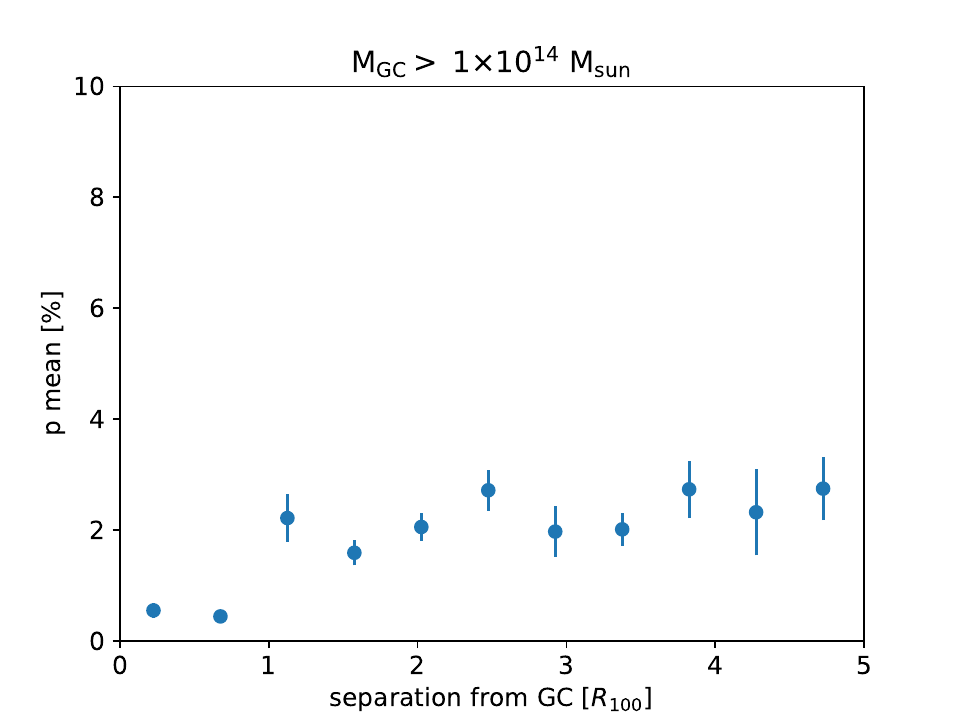}
   \caption{ Mean fractional polarization ($p$) at 144~MHz in bins of the impact parameter from the nearest galaxy cluster of mass $M > 1.0 \times 10^{14}\, \rm M_\odot$   along the LOS. The  case of the no GRM filter   sample restricted to the redshift range $z=0.5$--0.75 is shown. }
              \label{fig:clusters_p_highz}%
    \end{figure}
   \begin{figure}
   \centering
    \includegraphics[width=0.96\hsize]{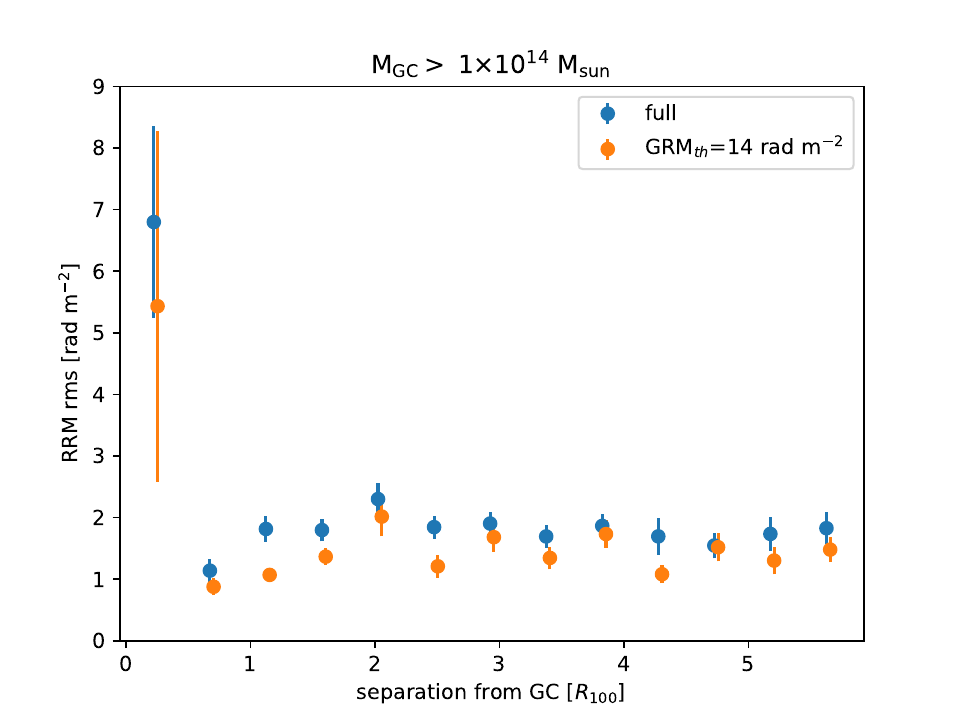}
   \caption{ RRM rms at 144~MHz in bins of the impact parameter from the nearest galaxy cluster of mass $M > 1.0 \times 10^{14}\, \rm M_\odot$   along the LOS. The full spectroscopic redshift  sample and  an RRM sample filtered by a GRM$_{\rm th}$ limit are shown. The cases are shifted  in separation for clarity. }
              \label{fig:clusters_grm}%
    \end{figure}
   \begin{figure*}
   \centering
    \includegraphics[width=0.49\hsize]{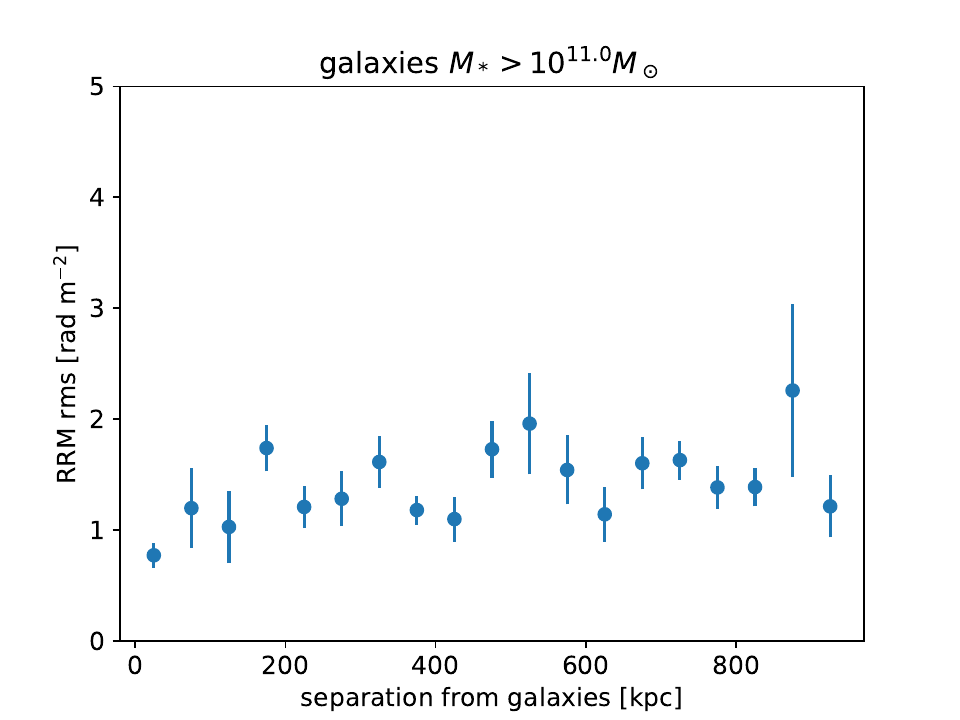}
    \includegraphics[width=0.49\hsize]{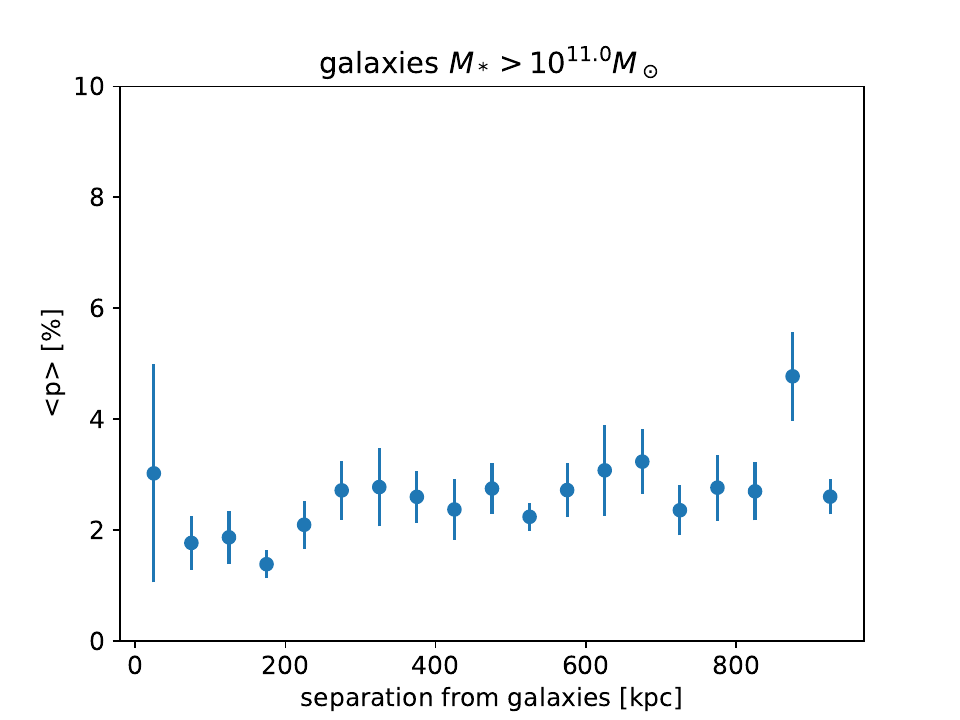}
    \includegraphics[width=0.49\hsize]{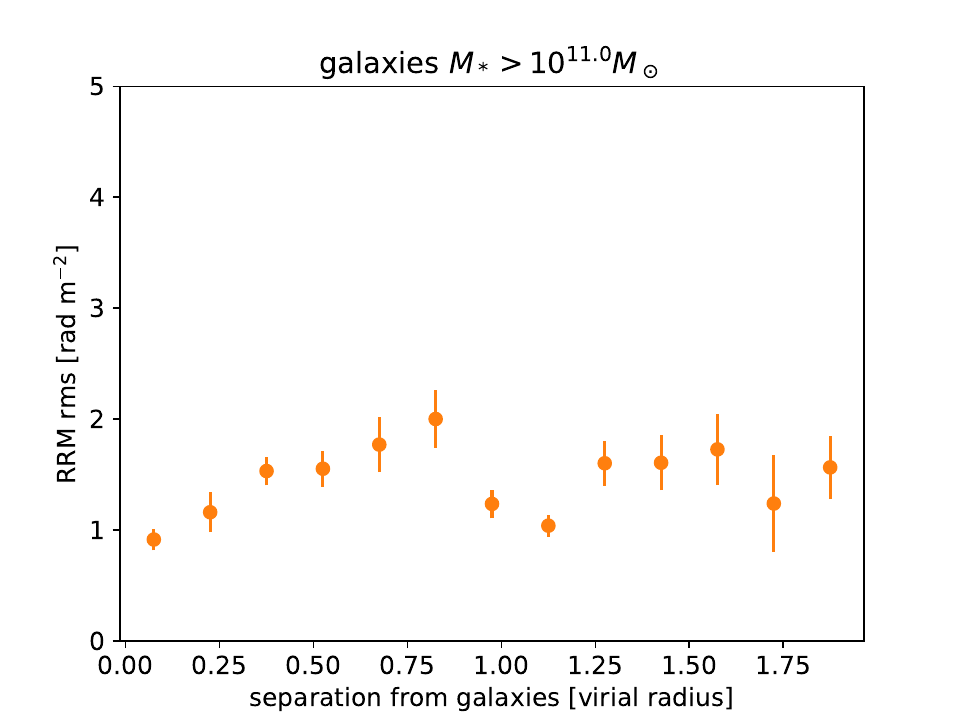}
    \includegraphics[width=0.49\hsize]{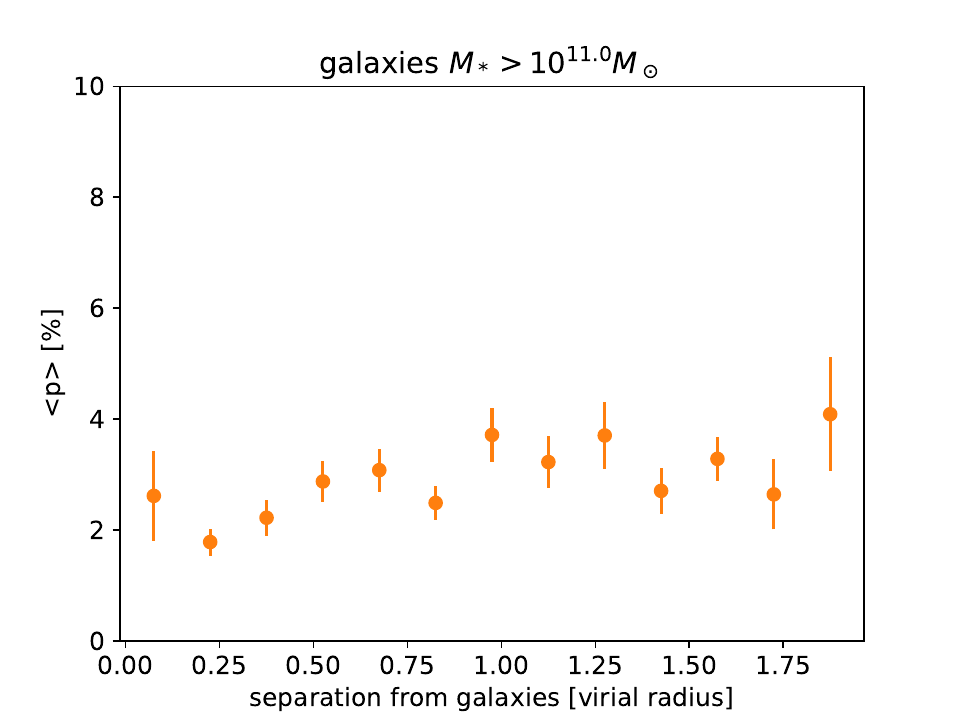}
   \caption{ RRM rms (top-left) and $<p>$  (top-right) of the LoTSS RRM sample with a GRM limit of 14~rad~m$^{-2}$ versus source separation in kpc from galaxies of stellar mass $M_* > 10^{11}\,M_\odot$.  The bin size is of  50 kpc. Bottom: as for top panels, except the separation is in virial radii, with bin size of 0.15 virial radii. }
              \label{fig:galaxies}%
    \end{figure*}
   \begin{figure*}
   \centering
    \includegraphics[width=0.49\hsize]{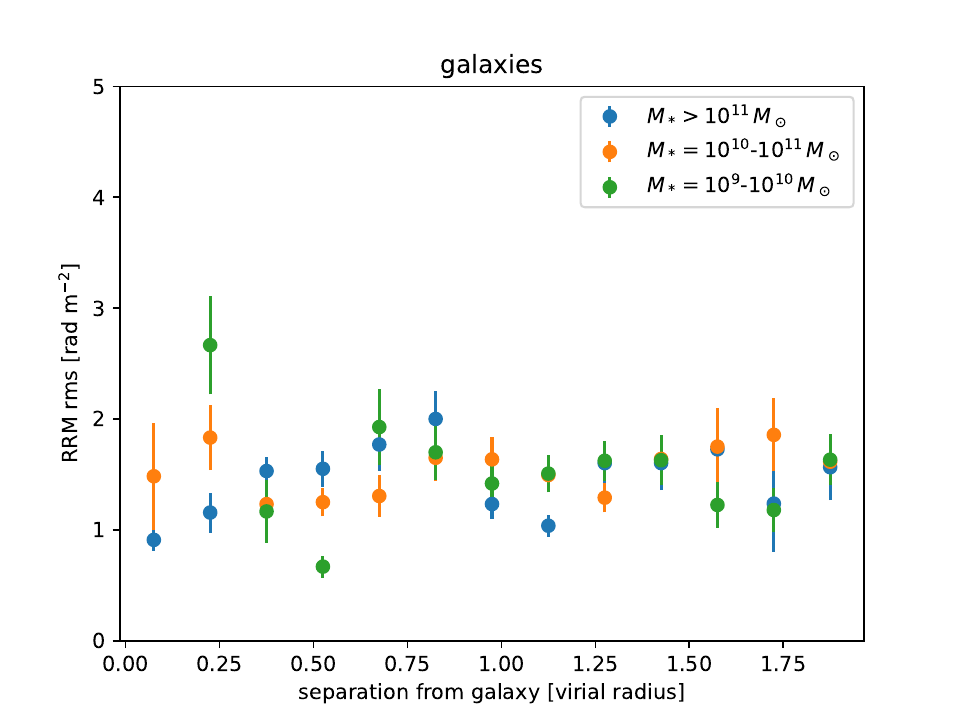}
    \includegraphics[width=0.49\hsize]{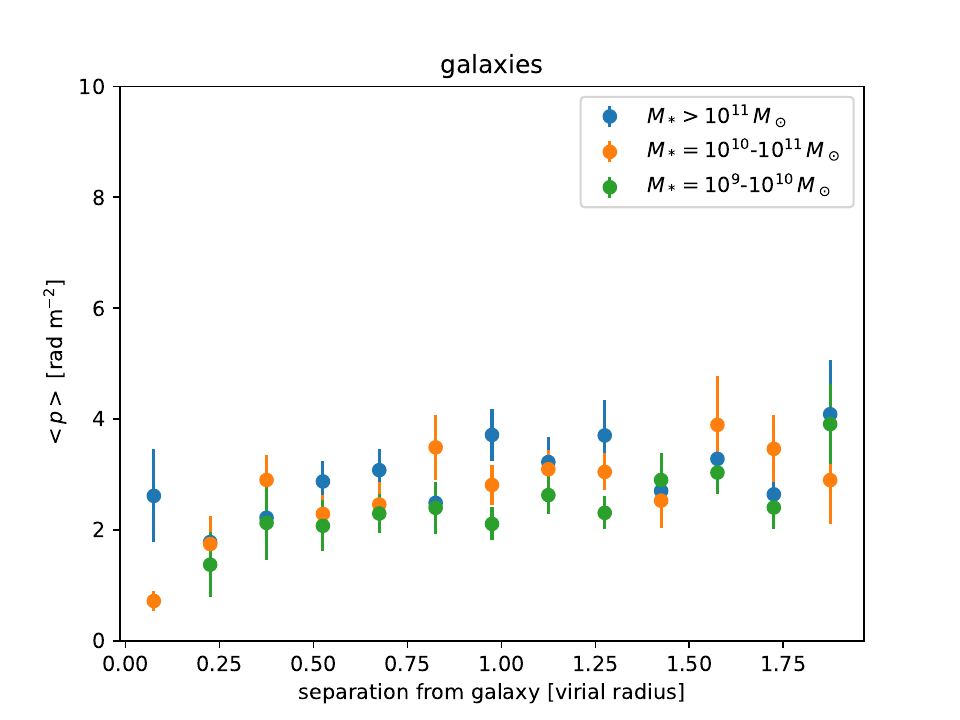}
   \caption{ RRM rms (left) and  $<p>$ (right)  of the LoTSS RRM sample with a GRM limit of 14~rad~m$^{-2}$ versus LOS projected separation in virial radii from galaxies of the three samples of stellar mass $M_* > 10^{11}\,M_\odot$, $M_* = 10^{10}$--$10^{11}\,M_\odot$, and $M_* = 10^{9}$--$10^{10}\,M_\odot$.  The bin size is of  0.15 virial radii. }
              \label{fig:galaxies_mass}%
    \end{figure*}
\section{Origin of  RRM at  low frequency }
\label{sec:tests}
In Paper I and II we addressed the origin of the RRMs of our sample at 144~MHz. Analysing the evolution with redshift   of the polarization fraction $p$ and the behaviour of RRM rms with $p$,  we found that a cosmic web origin is favoured, as opposed to an astrophysical origin. We also estimated the separation of our sources and their  LOS to  galaxy clusters and found that they tend to be  far from clusters. Thus, the source LOS tend to travel through   low density environments, where the  depolarization is lower.  

 We also used the results of differential RMs of  random pairs  (sources with  close projected separation, but  not physically associated and at different redshifts) and physical pairs (sources physically associated at the same redshift, such as two lobes of a radio galaxy) by \citet[][]{2022MNRAS.515..256P} and found that a cosmic web contribution  is dominant. The  astrophysical origin term is  estimated to be $\approx 10$ percent. 

All of the above points to an origin that is mostly from the cosmic web. The  RRM from the cosmic web is largely dominated by the filament contribution because their  RRM is much larger than   that of  voids, according to MHD cosmological simulations (Paper II).  

In this Section we further investigate the RRM origin at 144~MHz.

\subsection{Galaxy clusters}
\label{sec:clusters}

We test whether the galaxy clusters can contribute to the RRMs we measure. 
If the RRMs are of cosmic origin it is expected that the RRM rms would be larger in lines of sight passing through a cluster, instead of solely through filaments and voids.

We first analyse the full spectroscopic redshift sample with no GRM$_{\rm th}$ filter to investigate the origin  of the RRMs used in our two previous papers.

For each RRM source we found the galaxy cluster with the smallest impact parameter (or projected separation from the source LOS) in $R_{100}$ units\footnote{$R_{y}$ is the cluster radius within  which the mean matter density is  $\bar{\rho}_M = y\,\rho_c$, where  $\rho_c$ is the critical density of the Universe.  }. We then  computed the RRM rms as a function of this separation.  
We used  the galaxy cluster  catalogue of \citet{2015ApJ...807..178W}. 
The catalogue   contains 158,103 records in the redshift range of 0.05--0.75  with an error of up to 0.018. The cluster  masses ($M_{500}$\footnote{$M_y$ is the mass within $R_y$.}) are as low as $2.0\times 10^{12}\, \rm M_\odot$  and the catalogue  is 95 percent  complete for masses greater than $1.0\times 10^{14}\, \rm M_\odot$. We selected sources of our RRM sample  that are in the galaxy cluster catalogue  redshift range, for a total of 707 sources. The RRM rms of this sample is  $1.93\pm 0.06$~rad~m$^{-2}$.  We only used  massive clusters (mass $M > 1.0 \times 10^{14}\, \rm M_\odot$) that are expected to give the largest RRM  effect. The rms is computed excluding 2-$\sigma$ outliers.\footnote{  Outliers were not flagged out when the number of sources in a bin was 5 or less, to avoid false outliers due to the poor statistics.} 

The rms as a function of separation is computed in bins of 0.45~$R_{100}$ (which is approximately $R_{500}$). Figure~\ref{fig:clusters}, top left panel, shows the RRM rms as a function of the minimum projected separation from clusters.  We find an obvious excess of $6.8\pm 1.5$ ~rad~m$^{-2}$ in the first bin (out to 0.45 $\rh$), which contains 9 sources, whilst  beyond  0.45 $\rh$ there is no obvious trend, sitting at  a mean value of $1.75\pm 0.06$~rad~m$^{-2}$.  The latter is $9$ percent lower than the RRM rms of the entire sample and is an estimate of the astrophysical term contribution by intervening galaxy clusters to our RRM rms, for the full sample. 
Note that the value of 6.8~rad~m$^{-2}$ is consistent with the rms of $\approx 7$~rad~m$^{-2}$ of extragalactic sources observed at 1.4 GHz \citep[e.g.,][]{2015A&A...575A.118O, 2019MNRAS.485.1293S}.

Figure~\ref{fig:clusters}, bottom left panel, reports the mean polarization fraction  in the same bins. The  first bin has a significantly low value of $p = 0.62 \pm 0.12$ percent that is close to  the limit of detection of  our observations. The few sources whose LOS go through  clusters at 144~MHz are thus  highly depolarized and on the verge of not being detected. They are likely the few sources that survive such depolarization and this can explain why so few are detected in polarization.  The polarization fraction rises to $\approx 2$ percent just beyond 0.45 $\rh$ and then it increases nearly linearly with the separation. This increase might indicate that the environment density is decreasing with the separation to clusters leading to lower depolarization. A small dip can be seen at $\approx 3\,\rh \approx 6\, R_{500}$, the origin of which is still unclear. 

Our result of high depolarization of sources in the background of galaxy clusters is consistent with the results obtained at 1.4 GHz \citep{2010A&A...514A..71V, 2011A&A...530A..24B, 2022A&A...665A..71O}. However, while at 1.4 GHz high depolarization is restricted to the inner regions of the cluster,  here at low frequency it stretches out to  the cluster outskirts, to $\approx R_{500}$.

We checked whether there is a redshift dependence on how far out the depolarization stretches. We  split the sources  in two redshift bins, lower and higher than  0.5, where we find 4 and 5 sources in the first bin. While the lower redshift bin has the depolarization still limited to the first bin out to $0.45 \rh$, in the higher redshift bin ($z=0.5$--0.75) the depolarization stretches out to the second separation bin (Fig. \ref{fig:clusters_p_highz}), out to $0.9 \rh \approx 2\,R_{500}$, which is approximately the virial radius. That means that strong depolarization effects are produced by the entire cluster. 
However,  we verified that the  RRM excess is still limited to the first separation bin. 

Figure~\ref{fig:clusters}, bottom right panel,  shows the RRM rms when clusters of mass  $M < 1.0 \times 10^{14}\, \rm M_\odot$ are used, which  comprises  poor clusters and groups.  No excess   in the first bin out to 0.45 $\rh$ shows up in such a case.  The fractional polarization  $p$ raises  to $2.7\pm 0.7$ percent. These  results mean that groups  and poor clusters  give a marginal  effect at this frequency. 

The flat trend beyond 0.45 $\rh$, as opposed to the excess within massive clusters,  supports a cosmic web origin outside cluster environments. If the RRM rms  depended on the local environment, then a decrease with the separation would be observed because of a decrease of density, while a dependence on all the cosmic filaments in the source foreground would result in a term uncorrelated with the separation from the cluster. 

The source distribution as a function of the separation from  clusters ($x$) (Fig.~\ref{fig:clusters}, top right panel) is well approximated by a Rayleigh distribution, which is the distribution of the magnitude of a 2D-vector, as it is the projected separation on the plane of the sky:
\begin{equation}
    P(x)=A\, \frac{x}{\sigma_r^2} \, e^{ -\frac{x^2}{2\sigma_r^2}},
    \label{eq:rayleigh}
\end{equation}
where $\sigma_r$ is  the distribution spread, and  $A$ is a normalisation term. The best fit (solid line) gives $\sigma_r = 2.42\pm 0.05\,\,\rh$. The cumulative distribution function is : 
\begin{equation}
    F(x)= 1 -e^{ -\frac{x^2}{2\sigma_r^2}},
    \label{eq:rayleigh_cum}
\end{equation}

We also measured the RRM rms as a function of the separation from galaxy clusters for  the sample filtered  by the  GRM$_{\rm th}$ limit of 14~rad~m$^{-2}$ (Fig.~\ref{fig:clusters_grm}), which is the sample used for the analysis in this work. The RRM rms of the sources in the redshift range of the cluster catalogue is $1.56\pm 0.06$~rad~m$^{-2}$. 
The RRM rms of the first bin is of  $5.4\pm 2.8$~rad~m$^{-2}$,  which is consistent with that of the full sample case discussed above.

The mean RRM rms excluding the first bin gets down to $1.26\pm 0.05$~rad~m$^{-2}$. This is $19\pm 4$ percent lower than that of the sample and can be considered as the fraction of the astrophysical RRM  component from intervening clusters in the sample used in the analysis of this work. 

It is worth noticing a hint of a secondary excess at $\approx 2 \rh \approx 4 R_{500}$ (2.1 $\sigma$ and 2.4 $\sigma$ significance for the no GRM filter and GRM$_{\rm th} = 14$ rad m$^{-2}$ samples). This is possibly related to the presence of companion clusters, as previously found in the density and metallicity profiles obtained from  simulations \citep{2022A&A...663L...6A, 2023A&A...675A.188A}. It is unlikely related to the source excess found close to the cluster virial radius \citep{2024arXiv240216946I}, which is closer than   $ 2 R_{100}$.  

The mean $p$ of the first bin is $0.77\pm 0.19$ percent, again showing a small polarization fraction as in the unfiltered  sample.  The Rayleigh distribution spread   is  similar to that of the no filter case, at $\sigma_r = 2.51\pm 0.06 \,\rh$.

\subsection{Galaxies and CGM}
\label{sec:galaxies}

Here we analyse whether galaxies or their CGM give a contribution to the RRM sample  we use in this work. This is motivated by the recent finding that nearby spiral  galaxies can produce  a measurable RRM excess at 144~MHz, if the LOS of a background  extragalactic source goes through their  extraplanar, magnetised outflows from the inner part of the galaxies close to their minor axis  \citep{2023A&A...670L..23H}. In such galaxies the excess RRM is detectable out to 100 kpc. 

We  analyse the  RRM spectroscopic redshift sample with  a GRM$_{\rm th}$ filter set to 14~rad~m$^{-2}$.

For each RRM source we estimated the impact parameter of the LOS to the galaxy with the smallest  impact parameter in kpc  units, and then we computed the RRM rms as a function of that LOS  separation.  
We considered  the galaxy   catalogue of\footnote{https://cdsarc.u-strasbg.fr/viz-bin/cat/J/ApJS/242/8} \citet{2019ApJS..242....8Z}.  The catalogue is based on the photometric survey Dark Energy Spectroscopic Instrument (DESI)  Legacy Surveys and contains photometric redshifts and stellar masses ($M_*$). It  consists of   approximately $3.03\times 10^8$ records with photometric redshifts up to $z \approx 1$ and errors that depend mostly on their brightness. The median redshift errors for a few $M_*$ limited samples are  reported in Table~\ref{tab:gal_sample}.   The galaxy catalogue  is complete for $M_* > 10^{11}\,M_\odot$ (see Fig. 12 of \citealt{2019ApJS..242....8Z}). We selected sources of our RRM sample    that are in the galaxy catalogue redshift range, a total of 550 sources. The RRM rms of such a subsample is $1.53 \pm 0.05$~rad~m$^{-2}$. We only used  massive galaxies (stellar mass $M_* > 10^{11}\, \rm M_\odot$) that are expected to give the largest effect. The main features of this subsample are reported in Table~\ref{tab:gal_sample}. The RRM rms is computed excluding outliers, as for Sec.~\ref{sec:clusters}.  

\begin{table}
	\centering
	\caption{Main features of the galaxy subsamples used in Sec.~\ref{sec:galaxies}. The columns are: (1) the galaxy $M_*$ range of the subsample; (2)
 subsample  median redshift error ($\sigma_z$); (3) median $M_*$; (4) subsample number of galaxies in the RRM sample footprint ($N_{\rm gal}$). 
    } 
	\label{tab:gal_sample}
	\begin{tabular}{cccc}
	  \hline 
   (1)  &  (2)  &  (3)  &  (4)  \\
 $M_*$ range &   median $\sigma_z$  & median $M_*$   & $N_{\rm gal}$ \\
   $[M_\odot$]   &    &   [$M_\odot$]   &    \\
 \hline
$> 10^{11} $            &    0.042        &     $10^{11.2}$   &  $4.7\times 10^6$ \\
$10^{10}$--$10^{11}$    &    0.060      &      $10^{10.5}$    &  $29\times 10^6$ \\
$10^{9}$--$10^{10}$     &   0.090       &     $10^{9.6}$    &  $30\times 10^6$ \\
\hline
	\end{tabular}
\end{table}

   \begin{figure}
   \centering
    \includegraphics[width=0.96\hsize]{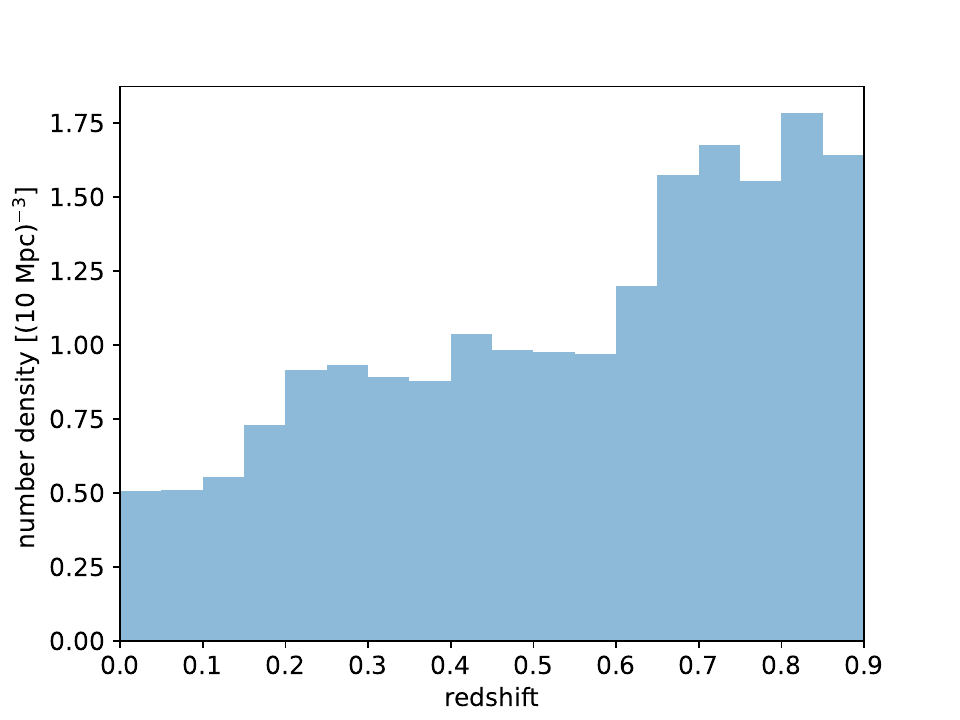}
   \caption{Galaxy number density as a function of redshift from the sample of galaxies with $M_* > 10^{11}$ M$_\odot$ obtained from the DESI Legacy Surveys photometric galaxy catalogue. }
              \label{fig:number_density}
    \end{figure}

   \begin{figure}
   \centering
    \includegraphics[width=0.96\hsize]{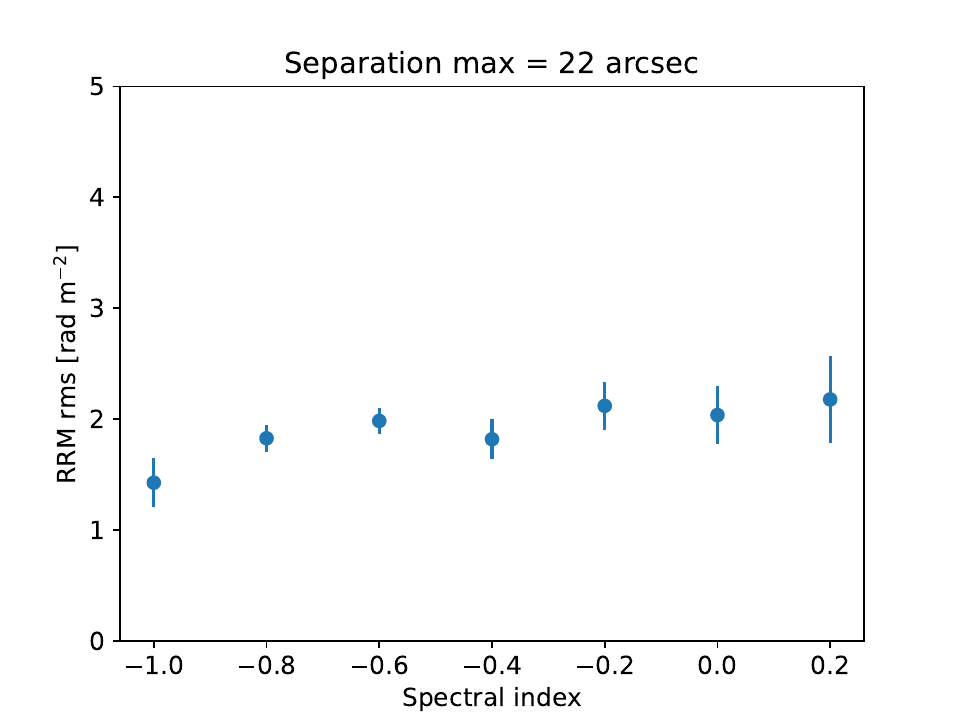}
    \includegraphics[width=0.96\hsize]{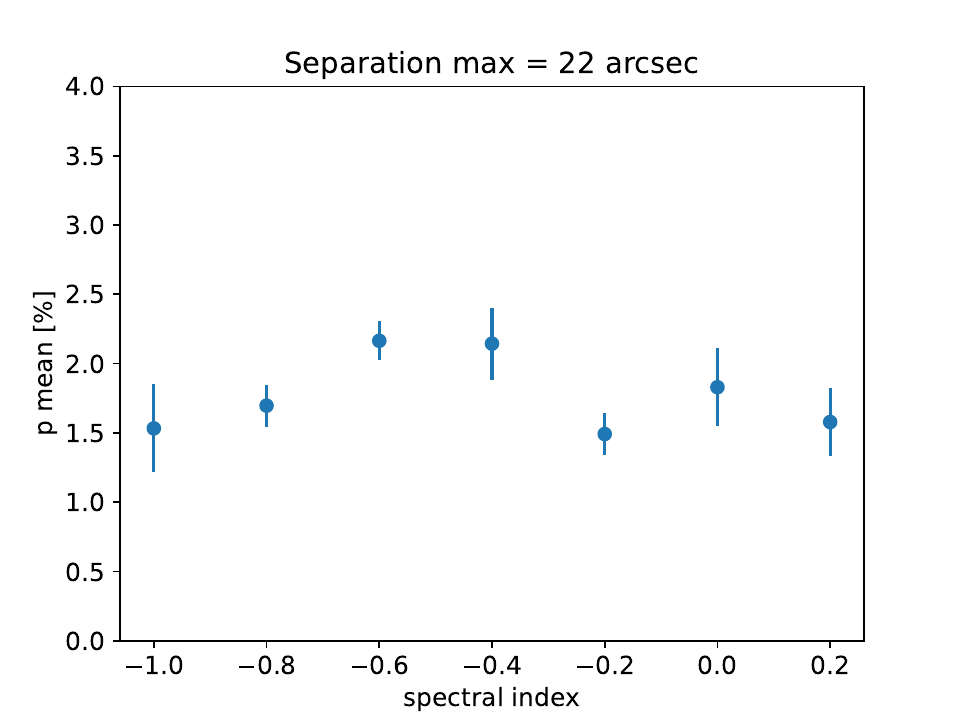}
    \includegraphics[width=0.96\hsize]{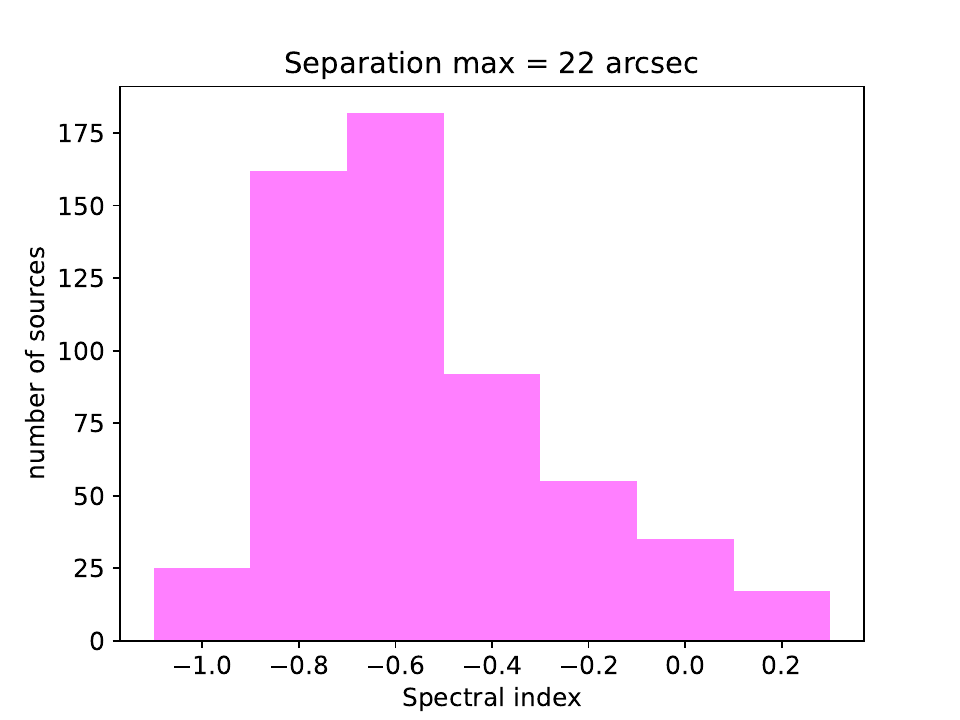}
   \caption{ RRM rms of the LoTSS RRM sample  as a function  of  spectral index obtained as described in the main text ({\it top}), polarization fraction ({\it mid}), and spectral index distribution of the sources we used for this analysis ({\it bottom}). We used bins of a width of 0.2. }
              \label{fig:spec}
    \end{figure}

   \begin{figure}
   \centering
    \includegraphics[width=0.96\hsize]{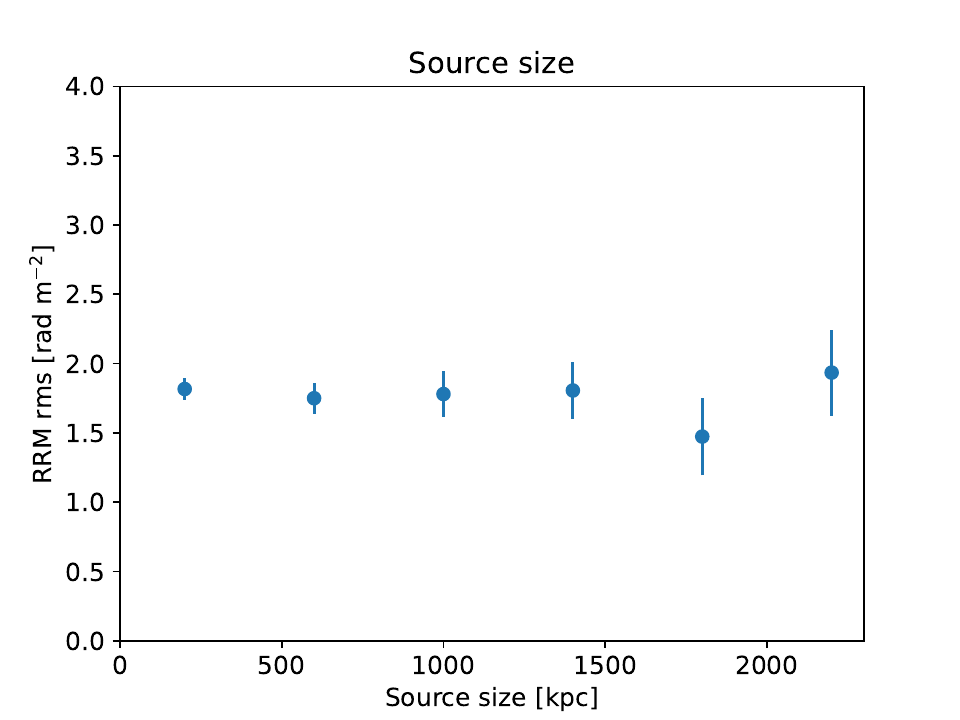}
    \includegraphics[width=0.96\hsize]{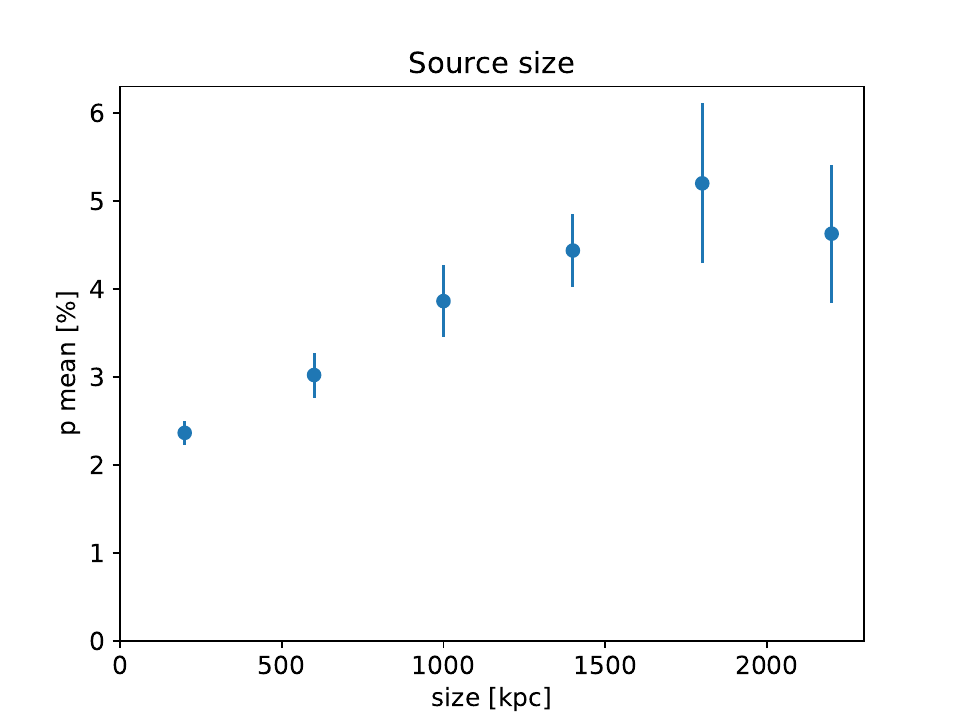}
    \includegraphics[width=0.96\hsize]{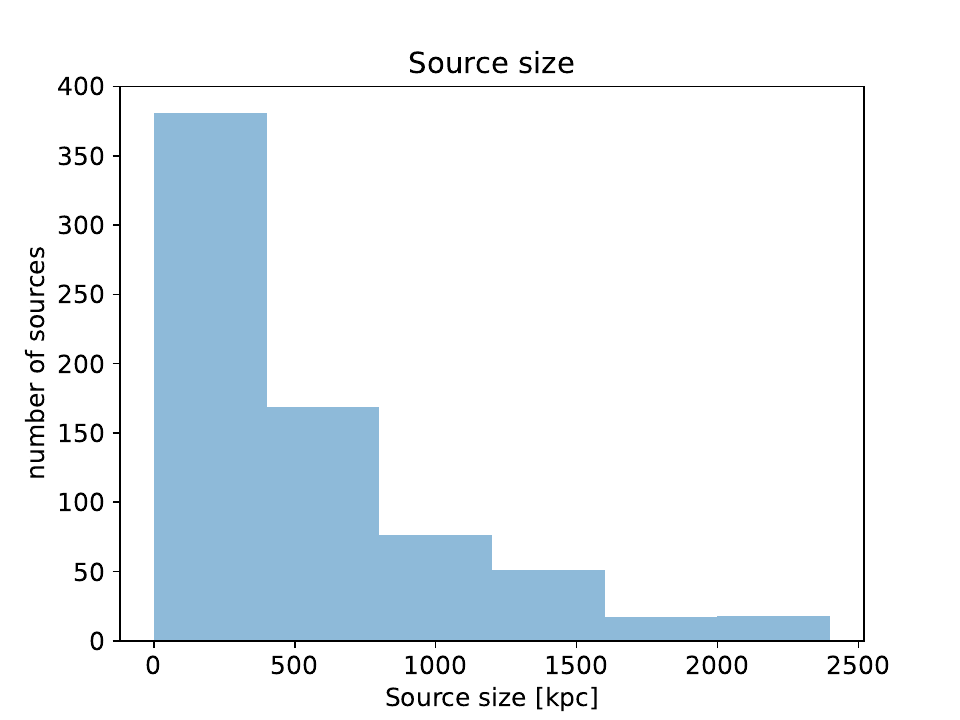}
   \caption{ RRM rms of the LoTSS RRM sample versus source linear size ({\it top}), mean fractional polarization ({\it mid}), and  linear size  distribution of  the sources we used for this analysis ({\it bottom}). We used bins of a width of 400 kpc. }
              \label{fig:size}%
    \end{figure}

Fig.~\ref{fig:galaxies}, top panels, shows the RRM rms and  the mean $p$ as a function of the separation   from  galaxies in kpc. The bin size is 50 kpc. We do not observe any RRM  excess in either the first bin out to 50 kpc nor in the second bin at 50--100 kpc. Also larger  separations  do not show any specific trend out to 1~Mpc. Our sample is not  spiral galaxy-dominated, neither are the LOS selected to  pass  close to   galaxy minor axis, thus our result of not finding any excess at short  impact parameters  is not inconsistent with the results of \citet[][]{2023A&A...670L..23H}.  The fractional polarization is at about 2 percent in the first bins and then it slightly increases towards larger separations up to some 3 percent. 

The CGM extension depends on the halo mass ($M_h$) and redshift of the galaxy and these stacking plots mix different fractional separation  within halos. Therefore, we also compute the same quantities in virial radius units (Fig.~\ref{fig:galaxies}, bottom panels), that we expect to be more related to the CGM extension than the physical separation. 
For each galaxy, we estimate $M_h$ from $M_*$ by the relation \citep{2020A&A...634A.135G}
\begin{equation}
    M_* = 2M_h \, A(z)\left[ \left(\frac{M_h}{M_A(z)}\right)^{-\beta(z)} + \left(\frac{M_h}{M_A(z)}\right)^{\gamma(z)} \right]^{-1}
\end{equation}
where  the best-fit parameters $A$, $M_A$, $\beta$, and $\gamma$ are listed in Table 1 of \citet[][]{2020A&A...634A.135G}{}{} as a function of redshift ranges. The virial radius is then estimated from  $r_v = \rh$ \citep{2014efxu.conf..362R} and using the equation 
\begin{equation}
    \frac{M_h}{4/3\, \pi r_v^3} = 100 \, \rho_c(z),
\end{equation}
where $\rho_c$ is the critical density of the Universe. 

Again, there is no RRM excess closest to the galaxies. Intriguingly, the RRM tends to increase out to $\approx 0.8\,r_v$. Further out, the RRM decreases to a mean value of $1.41\pm 0.06$~rad~m$^{-2}$, which is $8\pm5$ percent lower than that of the sample. This can be considered an additional contribution of astrophysical origin. The RRM rms adds up quadratically, thus we can estimate the total astrophysical  component contribution, intervening clusters and galaxies, at $21\pm 4$ percent.   The inner part of the RRM plot is difficult to interpret.  On average, it sits  about at the same value as the outer part. The first two bins are lower than average, while the two bins close to 0.8 $r_v$ are in excess. However, both the lower part and the excess are at low significance and  might just be a statistical fluctuation.  An intriguing possibility of the excess is as a signature of a shock at about the virial radius, perhaps infalling matter from the cosmic web surrounding the galaxy, or outflowing matter from the galaxy. Assuming plausible values of  galaxy far outskirts  for  magnetic field ($B=0.1$ $\mu$G), integration path ($l =50$--200 kpc), and electron number density ($n_e =10^{-4}$ cm$^{-3}$), we get RRMs of 0.4--1.6~rad~m$^{-2}$ at $z=0$ and  0.1--0.4~rad~m$^{-2}$ at $z=1$, which are consistent with the excess we see here.  We do not have plausible explanations for the first two lower bins, at this stage. Further investigations are required,  with a larger sample for smaller  errors and detailed  modelling, especially of the  smoothly decreasing RRM rms  towards short separations. However, this is beyond  the scope of this work.

The fractional polarization is at $\approx 2$ percent in the first bins, then it raises to 3--4 percent at $\approx 1.0\, r_v$, where it flattens.   The stronger depolarization at shorter separations might be due to a turbulent medium in the CGM. We argue that $<p>$  of background sources might be an efficient way to trace the CGM and its extension. However, further  investigations are required to confirm this.

We estimate the RRM rms and $<p>$ as a function of the separation from galaxies in $r_v $ units  for  two other galaxy subsamples, with $M_* = 10^{10}$--$10^{11}\, \rm M_\odot$ and $M_* = 10^{9}$--$10^{10}\, \rm M_\odot$ (Fig.~\ref{fig:galaxies_mass} and Table~\ref{tab:gal_sample}). These RRM profiles  do not  show the trend at separations within $r_v$ of the most massive subsample. The only possible feature is a hint of an excess ($\approx 2.5\,\sigma$) at a separation of $\approx 0.25\, r_v$ in the $M_* = 10^{9}$--$10^{10}\, \rm M_\odot$ subsample. With the caveat it has low significance, this excess   might have a different  origin compared to that found by \citet{2023A&A...670L..23H} who  do not find any excess in low mass galaxies.

As for the $M_* > 10^{11}\, \rm M_\odot$ case, the mean $p$ raises from the shortest  separations and  then it flattens out for the $M_* = 10^{10}$--$10^{11}\, \rm M_\odot$  subsample.  We can see no obvious trend for the  $M_* = 10^{9}$--$10^{10}\, \rm M_\odot$ subsample.  

We repeated the analysis flagging the galaxies with redshift error higher than either   3 times or 2 times  the median error, to avoid sources with too large uncertainties, and we obtained the same results. 

We also used our  same galaxy sample limited to galaxies  with $M_* > 10^{11}$ M$_\odot$ to investigate possible correlations with the RRM  wiggles of Fig.~\ref{fig:zdisp}, right panel, that approximately peak at redshift of  0.15, 0.36, and 0.56. The galaxy number density as a function of redshift is shown in Fig.~\ref{fig:number_density}. There are three dips   at redshifts  of 0.125, 0.375, and 0.575 (with an uncertainty of 0.25, half the bin size), that well match  the positions of the wiggles. The wiggles are interleaved by peaks of the galaxy number density, approximately matching the local minima of RRM rms vs redshift (Fig.~\ref{fig:zdisp}). The first minimum of RRM rms, at $z\approx 0$, corresponds to a raise of the number density. 
Thus, there is an anti-correlation. A RRM component of  local origin is unlikely, because of the lower source  density  at the wiggles peaks.  We do not have an obvious explanation for such a fascinating behaviour and further investigations will be needed.  

\subsection{Dependence of RRM on the source spectral index}
\label{sec:spec}

The spectral index of radio sources  can discriminate radio galaxies from blazars, steeper and flatter, respectively, and can also depend on the local environment  \citep[e.g.][]{1991MNRAS.249..343L}. The RRM is expected to be higher for more depolarized sources and hence to be stronger for blazars and for the more distant lobe  of a radio galaxy \citep{1988Natur.331..149L, 1988Natur.331..147G}. We thus would  expect the RRM to have a dependence on the spectral index in case that  our RRMs at low frequency would have a significant local component.

We explored the dependence of RRM on the spectral index of sources to assess whether there is a local origin. We used the spectral indexes measured by \citet{2018MNRAS.474.5008D}, who employed  data from the TIFR GMRT Sky Survey (TGSS) at 147~MHz \citep{2017A&A...598A..78I} and the NRAO VLA Sky Survey (NVSS) at 1.4 GHz \citep{1998AJ....115.1693C}. 

We cross-matched the spectral index catalogue with the position of our  sample of polarized sources with no GRM filter. We used a maximum  separation of 22 arcsec, because the NVSS beam-size is 45 arcsec and  Stokes I  and  polarized emission could be offset, finding a cross-match for 576 sources.\footnote{We only used the sources of \citet[][]{2018MNRAS.474.5008D} with safer spectral index, that is  the records with keyword S\_{code} labelled as S or M.}

Figure~\ref{fig:spec}, top panel, reports  RRM rms  versus  spectral index in bins of 0.2. Outliers beyond  2-$\sigma$ are flagged. The bottom panel shows the source distribution with the spectral index. There is no  obvious trend,  looking flat within the errors.   We checked three other cases:  setting no maximum separation of the cross-match, thus using all sources of the original RM catalogue; setting a maximum separation of 7 arcsec for  a safer cross-match; an RRM sample with a |GRM| limit of 14~rad~m$^{-2}$. In all cases, we obtain similar results. 

The mean fractional polarization  also shows no obvious trend  (Fig.~\ref{fig:spec}, mid panel).

\subsection{Dependence of RRM on the source linear size}

We also checked for a dependence of the RRM rms on the source linear size. The linear size information is  contained in the LoTSS RM catalogue for 725 sources of our unfiltered, spectroscopic redshift sample.  Figure~\ref{fig:size}, top and bottom panels, show  the RRM rms as a function of  the source linear size  and the corresponding distribution. Outliers beyond  2-$\sigma$ are flagged. The RRM rms behaviour is flat.

The mean fractional polarization increases  with the linear size (Fig.~\ref{fig:size}, mid panel). Giant radio galaxies, with sizes larger than 700 kpc,  have the  largest values. This is possibly because the largest radio galaxies have a higher chance to exit the densest regions of their local environment and the filaments spines, and are thus affected by smaller amounts of depolarization. 

We also tested the RRM sample limited to |GRM|~$<$~14~rad~m$^{-2}$, obtaining similar results. 

The lack of correlation of the RRM rms with source linear size and spectral index (Sect.~\ref{sec:spec}) indicates the absence of a clear dependence of the total RRM  on source properties.  This  again  points  to our RRMs   being dominated by the non-local, cosmic web component.

\section{Results}
\label{sec:zevo}

\subsection{Strength and Evolution with redshift of  filament magnetic fields}
\label{sec:fit}

\begin{table*}
	\centering
	\caption{Best-fit parameters of the filament magnetic field strength and  evolution with redshift  to the measured  RRM rms  at 144-MHz of the sample filtered with GRM$_{\rm th} =14$ rad m$^{-2}$. The  case with density taken from simulations and constant magnetic field strength is reported. Columns are the magnetogenesis model  studied ({\it astroph} and {\it astroph 2} are the first and second astrophysical models discussed in Section \ref{sec:simul}) and the fit parameters:  the slope $\alpha$ of the filament magnetic field  behaviour as a function of redshift; the   strength $B_{f,0}$ of the filament magnetic field at $z=0$;  the astrophysical component  term $A_{rrm}$;  the slope $\beta$ of the comoving magnetic field.  All cases are fit to the RRM rms computed in 20-source redshift bins. The astrophysical component term is assumed to be of shape $A_{rrm}/(1+z)^2$} 
	\label{tab:rrmf_fit}
	\begin{tabular}{lcccc}
	  \hline 
        model & $\alpha $ &  $B_{f,0}$ &  $A_{rrm}$  &   $\beta$  \\ 
           &   & [nG]  & [rad m$^{-2}$]  &   \\ 
       \hline   
       stochastic $\alpha_s$ = -1.0  & $2.1 \pm 0.4$ & $20 \pm 4$ & $1.14 \pm 0.10$ & $0.1 \pm 0.4$ \\ 
       stochastic $\alpha_s$ = 0.0 & $1.9 \pm 0.4$ & $24 \pm 5$ & $1.12 \pm 0.10$ & $-0.1 \pm 0.4$ \\ 
       stochastic $\alpha_s$ = 1.0 & $1.8 \pm 0.4$ & $24 \pm 5$ & $1.10 \pm 0.11$ & $-0.2 \pm 0.4$ \\ 
       stochastic $\alpha_s$ = 2.0 & $1.8 \pm 0.4$ & $25 \pm 5$ & $1.11 \pm 0.11$ & $-0.2 \pm 0.4$ \\ 
       astroph & $1.8 \pm 0.4$ & $27 \pm 6$ & $1.06 \pm 0.12$ & $-0.2 \pm 0.4$ \\ 
       astroph 2 & $1.9 \pm 0.4$ & $24 \pm 5$ & $1.02 \pm 0.11$ & $-0.1 \pm 0.4$ \\ 
       astroph 2 + stochastic $\alpha_s$ = -1.0 & $1.7 \pm 0.4$ & $27 \pm 6$ & $1.13 \pm 0.10$ & $-0.3 \pm 0.4$ \\ 
       uniform & $1.9 \pm 0.5$ & $24 \pm 6$ & $1.12 \pm 0.11$ & $-0.1 \pm 0.5$ \\ 
       \hline 
	\end{tabular}
\end{table*}

\begin{table*}
	\centering
	\caption{As for Table \ref{tab:rrmf_fit} except a shape  $A_{rrm}/(1+z)^3$ is used  as astrophysical component term.  } 
	\label{tab:rrmf_fit_arrm3}
	\begin{tabular}{lcccc}
	  \hline 
        model & $\alpha $ &  $B_{f,0}$ &  $A_{rrm}$  &   $\beta$  \\ 
           &   & [nG]  & [rad m$^{-2}$]  &   \\ 
        \hline    
       stochastic $\alpha_s$ = -1.0  & $1.6 \pm 0.4$ & $28 \pm 5$ & $1.20 \pm 0.12$ & $-0.4 \pm 0.4$ \\ 
       stochastic $\alpha_s$ = 0.0 & $1.4 \pm 0.5$ & $35 \pm 7$ & $1.17 \pm 0.13$ & $-0.6 \pm 0.5$ \\ 
       stochastic $\alpha_s$ = 1.0 & $1.4 \pm 0.4$ & $34 \pm 7$ & $1.14 \pm 0.13$ & $-0.6 \pm 0.4$ \\ 
       stochastic $\alpha_s$ = 2.0 & $1.3 \pm 0.5$ & $35 \pm 7$ & $1.15 \pm 0.13$ & $-0.7 \pm 0.5$ \\ 
       astroph & $1.3 \pm 0.4$ & $38 \pm 7$ & $1.07 \pm 0.14$ & $-0.7 \pm 0.4$ \\ 
       astroph 2 & $1.5 \pm 0.4$ & $32 \pm 5$ & $1.04 \pm 0.13$ & $-0.5 \pm 0.4$ \\ 
       astroph 2 + stochastic $\alpha_s$ = -1.0 & $1.2 \pm 0.5$ & $40 \pm 8$ & $1.16 \pm 0.13$ & $-0.8 \pm 0.5$ \\ 
       uniform & $1.3 \pm 0.4$ & $36 \pm 7$ & $1.14 \pm 0.13$ & $-0.7 \pm 0.4$ \\  
           \hline            
       \hline
	\end{tabular}
\end{table*}

\begin{table*}
	\centering
	\caption{As for Table \ref{tab:rrmf_fit} except a shape  $A_{rrm}/(1+z)$ is used  as astrophysical component term.  } 
	\label{tab:rrmf_fit_arrm1}
	\begin{tabular}{lcccc}
	  \hline 
        model  & $\alpha $ &  $B_{f,0}$ &  $A_{rrm}$  &   $\beta$  \\ 
           &   & [nG]  & [rad m$^{-2}$]  &   \\ 
        \hline    
        stochastic $\alpha_s$ = -1.0  & $2.6 \pm 0.5$ & $11 \pm 3$ & $1.07 \pm 0.08$ & $0.6 \pm 0.5$ \\ 
       stochastic $\alpha_s$ = 0.0 & $2.4 \pm 0.5$ & $14 \pm 4$ & $1.06 \pm 0.08$ & $0.4 \pm 0.5$ \\ 
       stochastic $\alpha_s$ = 1.0 & $2.4 \pm 0.5$ & $13 \pm 4$ & $1.06 \pm 0.09$ & $0.4 \pm 0.5$ \\ 
       stochastic $\alpha_s$ = 2.0 & $2.5 \pm 0.5$ & $13 \pm 4$ & $1.07 \pm 0.09$ & $0.5 \pm 0.5$ \\ 
       astroph & $2.3 \pm 0.5$ & $15 \pm 5$ & $1.03 \pm 0.09$ & $0.3 \pm 0.5$ \\ 
       astroph 2 & $2.5 \pm 0.4$ & $14 \pm 4$ & $1.01 \pm 0.09$ & $0.5 \pm 0.4$ \\ 
       astroph 2 + stochastic $\alpha_s$ = -1.0 & $2.4 \pm 0.5$ & $14 \pm 5$ & $1.08 \pm 0.08$ & $0.4 \pm 0.5$ \\ 
       uniform & $2.5 \pm 0.5$ & $13 \pm 4$ & $1.07 \pm 0.08$ & $0.5 \pm 0.5$ \\ 
       \hline
       \hline 
	\end{tabular}
\end{table*}

   \begin{figure*}
   \centering
    \includegraphics[width=0.96\hsize]{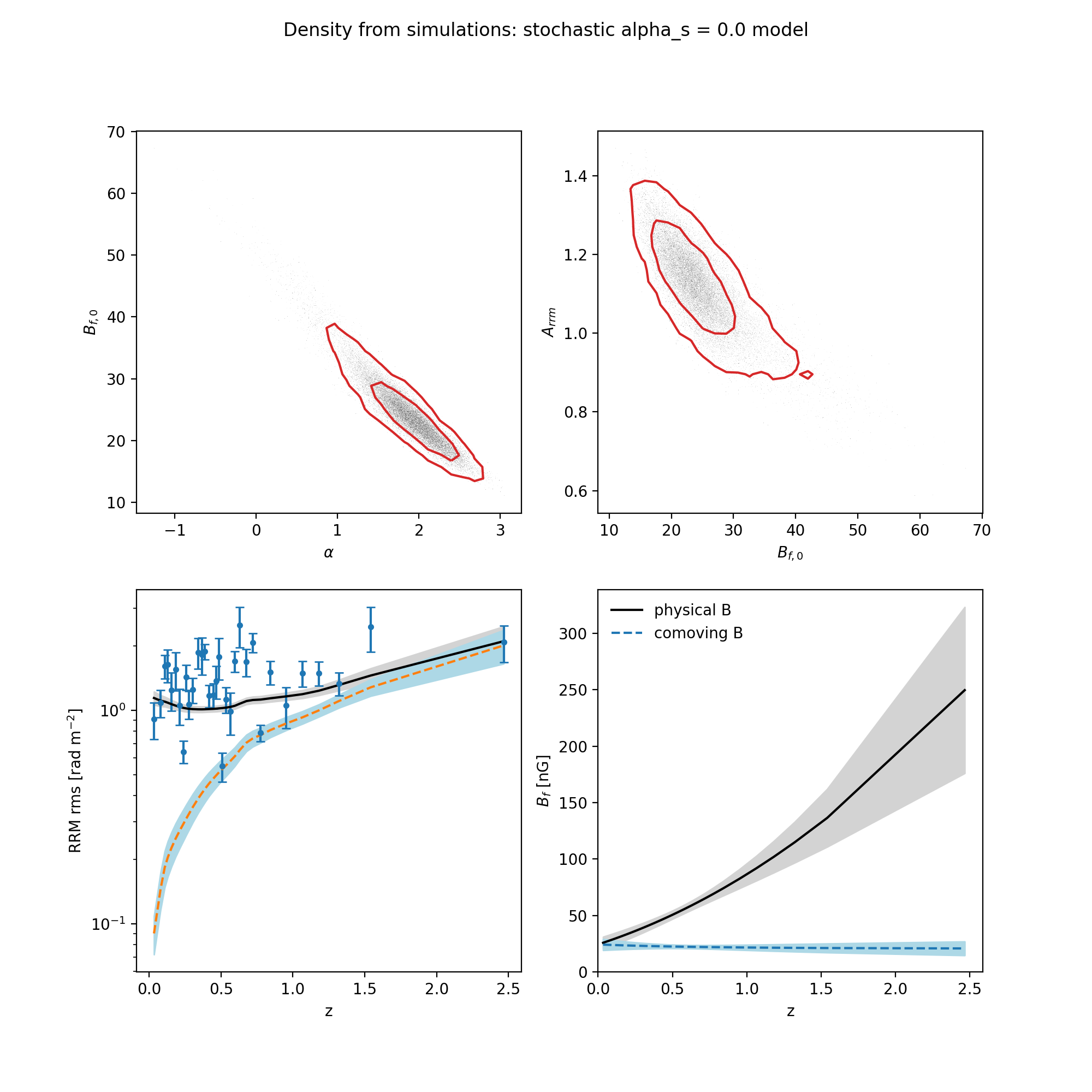}
   \caption{Best-fit results of equation~(\ref{eq:fitting}) to the RRM rms as a function of redshift computed from the GRM filtered sample.  The gas density is taken from the LOS extracted from the MHD simulation of the primordial stochastic model with $\alpha_s=0.0$. The case with $B_f$ independent of $\delta_g$ is assumed here.  {Top-left and top-right}: 2D distributions (dots), and 1-sigma and 2-sigma confidence level  contours (solid lines) of the fit parameters $\alpha$, $B_{f,0}$, and $A_{rrm}$.  { Bottom-left}: RRM rms measured in redshift bins (circles) and best-fit curve (solid) and its error range (grey-shaded area). The fit component of the sole filaments is also shown (dashed line). { Bottom-right}:  Evolution with redshift $z$ of the best-fit filament physical (solid line) and comoving magnetic field strength (dashed). The error range is also shown (shaded areas).  }
              \label{fig:mcmc_fit}%
    \end{figure*}

Following the approach of Paper II, we estimate the average strength of the magnetic field in filaments and its evolution with redshift by Bayesian fitting  of a model of the RRM rms to that measured with our RRM sample filtered by GRM$_{\rm th} =14$~rad~m$^{-2}$.  The RRM rms  model is assumed to be (see Paper II): 
\begin{eqnarray}
    \left<RRM^2\right>^{1/2} &=& \frac{A_{rrm}}{(1+z)^2} +  \left<RRM_f^2\right>^{1/2} \label{eq:fitting}\\
    RRM_f &=& 0.812\, \int_z^0 \frac{n_e\,B_\parallel}{(1+z)^2}\, dl
\end{eqnarray}
 where $n_e$ [cm$^{-3}$] is the electron number density, $B_\parallel$ [$\mu$G]  is the  magnetic field parallel to the LOS, and $dl$ [pc] is the differential path  length. All units are physical (proper) units. The term  RRM$_f$ is the cosmic  filament component and  is computed for each of the 100 LOS we extracted  from each magnetogenesis scenario. The term $A_{rrm}/(1+z)^2$ accounts for an astrophysical component constant with redshift, either local or by intervening objects.  

 The gas density is taken from the LOS extracted from the MHD cosmological simulations.  We only consider positions along the LOS where the gas contrast density ($\delta_g = \rho_g / <\rho_g>$, with $\rho_g$ the gas density) is greater than 1,  which is a typical value to separate filaments from voids and because the underdensities give a negligible contribution to the RRM of the IGM (see Paper II).
 
As discussed in Paper II, and here in Sect.~\ref{sec:tests}, the LOS of our polarized sources at low frequency tend to avoid galaxy clusters.
To account for this, in Paper II we set a density contrast limit above which all LOS points were flagged. However, this does not account for the few sight lines that approach clusters with an impact parameter closer than $R_{500}$ and that can contribute to the  astrophysical component. Here we implement a procedure a little more sophisticated to mimic that the sources  have different projected separations from  clusters: 
\begin{itemize}
    \item We search for the matter density maxima in each LOS and select those with matter density contrast above that at the cluster virial  radius, which is $\rho_M/\rho_c \approx 50$ or $\rho_M/<\rho_M> \approx 160$, where $\rho_M$ is the density of the matter and $\rho_c$ is the critical density of the Universe;  
    \item For each maximum, we draw a separation ($x$) from the distribution of Eqs. (\ref{eq:rayleigh}) and (\ref{eq:rayleigh_cum}), setting  $\sigma_r = 2.51\, \rh$ of the GRM$_{\rm th} =14$~rad~m$^{-2}$ sample. Then we find the corresponding gas density contrast ($\delta_{g,th}$) at that   separation  from a cluster (see Appendix~\ref{app:deltag_R200}); 
    \item All pixels of the LOS with gas density contrast greater than $\delta_{g,th}$ and within a distance from the peak of 3 cMpc (comoving), are flagged and excluded from the RRM computation.  The limit of 3 cMpc allows including clusters of any mass. 
\end{itemize}

The strength of the proper magnetic field in filaments is  assumed to follow the power law: 
\begin{equation}
    B_f = B_{f,0} \, (1+z)^\alpha
\end{equation}
where $B_{f,0}$ is the strength at $z=0$. 
The comoving field follows the power law:
\begin{equation}
    cB_f = B_{f,0} \, (1+z)^\beta
\end{equation}
where\footnote{It comes from  $B \propto \rho_g^{2/3}$ and hence $B \propto (1+z)^2$, for a  field  frozen to the plasma. }
\begin{equation}
    \beta = \alpha -2
\end{equation}
The field direction is set randomly and changes each time the LOS enters a new filament, that is wherever  $\delta_g$ is above 1. To increase the statistics, we produce 120 realisation of the field directions for each LOS.  Thus, there are 12,000 RRM$_f$ realisations for each cosmological model, 100 LOS times 120 field realisations. 

For each realisation we compute the RRM$_f$ as a function of $z$ assuming a single value of $B_{f,0}$. The slope $\alpha$ is set in the range of [-5, 5], spaced by 0.5. For each value of $\alpha$, the RRM$_f$ rms is computed with the 12,000 realizations, 2-$\sigma$ outliers are excluded. RRM$_f$ are smoothed on a scale $\Delta z=0.1$. RRM$_f$ have a linear dependence on $B_{f,0}$. The RRM$_f$ for a generic value of $\alpha$ is obtained by linearly interpolating the two closest values of $\alpha$ that the RRM$_f$ have been computed for. This gives the functional dependence on the two parameters that a Bayesian fitting needs. 

Our  Bayesian fit is a three-free-parameter fit that we conduct with the package {\small EMCEE}\footnote{https://pypi.org/project/emcee/}  \citep{2013PASP..125..306F}. We set priors of $B_{f,0} < 250$~nG \citep[][]{2021A&A...652A..80L}, $B_{f,0} \geqq 0$, and $A_{rrm} \geqq 0$.  
Best-fit results  for all magnetogenesis scenarios considered here  are reported in Table~\ref{tab:rrmf_fit} and the fit plots for an example case, primordial stochastic model with spectrum slope of $\alpha_s =0.0$, are shown in Fig.~\ref{fig:mcmc_fit}. 
The magnetic field strength at $z=0$ is in the range  $B_{f,0} =20$--$27\pm 5$~nG, which is the range of results for the different magnetogenesis models and the typical statistical error. The slope of the physical magnetic field is in the range  $\alpha=1.7$--$2.1\pm 0.4$. The field strength thus   increases  at a rate consistent with a comoving field invariant with redshift, the comoving slope being  $\beta = [-0.3$,~$0.1]\pm 0.4$. This is steeper than the behaviour found in Paper II and it  comes from the steeper RRM rms versus redshift of the RRM sample we use here.
The term $A_{rrm}$ is in the range 1.02--$1.14 \pm 0.11$~rad~m$^{-2}$. 

We  fitted Eq. (\ref{eq:fitting}) also using two other shapes  of the astrophysical term instead of $A_{rrm}/(1+z)^2$:  (1) a cubed inverse dependence on redshift $A_{rrm}/(1+z)^3$, which implies that the astrophysical RRM  decreases with redshift as $(1+z)^{-1}$, and  (2) an inverse dependence $A_{rrm}/(1+z)$, which implies that the astrophysical RRM increases with redshift as $(1+z)$. The results are reported in the Table~\ref{tab:rrmf_fit_arrm3} and~\ref{tab:rrmf_fit_arrm1}.  They are similar, consistent within the errors,  to the previous case with: (1)  $B_{f,0}= 28$--$40 \pm 7$~nG, $\alpha=1.2$--$1.6\pm 0.4$, and $A_{rrm} = 1.04$--$1.20 \pm 0.13$~rad~m$^{-2}$, the strength is higher and the slope shallower compared to the standard case, possibly because of the degeneracy between these two quantities;  (2) $B_{f,0}= 11$--$15 \pm 4$~nG, $\alpha=2.3$--$2.6\pm 0.5$, and $A_{rrm} = 1.01$--$1.08\pm 0.09$~rad~m$^{-2}$, the strength is smaller and the slope steeper. 

The term $A_{rrm}/(1+z)^2$ of Eq. (\ref{eq:fitting}) is meant to account for the local or intervening astrophysical objects component. $A_{rrm}$ is in the range 1.02--$1.14 \pm 0.11 $~rad~m$^{-2}$. If we assign to each source a contribution according to its redshift, the RRM rms because of the  $A_{rrm}/(1+z)^2$ term is 0.56--$0.60\pm 0.06$~rad~m$^{-2}$. Compared to the total  rms of $1.54 \pm 0.05$~rad~m$^{-2}$, this corresponds to a fraction of 35--$39\pm 4$ percent, which is larger than the $21\pm 4$ percent we estimated from clusters and CGM. The reasons we can identify are multiple:  we are still missing an additional astrophysical  component that contributes an additional $\approx 15$ percent, either local or intervening;  the astrophysical RM is not constant with redshift, as we assumed, and the corresponding astrophysical term does not depend on $(1+z)^{-2}$;  a statistical difference (there is a tension at 2--3-$\sigma$). To start investigating the second of these options, we fit with  different shapes of the astrophysical component term.  The case with an astrophysical term decreasing with redshift, $A_{rrm}/(1+z)^3$, gives an RRM rms contribution of 0.44--$0.51\pm 0.06$~rad~m$^{-2}$.  This is a fraction of 29--$33\pm 4$ percent, which is  closer to the observed $21 \pm 4$ percent than the previous, invariant with redshift case. The case with an astrophysical RRM increasing with redshift,  $A_{rrm}/(1+z)$, gives a fraction of 46--$49\pm 4$ percent, even larger than the invariant case. Even though the decreasing  astrophysical question  RRM shape is closest to the observed value, all shapes are in some tension. 

In the previous fitting we assumed $B$ as constant with the gas density. We perform a second series of fits using a    model of the filament field where $B \propto \rho_g^{2/3}$, which corresponds to a magnetic field frozen to the plasma, a legitimate condition in the low density environment of filaments: 
\begin{equation}
    B_f = B_{f,0}^{10} \left(\frac{\delta_g}{10} \right)^{2/3} \, (1+z)^\alpha
    \label{eq:Bdens}
\end{equation}
where $B_{f,0}^{10} $ is the field strength at $z=0$ and  a gas density contrast of $\delta_g =10$, which is the typical value for  cosmic filaments. 

The results of the Bayesian fitting are reported in Table~\ref{tab:rrmf_fit_Bdens} of Appendix~\ref{app:fitting}. We  fitted the case with astrophysical  component case of  $A_{rrm}/(1+z)^2$, which is mid way of the three shapes we tested. The ranges of the values are larger than in the previous fit, possibly because the computed RRMs now depend on $\rho_g^{5/3}$. The field strength at $\delta_g=10$ and $z=0$ ranges in  $B_{f,0}^{10}  = 4$-$10\pm 2$~nG  The slope sits at $\alpha=2.0$--$2.7\pm 0.7$. The slope of the comoving field ranges in $\beta = [0.0$, $0.7]\pm 0.7$.

\subsection{RRM of filaments from cosmological simulations}
\label{sec:synth_rrm}

   \begin{figure*}
   \centering
    \includegraphics[width=0.49\hsize]{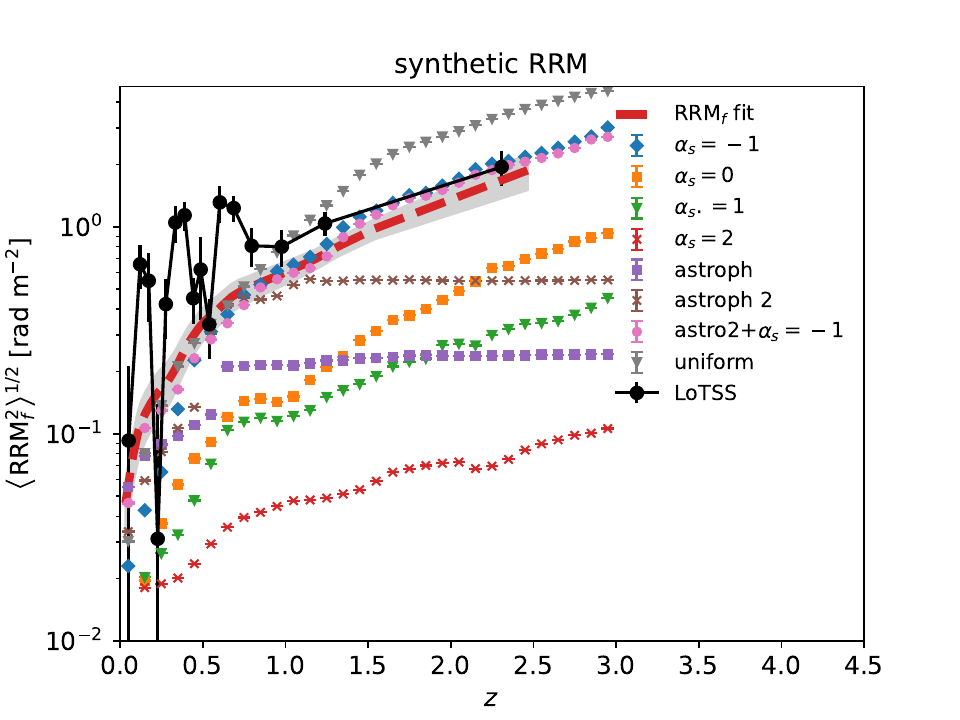}
    \includegraphics[width=0.49\hsize]{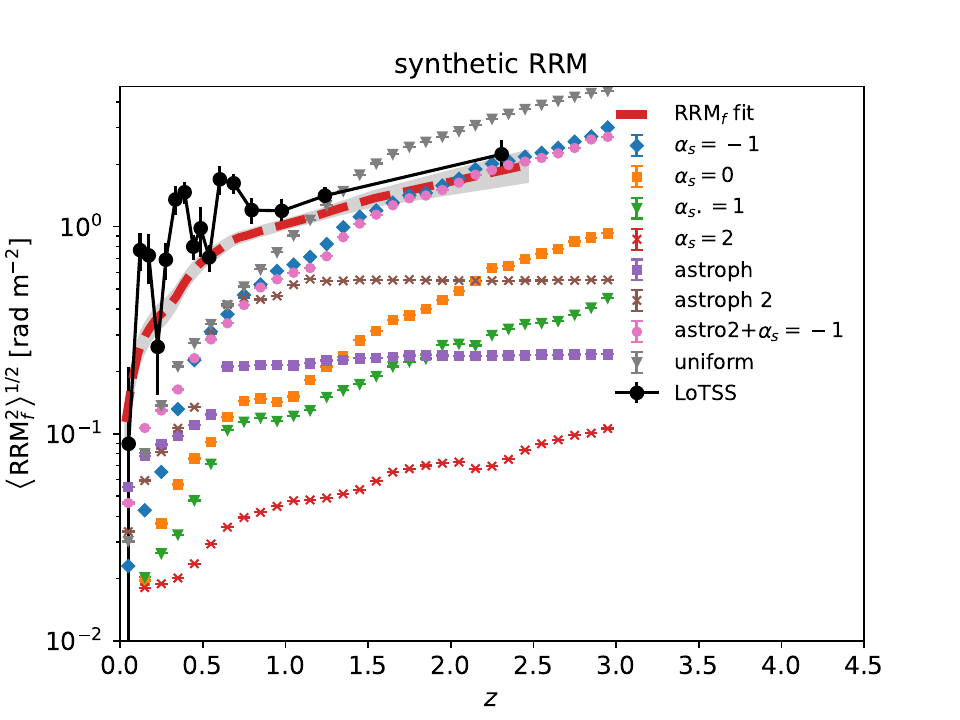}
   \caption{Filaments RRM$_f$ rms as a function of redshift of the cosmological models used in this work. The RRMs are computed using  the gas densities and  magnetic fields from the MHD cosmological simulations ({\it astroph} and {\it astroph 2} are the first and second astrophysical models discussed in Sect.~\ref{sec:simul}). The error bars of the simulated LOS are reported, albeit smaller than the markers (the typical mean fractional error is 0.07 percent). The measured RRM rms is also shown for comparison, with the $A_{rrm}/(1+z)$ (left panel) or  $A_{rrm}/(1+z)^3$  (right panel)   term subtracted off to only display the filament component. The corresponding filament component of the best fit is also shown (red, thick, dashed line) along with its uncertainty (shaded area). }
              \label{fig:rm_synth}%
    \end{figure*}

Here we compute the RRM$_f$ rms of filaments of the magnetogenesis scenarios as a function of redshift using the magnetic field distributions directly produced by  the cosmological MHD simulations. For each scenario, we computed  the RRM versus redshift for each of the 100 LOS and then computed the rms.  To account for the LOFAR polarized sources sight lines avoiding clusters,  we apply the same procedure defined  in Sect.~\ref{sec:fit} to flag points at high density contrast. The results are shown in Fig.~\ref{fig:rm_synth}. The  RRM  rms measured from our sample in 40-source bins  is also reported for comparison.  The latter has either the term $A_{rrm}/(1+z)$ (left panel) or  $A_{rrm}/(1+z)^3$ (right panel)  subtracted off to only display  the cosmic filament component. The term  $A_{rrm}$ is set to 1.08~rad~m$^{-2}$ and  1.20~rad~m$^{-2}$, respectively,  that are  the top of the ranges of the corresponding sets of fittings. We also show the corresponding filament component of the best fit. 

The comparison of the two panels of Fig.~\ref{fig:rm_synth} shows a better match between the observed data and the simulations for the model with astrophysical  RRM increasing with redshift, that is the shape $A_{rrm}/(1+z)$, especially  at low redshift. 

A common feature of all primordial scenarios, either uniform or stochastic, is that their large filling factor for magnetic fields produce significant RRM already at high redshift, and their RRM rms show some slope. Their amplitude  mostly depends on the initial normalisation.  
The astrophysical models produce cosmic fields late, injected  by outflows and AGNs activity with a maximum at $z \sim 2$, and have little power at high redshift. Their RRM$_f$ rms flattens and does not match the slope of the observed data. 

At low redshift both the astrophysical models and the primordial models with highest power match the filament component of the fit to the observed data. 

The slope of the observed RRM rms thus tends to favour  primordial scenarios at high redshift. The stochastic models with $\alpha_s \geq 0$ show RRM rms smaller than that measured already with the current normalisation, which is the highest consistent with CMB data \citep{2019JCAP...11..028P, 2019MNRAS.484..185P}, and look disfavoured by our sample.  The model with $\alpha_s=-1.0$ and a normalisation that is 20 percent of the current limit from CMB data (i.e. $0.37~\rm nG$ compared to the upper limit of $1.87 \rm ~nG$ inferred by CMB modelling), looks to match the observed RRM rms.  Its  RRM$_f$ rms, also seems to match   the observed data at low redshift. 
These  considerations  differ from the Paper II results, where the unfiltered sample gave an RRM rms flat at high redshift.  We further elaborate on these differences in Sect.~\ref{sec:disc}.

\section{Discussion and Conclusions}
\label{sec:disc}

\subsection{Origin of the RRM at low frequency}

We have investigated possible correlations of the RRMs measured with LOFAR at 144~MHz with  astrophysical sources to check whether their origin is the IGM or due to astrophysical sources.

We measured the RRM rms as a function of the projected separation of the source sight line from intervening galaxy clusters. We found an obvious excess at separations shorter than $R_{500}$ for clusters of mass $M >10^{14}$ M$_\odot$. This excess contributes for  19 percent to the total RRM rms. These are few sources (less than 2 percent)  the  polarization fraction of which, some 0.6 percent, is much smaller  than that of the bulk of the sources and close to the LoTSS detection limit. The sources whose sight lines pass through clusters are thus highly depolarized and only few of them survive complete depolarization. At redshift beyond 0.5 the strong depolarization stretches  out to $\approx \rh$.   No  obvious RRM excess is found beyond $R_{500}$, which supports an IGM origin. The only exception is a possible small excess at a separation from  cluster centres of some $2 R_{100} \approx 4 R_{500}$, which might be associated to companion clusters typically found  at that separation from rich clusters. We also find that the polarization fraction increases with the separation from clusters, which indicates the density of the filament environment decreases with the separation from the cluster. We find no obvious excess for poor clusters and galaxy groups of mass $M < 10^{14}$ M$_\odot$, which indicates  that only rich clusters can leave an RRM signature at this low frequency.

A similar analysis of the RRM rms as a function of the projected separation from massive galaxies ($M_* >10^{11}$ M$_\odot$) finds a possible hint of a shock in the CGM  at the galaxy virial radius, which could possibly be produced by matter infalling from the cosmic web or outflows from the galaxies. This is small at some 0.5~rad~m$^{-2}$, and further investigations with larger RM samples or at higher frequencies are required to confirm it. POSSUM \citep{2010AAS...21547013G} or APERTIF \citep{2022A&A...663A.103A} are well suited for conducting such an analysis. The RRM rms beyond the virial radius is 8 percent smaller than  that of the entire sample, thus we find the galaxy CGM gives an additional contribution of astrophysical  origin to the RRMs at this frequency.  The polarization fraction shows a clear, progressive decrement from the viral radius towards the galaxy centre and we argue the polarization fraction of extragalactic background polarized sources can be a good tracer of the CGM. No obvious  RRM trend is found for lower mass galaxies ($M_* < 10^{11}$ M$_\odot$), as well as for the fractional polarization of our lowest mass galaxy subsample ($M_* = 10^{9}$--$10^{10}$ M$_\odot$).

A step forward in such an analysis can be done following \citet{2023A&A...670L..23H} by only selecting star forming galaxies with large extraplanar outflows from the galaxy  central regions. Also, the outflows orientation must be known, to select only cases where the LOS  passes  close to the galaxy minor axis and intercept the outflows. To our knowledge, such data are not yet available for the catalogue  we use and such an analysis has to be postponed.  The use of Mg~II absorption lines in the sight lines of quasars can help identify galaxies with outflows \citep{2007ApJ...669L...5B, 2012ApJ...760...49L, 2014ApJ...792L..12K, 2019MNRAS.490.4368S} that have been suggested to harbor galaxy-scale extraplanar magnetic fields \citep{2010ApJ...711..380B, 2013MNRAS.434.3566J, 2014ApJ...795...63F, 2016ApJ...829..133K, 2020ApJ...890..132M}. This requires large spectroscopic datasets.   A method as  in \citet{2015ApJS..219....8C} can be used to separate star forming from quiescent galaxies. Overall, these are complex and extensive investigations that require  their own dedicated work.

We also analysed the RRM rms as a function of  source properties, such as size  and spectral index, and we find no obvious correlation.  This supports that the RRMs at this frequency have a marginal component of local origin. The polarization fraction increases with the source size, which  possibly is because larger sources extend beyond the centres of clusters and filament spines and are affected by lower depolarization. 

We thus conclude that the RRMs are dominated by the IGM /filaments component and only a contribution of $\approx 21$ percent is of astrophysical object origin, that is galaxy clusters and galaxy CGM  intervening along the source LOS. We note that the shape of the astrophysical contribution that better matches the RRM rms predicted by our MHD simulations implies that a further 25--30 percent of local origin might add up to the astrophysical contribution, but this is still uncertain. It is also worth noticing that, while the fractional polarization is affected  by local effects,  the RRM  rms  is not, except when the depolarization is high. 

]\subsection{Magnetic fields in cosmic filaments}

In this work we have  filtered the LoTSS DR2 RMs, only selecting those with low GRM, to further reduce the residual GRM contamination after its subtraction. This produces a sample with an RRM rms $\approx 20$ percent smaller than that of the unfiltered sample. There is  negligible dependence of the RRM rms on the Galactic latitude and on GRM, which  indicates little  residual GRM contamination. The result is a smaller RRM rms which increases with redshift more steeply than in the original unfiltered sample.  

The RRM rms as a function of redshift shows wiggles, the nature of which is still uncertain. They are unlikely to be due to GRM residuals. They are anti-correlated with the galaxy number density and thus unlikely are of local origin. Understanding their origin requires further investigation. They are associated with structures on scales of $\Delta z \approx  0.2$, that is $\approx 800$--900~Mpc. The wiggles  thus are unlikely to appear in our MHD simulations whose boxes have a linear size of 42.5~Mpc.

Our model fitting of this sample  finds a tension between the observed astrophysical component fractional contribution and that estimated from the fitting.  The latter is larger by a factor of 1.5--2.5, depending on the shape  of the term modelling the astrophysical component. We can identify a few  possible causes of such a tension: (1) we are still missing an additional astrophysical  component that contributes an additional $\approx 15$ percent; (2) the astrophysical  term has not the shapes we considered; (3) a statistical difference (it is at 2--4-$\sigma$). However, the data at low frequency are dominated by the IGM component and  are not best to investigate the properties of the RRMs of astrophysical origin. This calls for RM data at higher frequency, that is dominated by the astrophysical  component (see Paper I). Again, POSSUM and APERTIF can help in that. 

The comparison of  our data with the results of the MHD simulations of the magnetogenesis scenarios we consider in this work favours  $A_{rrm}/(1+z)$ as a shape for the astrophysical  component term, which means that the RRM of astrophysical origin increases with redshift. 

Assuming this astrophysical  term, the best-fitting results are pretty  independent of the magnetogenesis scenario used to draw the gas density.  The average physical magnetic field strength  in filaments  at $z=0$ is  of $B_{f,0}=11$--$15\pm 4$~nG. The slope as a function of $(1+z)$ is at $\alpha=2.3$--$2.6\pm 0.5$, while that of the comoving field is $\beta = 0.3$--$0.6\pm0.5$, which is consistent with a comoving field invariant with redshift. This is steeper than our previous result of a physical field invariant with redshift and it comes from the steeper RRM versus redshift we found for this new GRM filtered sample. 

The comparison with the RRM rms predicted by different magnetogenesis scenarios   favours primordial models, either uniform or stochastic with a power spectrum slope of $\alpha_s=-1.0$.  These models, having magnetic field power at high redshift, give an RRM rms increasing with redshift, as observed. Also at  low redshift there is a good match with the observed data. The astrophysical  scenarios produce magnetic fields late with little power at high redshift where their RRM rms flattens, inconsistent with the observations. A combined primordial and astrophysical model gives results similar to the primordial model and it is also favoured. 

The amplitudes of the simulated primordial stochastic models look  to be on the low side, i.e. even using the upper limits allowed by CMB constraints, the simulated RRM is significantly lower than the data. Only the   $\alpha=-1.0$ case, of  those  tested in this work,  reproduces  the LOFAR RRM data if   an amplitude   significantly  smaller than the corresponding upper limits from the CMB is used.
 However, in this work we only simulated non-helical fields. Helical fields of primordial models may in principle produce a different pattern of  RRMs \citep{2022ApJ...929..127M}, as well as be subject to different constraints from CMB analysis.  The power of the primordial uniform model is too high, but it  can be fixed with a lower initial normalisation of the order of 0.05~nG. 

Compared to the simulations used in Paper II, there are two main differences. First, the RRMs of the astrophysical models are higher, albeit  smaller than the observed RRM by LOFAR. The increase in the magnetic output  stems from  the use of a $64$ times finer mass resolution of the dark matter component and a $4$ times better spatial resolution, giving better resolved star forming regions and galaxy evolution. This resulted in a larger fraction of low-mass halos  (dwarf galaxies), which in turn has increased the magnetisation of voids and filaments and likely generated higher RRM even in the low density environments we explore in our analysis. Also, this model has been calibrated in more detail against both cosmic star formation history and the distribution of observable radio galaxies connected to the AGN model (Vazza et al., submitted). There are a couple of more  items to be considered: 
\begin{enumerate}
    \item The basic result that a purely astrophysical origin scenario for extragalactic magnetic fields under predicts the amount of observed RRM at all redshift is consistent with Paper II and other works that reported incompatibility between this model and the non-detection of the Inverse Compton Cascade from blazars \citep[][]{2022A&A...660A..80B, 2024ApJ...963..135T}.
    \item  We must note that the level of the RRM predicted by other simulations is higher than what we predict here and in Paper II with our cosmological simulations \citep[e.g.][]{2022MNRAS.515.5673A, 2023MNRAS.519.4030A, 2024arXiv240313418B}. This might signal that, despite how low the density where most of the RRM detected by LOFAR originates from, modern numerical simulations are not in agreement with their predictions in low-density environments, although they are able to reproduce the most basic properties of galaxy populations (e.g., stellar mass distribution and star formation history). This, in turn, strengthens the use of deep radio observations, like those  used  here, in giving strong constraints to feedback models applied to simulations of galaxy formation, in regimes which are otherwise very difficult to probe with other  observational techniques. 
\end{enumerate}

The second significant difference with simulations in Paper II is the significantly smaller RRMs of the  stochastic model with   $\alpha_s=1$. 
This is likely a resolution dependent effect. For the primordial models that have  most of the power at small spatial scales, any change in resolution introduces more field reversals along the line of sight. The increased magnetic energy at small scales affects, both, the dynamics of gas  already since the start of the simulation, as well as the small-scale distribution of magnetic fields at any epoch of the simulation. Combined with the 
removal of dense halos from this more resolved set of simulations, we think that this effect is plausibly responsible for a reduction of a factor  $\approx 2$ in the global RRM trend measured in this new set of simulations. 

In summary, these trends call for future larger simulations, with even better resolution and detail on galaxy formation processes, and also including more realistic magnetic fields topologies (e.g. helicity) to seek a possible convergence of these RRMs on the density regime that  is most relevant to compare with LOFAR observations. 

We also would like to highlight that reducing  the residual GRM contamination has significantly changed the redshift  evolution scenario from  flat  to a significant evolution. Even if we have done an important work to minimise bias and residuals, this  emphasises the need for even better GRM maps. A leap in this field can be obtained  by POSSUM, thanks to both its high RM density and sensitivity. 


\begin{acknowledgements}
We thank an anonymous  referee for their thoughtful comments that allowed us to improve  the paper. This work has been conducted   within the LOFAR Magnetism Key Science Project\footnote{https://lofar-mksp.org/} (MKSP). 
This work has made use of LoTSS DR2 data \citep{2022A&A...659A...1S}. 
EC and VV acknowledge this work has been conducted within the INAF program METEORA. 
VV acknowledges support from the Premio per Giovani Ricercatori “Gianni Tofani” II edizione, promoted by INAF-Osservatorio Astrofisico di Arcetri (DD n. 84/2023).
AB acknowledges support from ERC Stg DRANOEL n.~714245 and MIUR FARE grant "SMS". 
SPO acknowledges support from the Comunidad de Madrid Atracción de Talento program via grant 2022-T1/TIC-23797, and grant PID2023-146372OB-I00 funded by MICIU/AEI/10.13039/501100011033 and by ERDF, EU.
FV and SM have been supported by Fondazione Cariplo and Fondazione CDP, thorugh grant n° Rif: 2022-2088 CUP J33C22004310003 for "BREAKTHRU" project.
In this work we used the {\enzo} code (\hyperlink{http://enzo-project.org}{http://enzo-project.org}), the product of a collaborative effort of scientists at many universities and national laboratories.  
FV acknowledges the CINECA award  "IsB28\_RADGALEO" under the ISCRA initiative, for the availability of high-performance computing resources and support. 

LOFAR \citep{2013A&A...556A...2V} is the Low Frequency Array designed and constructed by ASTRON. It has observing, data processing, and data storage facilities in several countries, which are owned by various parties (each with their own funding sources), and which are collectively operated by the ILT foundation under a joint scientific policy. The ILT resources have benefited from the following recent major funding sources: CNRS-INSU, Observatoire de Paris and Universit\'e d'Orl\'eans, France; BMBF, MIWF-NRW, MPG, Germany; Science Foundation Ireland (SFI), Department of Business, Enterprise and Innovation (DBEI), Ireland; NWO, The Netherlands; The Science and Technology Facilities Council, UK; Ministry of Science and Higher Education, Poland; The Istituto Nazionale di Astrofisica (INAF), Italy.   LoTSS made use of the Dutch national e-infrastructure with support of the SURF Cooperative (e-infra 180169) and the LOFAR e-infra group. The J\"ulich LOFAR Long Term Archive and the German LOFAR network are both coordinated and operated by the J\"ulich Supercomputing Centre (JSC), and computing resources on the supercomputer JUWELS at JSC were provided by the Gauss Centre for Supercomputing e.V. (grant CHTB00) through the John von Neumann Institute for Computing (NIC).
LoTSS made use of the University of Hertfordshire high-performance computing facility and the LOFAR-UK computing facility located at the University of Hertfordshire and supported by STFC [ST/P000096/1], and of the Italian LOFAR IT computing infrastructure supported and operated by INAF, and by the Physics Department of Turin university (under an agreement with Consorzio Interuniversitario per la Fisica Spaziale) at the C3S Supercomputing Centre, Italy.
This work made use  of the Python packages NumPy \citep{2020Natur.585..357H}, Astropy \citep{2013A&A...558A..33A}, Matplotlib \citep{2007CSE.....9...90H}, and  EMCEE \citep{2013PASP..125..306F}. Some of the results in this paper have been derived using the healpy \citep{2019JOSS....4.1298Z} and HEALPix\footnote{http://healpix.sf.net} \citep{2005ApJ...622..759G} packages. 
\end{acknowledgements}

\bibliographystyle{aa}
\bibliography{b_stoch}

\begin{thebibliography}{109}
\expandafter\ifx\csname natexlab\endcsname\relax\def\natexlab#1{#1}\fi

\bibitem[{{Adebahr} {et~al.}(2022){Adebahr}, {Berger}, {Adams}, {Hess}, {de
  Blok}, {D{\'e}nes}, {Moss}, {Schulz}, {van der Hulst}, {Connor}, {Damstra},
  {Hut}, {Ivashina}, {Loose}, {Maan}, {Mika}, {Mulder}, {Norden}, {Oostrum},
  {Orr{\'u}}, {Ruiter}, {Smits}, {van Cappellen}, {van Leeuwen}, {Vermaas},
  {Voh}, \& {Ziemke}}]{2022A&A...663A.103A}
{Adebahr}, B., {Berger}, A., {Adams}, E.~A.~K., {et~al.} 2022, \aap, 663, A103

\bibitem[{{Aharonian} {et~al.}(2023){Aharonian}, {Aschersleben}, {Backes},
  {Martins}, {Batzofin}, {Becherini}, {Berge}, {Bi}, {Bouyahiaoui}, {Breuhaus},
  {Brose}, {Brun}, {Bruno}, {Bulik}, {Burger-Scheidlin}, {Bylund}, {Caroff},
  {Casanova}, {Celic}, {Cerruti}, {Chand}, {Chandra}, {Chen}, {Chibueze},
  {Chibueze}, {Cotter}, {de Bony}, {Egberts}, {Ernenwein}, {Fichet de
  Clairfontaine}, {Filipovic}, {Fontaine}, {F{\"u}ssling}, {Funk}, {Gabici},
  {Ghafourizadeh}, {Giavitto}, {Glawion}, {Glicenstein}, {Goswami}, {Grondin},
  {Haerer}, {Holch}, {Holler}, {Horns}, {Jamrozy}, {Jankowsky}, {Joshi},
  {Jung-Richardt}, {Kasai}, {Katarzy{\'n}ski}, {Khatoon}, {Kh{\'e}lifi},
  {Klu{\'z}niak}, {Komin}, {Kosack}, {Kostunin}, {Lang}, {Le Stum}, {Leitl},
  {Lemi{\`e}re}, {Lenain}, {Leuschner}, {Lohse}, {Luashvili}, {Lypova},
  {Mackey}, {Malyshev}, {Malyshev}, {Marandon}, {Marchegiani}, {Marcowith},
  {Mart{\'\i}-Devesa}, {Marx}, {Meyer}, {Mitchell}, {Moderski}, {Mohrmann},
  {Montanari}, {Moulin}, {Muller}, {Murach}, {Nakashima}, {Niemiec}, {Ohm},
  {Olivera-Nieto}, {de Ona Wilhelmi}, {Panny}, {Panter}, {Parsons}, {Peron},
  {Prokhorov}, {Prokoph}, {P{\"u}hlhofer}, {Punch}, {Quirrenbach},
  {Reichherzer}, {Reimer}, {Reimer}, {Reville}, {Rieger}, {Rowell}, {Rudak},
  {Ruiz-Velasco}, {Sahakian}, {Sanchez}, {Sasaki}, {Sch{\"u}ssler}, {Schutte},
  {Schwanke}, {Shapopi}, {Sol}, {Spencer}, {Steinmassl}, {Suzuki}, {Takahashi},
  {Tanaka}, {Taylor}, {Terrier}, {Thorpe-Morgan}, {Tsirou}, {Tsuji},
  {Uchiyama}, {van Eldik}, {Veh}, {Venter}, {Wagner}, {White}, {Wierzcholska},
  {Wong}, {Zacharias}, {Zargaryan}, {Zdziarski}, {Zouari}, {{\.Z}ywucka},
  {Meyer}, \& {Fermi-LAT Collaboration}}]{2023ApJ...950L..16A}
{Aharonian}, F., {Aschersleben}, J., {Backes}, M., {et~al.} 2023, \apjl, 950,
  L16

\bibitem[{{Amaral} {et~al.}(2021){Amaral}, {Vernstrom}, \&
  {Gaensler}}]{2021MNRAS.503.2913A}
{Amaral}, A.~D., {Vernstrom}, T., \& {Gaensler}, B.~M. 2021, \mnras, 503, 2913

\bibitem[{{Anderson} {et~al.}(2021){Anderson}, {Heald}, {Eilek}, {Lenc},
  {Gaensler}, {Rudnick}, {Van Eck}, {O'Sullivan}, {Stil}, {Chippendale},
  {Riseley}, {Carretti}, {West}, {Farnes}, {Harvey-Smith}, {McClure-Griffiths},
  {Bock}, {Bunton}, {Koribalski}, {Tremblay}, {Voronkov}, \&
  {Warhurst}}]{2021PASA...38...20A}
{Anderson}, C.~S., {Heald}, G.~H., {Eilek}, J.~A., {et~al.} 2021, \pasa, 38,
  e020

\bibitem[{{Anderson} {et~al.}(2024){Anderson}, {McClure-Griffiths}, {Rudnick},
  {Gaensler}, {O'Sullivan}, {Bradbury}, {Akahori}, {Baidoo}, {Bruggen},
  {Carretti}, {Duchesne}, {Heald}, {Jung}, {Kaczmarek}, {Leahy}, {Loi}, {Ma},
  {Osinga}, {Seta}, {Stuardi}, {Thomson}, {Van Eck}, {Vernstrom}, \&
  {West}}]{2024arXiv240720325A}
{Anderson}, C.~S., {McClure-Griffiths}, N.~M., {Rudnick}, L., {et~al.} 2024,
  arXiv e-prints, arXiv:2407.20325

\bibitem[{{Angelinelli} {et~al.}(2022){Angelinelli}, {Ettori}, {Dolag},
  {Vazza}, \& {Ragagnin}}]{2022A&A...663L...6A}
{Angelinelli}, M., {Ettori}, S., {Dolag}, K., {Vazza}, F., \& {Ragagnin}, A.
  2022, \aap, 663, L6

\bibitem[{{Angelinelli} {et~al.}(2023){Angelinelli}, {Ettori}, {Dolag},
  {Vazza}, \& {Ragagnin}}]{2023A&A...675A.188A}
{Angelinelli}, M., {Ettori}, S., {Dolag}, K., {Vazza}, F., \& {Ragagnin}, A.
  2023, \aap, 675, A188

\bibitem[{{Ar{\'a}mburo-Garc{\'\i}a} {et~al.}(2021){Ar{\'a}mburo-Garc{\'\i}a},
  {Bondarenko}, {Boyarsky}, {Nelson}, {Pillepich}, \&
  {Sokolenko}}]{2021MNRAS.505.5038A}
{Ar{\'a}mburo-Garc{\'\i}a}, A., {Bondarenko}, K., {Boyarsky}, A., {et~al.}
  2021, \mnras, 505, 5038

\bibitem[{{Ar{\'a}mburo-Garc{\'\i}a} {et~al.}(2022){Ar{\'a}mburo-Garc{\'\i}a},
  {Bondarenko}, {Boyarsky}, {Neronov}, {Scaife}, \&
  {Sokolenko}}]{2022MNRAS.515.5673A}
{Ar{\'a}mburo-Garc{\'\i}a}, A., {Bondarenko}, K., {Boyarsky}, A., {et~al.}
  2022, \mnras, 515, 5673

\bibitem[{{Ar{\'a}mburo-Garc{\'\i}a} {et~al.}(2023){Ar{\'a}mburo-Garc{\'\i}a},
  {Bondarenko}, {Boyarsky}, {Neronov}, {Scaife}, \&
  {Sokolenko}}]{2023MNRAS.519.4030A}
{Ar{\'a}mburo-Garc{\'\i}a}, A., {Bondarenko}, K., {Boyarsky}, A., {et~al.}
  2023, \mnras, 519, 4030

\bibitem[{{Astropy Collaboration} {et~al.}(2013){Astropy Collaboration},
  {Robitaille}, {Tollerud}, {Greenfield}, {Droettboom}, {Bray}, {Aldcroft},
  {Davis}, {Ginsburg}, {Price-Whelan}, {Kerzendorf}, {Conley}, {Crighton},
  {Barbary}, {Muna}, {Ferguson}, {Grollier}, {Parikh}, {Nair}, {Unther},
  {Deil}, {Woillez}, {Conseil}, {Kramer}, {Turner}, {Singer}, {Fox}, {Weaver},
  {Zabalza}, {Edwards}, {Azalee Bostroem}, {Burke}, {Casey}, {Crawford},
  {Dencheva}, {Ely}, {Jenness}, {Labrie}, {Lim}, {Pierfederici}, {Pontzen},
  {Ptak}, {Refsdal}, {Servillat}, \& {Streicher}}]{2013A&A...558A..33A}
{Astropy Collaboration}, {Robitaille}, T.~P., {Tollerud}, E.~J., {et~al.} 2013,
  \aap, 558, A33

\bibitem[{{Bernet} {et~al.}(2010){Bernet}, {Miniati}, \&
  {Lilly}}]{2010ApJ...711..380B}
{Bernet}, M.~L., {Miniati}, F., \& {Lilly}, S.~J. 2010, \apj, 711, 380

\bibitem[{{Bernet} {et~al.}(2008){Bernet}, {Miniati}, {Lilly}, {Kronberg}, \&
  {Dessauges-Zavadsky}}]{2008Natur.454..302B}
{Bernet}, M.~L., {Miniati}, F., {Lilly}, S.~J., {Kronberg}, P.~P., \&
  {Dessauges-Zavadsky}, M. 2008, \nat, 454, 302

\bibitem[{{Bertone} {et~al.}(2006){Bertone}, {Vogt}, \&
  {En{\ss}lin}}]{2006MNRAS.370..319B}
{Bertone}, S., {Vogt}, C., \& {En{\ss}lin}, T. 2006, \mnras, 370, 319

\bibitem[{{Blunier} \& {Neronov}(2024)}]{2024arXiv240313418B}
{Blunier}, J. \& {Neronov}, A. 2024, arXiv e-prints, arXiv:2403.13418

\bibitem[{{B{\"o}ckmann} {et~al.}(2023){B{\"o}ckmann}, {Br{\"u}ggen}, {Heesen},
  {Basu}, {O'Sullivan}, {Heywood}, {Jarvis}, {Scaife}, {Stil}, {Taylor},
  {Adams}, {Bowler}, \& {Tudorache}}]{2023arXiv230811391B}
{B{\"o}ckmann}, K., {Br{\"u}ggen}, M., {Heesen}, V., {et~al.} 2023, arXiv
  e-prints, arXiv:2308.11391

\bibitem[{{Bonafede} {et~al.}(2011){Bonafede}, {Govoni}, {Feretti}, {Murgia},
  {Giovannini}, \& {Br{\"u}ggen}}]{2011A&A...530A..24B}
{Bonafede}, A., {Govoni}, F., {Feretti}, L., {et~al.} 2011, \aap, 530, A24

\bibitem[{{Bondarenko} {et~al.}(2022){Bondarenko}, {Boyarsky}, {Korochkin},
  {Neronov}, {Semikoz}, \& {Sokolenko}}]{2022A&A...660A..80B}
{Bondarenko}, K., {Boyarsky}, A., {Korochkin}, A., {et~al.} 2022, \aap, 660,
  A80

\bibitem[{{Bouch{\'e}} {et~al.}(2007){Bouch{\'e}}, {Murphy}, {P{\'e}roux},
  {Davies}, {Eisenhauer}, {F{\"o}rster Schreiber}, \&
  {Tacconi}}]{2007ApJ...669L...5B}
{Bouch{\'e}}, N., {Murphy}, M.~T., {P{\'e}roux}, C., {et~al.} 2007, \apjl, 669,
  L5

\bibitem[{{Brentjens} \& {de Bruyn}(2005)}]{2005A&A...441.1217B}
{Brentjens}, M.~A. \& {de Bruyn}, A.~G. 2005, \aap, 441, 1217

\bibitem[{{Brown} {et~al.}(2017){Brown}, {Vernstrom}, {Carretti}, {Dolag},
  {Gaensler}, {Staveley-Smith}, {Bernardi}, {Haverkorn}, {Kesteven}, \&
  {Poppi}}]{2017MNRAS.468.4246B}
{Brown}, S., {Vernstrom}, T., {Carretti}, E., {et~al.} 2017, \mnras, 468, 4246

\bibitem[{{Burn}(1966)}]{1966MNRAS.133...67B}
{Burn}, B.~J. 1966, \mnras, 133, 67

\bibitem[{{Carretti} {et~al.}(2023){Carretti}, {O'Sullivan}, {Vacca}, {Vazza},
  {Gheller}, {Vernstrom}, \& {Bonafede}}]{2023MNRAS.518.2273C}
{Carretti}, E., {O'Sullivan}, S.~P., {Vacca}, V., {et~al.} 2023, \mnras, 518,
  2273

\bibitem[{{Carretti} {et~al.}(2022){Carretti}, {Vacca}, {O'Sullivan}, {Heald},
  {Horellou}, {R{\"o}ttgering}, {Scaife}, {Shimwell}, {Shulevski}, {Stuardi},
  \& {Vernstrom}}]{2022MNRAS.512..945C}
{Carretti}, E., {Vacca}, V., {O'Sullivan}, S.~P., {et~al.} 2022, \mnras, 512,
  945

\bibitem[{{Chang} {et~al.}(2015){Chang}, {van der Wel}, {da Cunha}, \&
  {Rix}}]{2015ApJS..219....8C}
{Chang}, Y.-Y., {van der Wel}, A., {da Cunha}, E., \& {Rix}, H.-W. 2015, \apjs,
  219, 8

\bibitem[{{Cho}(2014)}]{2014ApJ...797..133C}
{Cho}, J. 2014, \apj, 797, 133

\bibitem[{{Condon} {et~al.}(1998){Condon}, {Cotton}, {Greisen}, {Yin},
  {Perley}, {Taylor}, \& {Broderick}}]{1998AJ....115.1693C}
{Condon}, J.~J., {Cotton}, W.~D., {Greisen}, E.~W., {et~al.} 1998, \aj, 115,
  1693

\bibitem[{{de Gasperin} {et~al.}(2018){de Gasperin}, {Intema}, \&
  {Frail}}]{2018MNRAS.474.5008D}
{de Gasperin}, F., {Intema}, H.~T., \& {Frail}, D.~A. 2018, \mnras, 474, 5008

\bibitem[{{Dickey} {et~al.}(2022){Dickey}, {West}, {Thomson}, {Landecker},
  {Bracco}, {Carretti}, {Han}, {Hill}, {Ma}, {Mao}, {Ordog}, {Brown},
  {Douglas}, {Erceg}, {Jeli{\'c}}, {Kothes}, \&
  {Wolleben}}]{2022ApJ...940...75D}
{Dickey}, J.~M., {West}, J., {Thomson}, A. J.~M., {et~al.} 2022, \apj, 940, 75

\bibitem[{{Dietl} {et~al.}(2024){Dietl}, {Pacaud}, {Reiprich}, {Veronica},
  {Migkas}, {Spinelli}, {Dolag}, \& {Seidel}}]{2024arXiv240117281D}
{Dietl}, J., {Pacaud}, F., {Reiprich}, T.~H., {et~al.} 2024, arXiv e-prints,
  arXiv:2401.17281

\bibitem[{{Donnert} {et~al.}(2009){Donnert}, {Dolag}, {Lesch}, \&
  {M{\"u}ller}}]{2009MNRAS.392.1008D}
{Donnert}, J., {Dolag}, K., {Lesch}, H., \& {M{\"u}ller}, E. 2009, \mnras, 392,
  1008

\bibitem[{{Eroshenko}(2023)}]{2023arXiv231101207E}
{Eroshenko}, Y.~N. 2023, arXiv e-prints, arXiv:2311.01207

\bibitem[{{Farnes} {et~al.}(2014){Farnes}, {O'Sullivan}, {Corrigan}, \&
  {Gaensler}}]{2014ApJ...795...63F}
{Farnes}, J.~S., {O'Sullivan}, S.~P., {Corrigan}, M.~E., \& {Gaensler}, B.~M.
  2014, \apj, 795, 63

\bibitem[{{Foreman-Mackey} {et~al.}(2013){Foreman-Mackey}, {Hogg}, {Lang}, \&
  {Goodman}}]{2013PASP..125..306F}
{Foreman-Mackey}, D., {Hogg}, D.~W., {Lang}, D., \& {Goodman}, J. 2013, \pasp,
  125, 306

\bibitem[{{Fujimoto} {et~al.}(1971){Fujimoto}, {Kawabata}, \&
  {Sofue}}]{1971PThPS..49..181F}
{Fujimoto}, M., {Kawabata}, K., \& {Sofue}, Y. 1971, Progress of Theoretical
  Physics Supplement, 49, 181

\bibitem[{{Gaensler} {et~al.}(2010){Gaensler}, {Landecker}, {Taylor}, \&
  {POSSUM Collaboration}}]{2010AAS...21547013G}
{Gaensler}, B.~M., {Landecker}, T.~L., {Taylor}, A.~R., \& {POSSUM
  Collaboration}. 2010, in American Astronomical Society Meeting Abstracts,
  Vol. 215, American Astronomical Society Meeting Abstracts \#215, 470.13

\bibitem[{{Garrington} {et~al.}(1988){Garrington}, {Leahy}, {Conway}, \&
  {Laing}}]{1988Natur.331..147G}
{Garrington}, S.~T., {Leahy}, J.~P., {Conway}, R.~G., \& {Laing}, R.~A. 1988,
  \nat, 331, 147

\bibitem[{{Gaspari} {et~al.}(2019){Gaspari}, {Eckert}, {Ettori}, {Tozzi},
  {Bassini}, {Rasia}, {Brighenti}, {Sun}, {Borgani}, {Johnson}, {Tremblay},
  {Stone}, {Temi}, {Yang}, {Tombesi}, \& {Cappi}}]{2019ApJ...884..169G}
{Gaspari}, M., {Eckert}, D., {Ettori}, S., {et~al.} 2019, \apj, 884, 169

\bibitem[{{Girelli} {et~al.}(2020){Girelli}, {Pozzetti}, {Bolzonella},
  {Giocoli}, {Marulli}, \& {Baldi}}]{2020A&A...634A.135G}
{Girelli}, G., {Pozzetti}, L., {Bolzonella}, M., {et~al.} 2020, \aap, 634, A135

\bibitem[{{G{\'o}rski} {et~al.}(2005){G{\'o}rski}, {Hivon}, {Banday},
  {Wandelt}, {Hansen}, {Reinecke}, \& {Bartelmann}}]{2005ApJ...622..759G}
{G{\'o}rski}, K.~M., {Hivon}, E., {Banday}, A.~J., {et~al.} 2005, \apj, 622,
  759

\bibitem[{{Hammond} {et~al.}(2012){Hammond}, {Robishaw}, \&
  {Gaensler}}]{2012arXiv1209.1438H}
{Hammond}, A.~M., {Robishaw}, T., \& {Gaensler}, B.~M. 2012, arXiv e-prints,
  arXiv:1209.1438

\bibitem[{{Harris} {et~al.}(2020){Harris}, {Millman}, {van der Walt},
  {Gommers}, {Virtanen}, {Cournapeau}, {Wieser}, {Taylor}, {Berg}, {Smith},
  {Kern}, {Picus}, {Hoyer}, {van Kerkwijk}, {Brett}, {Haldane}, {del R{\'\i}o},
  {Wiebe}, {Peterson}, {G{\'e}rard-Marchant}, {Sheppard}, {Reddy}, {Weckesser},
  {Abbasi}, {Gohlke}, \& {Oliphant}}]{2020Natur.585..357H}
{Harris}, C.~R., {Millman}, K.~J., {van der Walt}, S.~J., {et~al.} 2020, \nat,
  585, 357

\bibitem[{{Heesen} {et~al.}(2023){Heesen}, {O'Sullivan}, {Br{\"u}ggen}, {Basu},
  {Beck}, {Seta}, {Carretti}, {Krause}, {Haverkorn}, {Hutschenreuter},
  {Bracco}, {Stein}, {Bomans}, {Dettmar}, {Chy{\.z}y}, {Heald}, {Paladino}, \&
  {Horellou}}]{2023A&A...670L..23H}
{Heesen}, V., {O'Sullivan}, S.~P., {Br{\"u}ggen}, M., {et~al.} 2023, \aap, 670,
  L23

\bibitem[{{Hoang} {et~al.}(2023){Hoang}, {Br{\"u}ggen}, {Zhang}, {Bonafede},
  {Liu}, {Liu}, {Shimwell}, {Botteon}, {Brunetti}, {Bulbul}, {Gennaro},
  {O'Sullivan}, {Pasini}, {R{\"o}ttgering}, {Vernstrom}, \& {van
  Weeren}}]{2023MNRAS.523.6320H}
{Hoang}, D.~N., {Br{\"u}ggen}, M., {Zhang}, X., {et~al.} 2023, \mnras, 523,
  6320

\bibitem[{{Huang} {et~al.}(2023){Huang}, {Dai}, {Zhang}, {Liu}, \&
  {Wang}}]{2023arXiv230605970H}
{Huang}, Y.-Y., {Dai}, C.-y., {Zhang}, H.-M., {Liu}, R.-Y., \& {Wang}, X.-Y.
  2023, arXiv e-prints, arXiv:2306.05970

\bibitem[{{Hunter}(2007)}]{2007CSE.....9...90H}
{Hunter}, J.~D. 2007, Computing in Science and Engineering, 9, 90

\bibitem[{{Hutschenreuter} {et~al.}(2022){Hutschenreuter}, {Anderson}, {Betti},
  {Bower}, {Brown}, {Br{\"u}ggen}, {Carretti}, {Clarke}, {Clegg}, {Costa},
  {Croft}, {Van Eck}, {Gaensler}, {de Gasperin}, {Haverkorn}, {Heald}, {Hull},
  {Inoue}, {Johnston-Hollitt}, {Kaczmarek}, {Law}, {Ma}, {MacMahon}, {Mao},
  {Riseley}, {Roy}, {Shanahan}, {Shimwell}, {Stil}, {Sobey}, {O'Sullivan},
  {Tasse}, {Vacca}, {Vernstrom}, {Williams}, {Wright}, \&
  {En{\ss}lin}}]{2022A&A...657A..43H}
{Hutschenreuter}, S., {Anderson}, C.~S., {Betti}, S., {et~al.} 2022, \aap, 657,
  A43

\bibitem[{{Hutschenreuter} \& {En{\ss}lin}(2020)}]{2020A&A...633A.150H}
{Hutschenreuter}, S. \& {En{\ss}lin}, T.~A. 2020, \aap, 633, A150

\bibitem[{{Ilani} {et~al.}(2024){Ilani}, {Hou}, \&
  {Keshet}}]{2024arXiv240216946I}
{Ilani}, G., {Hou}, K.-C., \& {Keshet}, U. 2024, arXiv e-prints,
  arXiv:2402.16946

\bibitem[{{Intema} {et~al.}(2017){Intema}, {Jagannathan}, {Mooley}, \&
  {Frail}}]{2017A&A...598A..78I}
{Intema}, H.~T., {Jagannathan}, P., {Mooley}, K.~P., \& {Frail}, D.~A. 2017,
  \aap, 598, A78

\bibitem[{{Joshi} \& {Chand}(2013)}]{2013MNRAS.434.3566J}
{Joshi}, R. \& {Chand}, H. 2013, \mnras, 434, 3566

\bibitem[{{Kacprzak} {et~al.}(2014){Kacprzak}, {Martin}, {Bouch{\'e}},
  {Churchill}, {Cooke}, {LeReun}, {Schroetter}, {Ho}, \&
  {Klimek}}]{2014ApJ...792L..12K}
{Kacprzak}, G.~G., {Martin}, C.~L., {Bouch{\'e}}, N., {et~al.} 2014, \apjl,
  792, L12

\bibitem[{{Kennicutt}(1998)}]{1998ApJ...498..541K}
{Kennicutt}, Jr., R.~C. 1998, \apj, 498, 541

\bibitem[{{Kim} {et~al.}(2016){Kim}, {Lilly}, {Miniati}, {Bernet}, {Beck},
  {O'Sullivan}, \& {Gaensler}}]{2016ApJ...829..133K}
{Kim}, K.~S., {Lilly}, S.~J., {Miniati}, F., {et~al.} 2016, \apj, 829, 133

\bibitem[{{Kravtsov}(2003)}]{2003ApJ...590L...1K}
{Kravtsov}, A.~V. 2003, \apjl, 590, L1

\bibitem[{{Kronberg}(1994)}]{1994RPPh...57..325K}
{Kronberg}, P.~P. 1994, Reports on Progress in Physics, 57, 325

\bibitem[{{Kronberg} {et~al.}(2008){Kronberg}, {Bernet}, {Miniati}, {Lilly},
  {Short}, \& {Higdon}}]{2008ApJ...676...70K}
{Kronberg}, P.~P., {Bernet}, M.~L., {Miniati}, F., {et~al.} 2008, \apj, 676, 70

\bibitem[{{Kronberg} \& {Perry}(1982)}]{1982ApJ...263..518K}
{Kronberg}, P.~P. \& {Perry}, J.~J. 1982, \apj, 263, 518

\bibitem[{{Kronberg} {et~al.}(1977){Kronberg}, {Reinhardt}, \&
  {Simard-Normandin}}]{1977A&A....61..771K}
{Kronberg}, P.~P., {Reinhardt}, M., \& {Simard-Normandin}, M. 1977, \aap, 61,
  771

\bibitem[{{Laing}(1988)}]{1988Natur.331..149L}
{Laing}, R.~A. 1988, \nat, 331, 149

\bibitem[{{Laing} {et~al.}(2008){Laing}, {Bridle}, {Parma}, \&
  {Murgia}}]{2008MNRAS.391..521L}
{Laing}, R.~A., {Bridle}, A.~H., {Parma}, P., \& {Murgia}, M. 2008, \mnras,
  391, 521

\bibitem[{{Lamee} {et~al.}(2016){Lamee}, {Rudnick}, {Farnes}, {Carretti},
  {Gaensler}, {Haverkorn}, \& {Poppi}}]{2016ApJ...829....5L}
{Lamee}, M., {Rudnick}, L., {Farnes}, J.~S., {et~al.} 2016, \apj, 829, 5

\bibitem[{{Liu} \& {Pooley}(1991)}]{1991MNRAS.249..343L}
{Liu}, R. \& {Pooley}, G. 1991, \mnras, 249, 343

\bibitem[{{Locatelli} {et~al.}(2021){Locatelli}, {Vazza}, {Bonafede}, {Banfi},
  {Bernardi}, {Gheller}, {Botteon}, \& {Shimwell}}]{2021A&A...652A..80L}
{Locatelli}, N., {Vazza}, F., {Bonafede}, A., {et~al.} 2021, \aap, 652, A80

\bibitem[{{Lundgren} {et~al.}(2012){Lundgren}, {Brammer}, {van Dokkum},
  {Bezanson}, {Franx}, {Fumagalli}, {Momcheva}, {Nelson}, {Skelton}, {Wake},
  {Whitaker}, {da Cunha}, {Erb}, {Fan}, {Kriek}, {Labb{\'e}}, {Marchesini},
  {Patel}, {Rix}, {Schmidt}, \& {van der Wel}}]{2012ApJ...760...49L}
{Lundgren}, B.~F., {Brammer}, G., {van Dokkum}, P., {et~al.} 2012, \apj, 760,
  49

\bibitem[{{Malik} {et~al.}(2020){Malik}, {Chand}, \&
  {Seshadri}}]{2020ApJ...890..132M}
{Malik}, S., {Chand}, H., \& {Seshadri}, T.~R. 2020, \apj, 890, 132

\bibitem[{{Martin} {et~al.}(2023){Martin}, {Darvish}, {Lin}, {Cen},
  {Matuszewski}, {Morrissey}, {Neill}, \& {Moore}}]{2023NatAs...7.1390M}
{Martin}, D.~C., {Darvish}, B., {Lin}, Z., {et~al.} 2023, Nature Astronomy, 7,
  1390

\bibitem[{{Mtchedlidze} {et~al.}(2022){Mtchedlidze},
  {Dom{\'\i}nguez-Fern{\'a}ndez}, {Du}, {Brandenburg}, {Kahniashvili},
  {O'Sullivan}, {Schmidt}, \& {Br{\"u}ggen}}]{2022ApJ...929..127M}
{Mtchedlidze}, S., {Dom{\'\i}nguez-Fern{\'a}ndez}, P., {Du}, X., {et~al.} 2022,
  \apj, 929, 127

\bibitem[{{Neronov} \& {Vovk}(2010)}]{2010Sci...328...73N}
{Neronov}, A. \& {Vovk}, I. 2010, Science, 328, 73

\bibitem[{{Oppermann} {et~al.}(2015){Oppermann}, {Junklewitz}, {Greiner},
  {En{\ss}lin}, {Akahori}, {Carretti}, {Gaensler}, {Goobar}, {Harvey-Smith},
  {Johnston-Hollitt}, {Pratley}, {Schnitzeler}, {Stil}, \&
  {Vacca}}]{2015A&A...575A.118O}
{Oppermann}, N., {Junklewitz}, H., {Greiner}, M., {et~al.} 2015, \aap, 575,
  A118

\bibitem[{{Oren} \& {Wolfe}(1995)}]{1995ApJ...445..624O}
{Oren}, A.~L. \& {Wolfe}, A.~M. 1995, \apj, 445, 624

\bibitem[{{Osinga} {et~al.}(2022){Osinga}, {van Weeren}, {Andrade-Santos},
  {Rudnick}, {Bonafede}, {Clarke}, {Duncan}, {Giacintucci}, {Mroczkowski}, \&
  {R{\"o}ttgering}}]{2022A&A...665A..71O}
{Osinga}, E., {van Weeren}, R.~J., {Andrade-Santos}, F., {et~al.} 2022, \aap,
  665, A71

\bibitem[{{O'Sullivan} {et~al.}(2023){O'Sullivan}, {Shimwell}, {Hardcastle},
  {Tasse}, {Heald}, {Carretti}, {Br{\"u}ggen}, {Vacca}, {Sobey}, {Van Eck},
  {Horellou}, {Beck}, {Bilicki}, {Bourke}, {Botteon}, {Croston}, {Drabent},
  {Duncan}, {Heesen}, {Ideguchi}, {Kirwan}, {Lawlor}, {Mingo},
  {Nikiel-Wroczy{\'n}ski}, {Piotrowska}, {Scaife}, \& {van
  Weeren}}]{2023MNRAS.519.5723O}
{O'Sullivan}, S.~P., {Shimwell}, T.~W., {Hardcastle}, M.~J., {et~al.} 2023,
  \mnras, 519, 5723

\bibitem[{{Padmanabhan} \& {Loeb}(2023)}]{2023ApJ...946L..18P}
{Padmanabhan}, H. \& {Loeb}, A. 2023, \apjl, 946, L18

\bibitem[{{Paoletti} {et~al.}(2019){Paoletti}, {Chluba}, {Finelli}, \&
  {Rubi{\~n}o-Mart{\'\i}n}}]{2019MNRAS.484..185P}
{Paoletti}, D., {Chluba}, J., {Finelli}, F., \& {Rubi{\~n}o-Mart{\'\i}n}, J.~A.
  2019, \mnras, 484, 185

\bibitem[{{Paoletti} \& {Finelli}(2019)}]{2019JCAP...11..028P}
{Paoletti}, D. \& {Finelli}, F. 2019, \jcap, 2019, 028

\bibitem[{{Planck Collaboration} {et~al.}(2016{\natexlab{a}}){Planck
  Collaboration}, {Adam}, {Ade}, {Alves}, {Ashdown}, {Aumont}, {Baccigalupi},
  {Banday}, {Barreiro}, {Bartolo}, {Battaner}, {Benabed}, {Benoit-L{\'e}vy},
  {Bernard}, {Bersanelli}, {Bielewicz}, {Bonavera}, {Bond}, {Borrill},
  {Bouchet}, {Boulanger}, {Bucher}, {Burigana}, {Butler}, {Calabrese},
  {Cardoso}, {Catalano}, {Chiang}, {Christensen}, {Colombo}, {Combet},
  {Couchot}, {Crill}, {Curto}, {Cuttaia}, {Danese}, {Davis}, {de Bernardis},
  {de Rosa}, {de Zotti}, {Delabrouille}, {Dickinson}, {Diego}, {Dolag},
  {Dor{\'e}}, {Ducout}, {Dupac}, {Elsner}, {En{\ss}lin}, {Eriksen},
  {Ferri{\`e}re}, {Finelli}, {Forni}, {Frailis}, {Fraisse}, {Franceschi},
  {Galeotta}, {Ganga}, {Ghosh}, {Giard}, {Gjerl{\o}w}, {Gonz{\'a}lez-Nuevo},
  {G{\'o}rski}, {Gregorio}, {Gruppuso}, {Gudmundsson}, {Hansen}, {Harrison},
  {Hern{\'a}ndez-Monteagudo}, {Herranz}, {Hildebrandt}, {Hobson}, {Hornstrup},
  {Hurier}, {Jaffe}, {Jaffe}, {Jones}, {Juvela}, {Keih{\"a}nen}, {Keskitalo},
  {Kisner}, {Knoche}, {Kunz}, {Kurki-Suonio}, {Lamarre}, {Lasenby}, {Lattanzi},
  {Lawrence}, {Leahy}, {Leonardi}, {Levrier}, {Liguori}, {Lilje},
  {Linden-V{\o}rnle}, {L{\'o}pez-Caniego}, {Lubin}, {Mac{\'\i}as-P{\'e}rez},
  {Maggio}, {Maino}, {Mandolesi}, {Mangilli}, {Maris}, {Martin},
  {Mart{\'\i}nez-Gonz{\'a}lez}, {Masi}, {Matarrese}, {Melchiorri}, {Mennella},
  {Migliaccio}, {Miville-Desch{\^e}nes}, {Moneti}, {Montier}, {Morgante},
  {Munshi}, {Murphy}, {Naselsky}, {Nati}, {Natoli}, {N{\o}rgaard-Nielsen},
  {Oppermann}, {Orlando}, {Pagano}, {Pajot}, {Paladini}, {Paoletti}, {Pasian},
  {Perotto}, {Pettorino}, {Piacentini}, {Piat}, {Pierpaoli}, {Plaszczynski},
  {Pointecouteau}, {Polenta}, {Ponthieu}, {Pratt}, {Prunet}, {Puget}, {Rachen},
  {Reinecke}, {Remazeilles}, {Renault}, {Renzi}, {Ristorcelli}, {Rocha},
  {Rossetti}, {Roudier}, {Rubi{\~n}o-Mart{\'\i}n}, {Rusholme}, {Sandri},
  {Santos}, {Savelainen}, {Scott}, {Spencer}, {Stolyarov}, {Stompor}, {Strong},
  {Sudiwala}, {Sunyaev}, {Suur-Uski}, {Sygnet}, {Tauber}, {Terenzi},
  {Toffolatti}, {Tomasi}, {Tristram}, {Tucci}, {Valenziano}, {Valiviita}, {Van
  Tent}, {Vielva}, {Villa}, {Wade}, {Wandelt}, {Wehus}, {Yvon}, {Zacchei}, \&
  {Zonca}}]{2016A&A...596A.103P}
{Planck Collaboration}, {Adam}, R., {Ade}, P.~A.~R., {et~al.}
  2016{\natexlab{a}}, \aap, 596, A103

\bibitem[{{Planck Collaboration} {et~al.}(2016{\natexlab{b}}){Planck
  Collaboration}, {Ade}, {Aghanim}, {Arnaud}, {Ashdown}, {Aumont},
  {Baccigalupi}, {Banday}, {Barreiro}, {Bartlett}, {Bartolo}, {Battaner},
  {Battye}, {Benabed}, {Beno{\^\i}t}, {Benoit-L{\'e}vy}, {Bernard},
  {Bersanelli}, {Bielewicz}, {Bock}, {Bonaldi}, {Bonavera}, {Bond}, {Borrill},
  {Bouchet}, {Boulanger}, {Bucher}, {Burigana}, {Butler}, {Calabrese},
  {Cardoso}, {Catalano}, {Challinor}, {Chamballu}, {Chary}, {Chiang}, {Chluba},
  {Christensen}, {Church}, {Clements}, {Colombi}, {Colombo}, {Combet},
  {Coulais}, {Crill}, {Curto}, {Cuttaia}, {Danese}, {Davies}, {Davis}, {de
  Bernardis}, {de Rosa}, {de Zotti}, {Delabrouille}, {D{\'e}sert}, {Di
  Valentino}, {Dickinson}, {Diego}, {Dolag}, {Dole}, {Donzelli}, {Dor{\'e}},
  {Douspis}, {Ducout}, {Dunkley}, {Dupac}, {Efstathiou}, {Elsner},
  {En{\ss}lin}, {Eriksen}, {Farhang}, {Fergusson}, {Finelli}, {Forni},
  {Frailis}, {Fraisse}, {Franceschi}, {Frejsel}, {Galeotta}, {Galli}, {Ganga},
  {Gauthier}, {Gerbino}, {Ghosh}, {Giard}, {Giraud-H{\'e}raud}, {Giusarma},
  {Gjerl{\o}w}, {Gonz{\'a}lez-Nuevo}, {G{\'o}rski}, {Gratton}, {Gregorio},
  {Gruppuso}, {Gudmundsson}, {Hamann}, {Hansen}, {Hanson}, {Harrison}, {Helou},
  {Henrot-Versill{\'e}}, {Hern{\'a}ndez-Monteagudo}, {Herranz}, {Hildebrandt},
  {Hivon}, {Hobson}, {Holmes}, {Hornstrup}, {Hovest}, {Huang}, {Huffenberger},
  {Hurier}, {Jaffe}, {Jaffe}, {Jones}, {Juvela}, {Keih{\"a}nen}, {Keskitalo},
  {Kisner}, {Kneissl}, {Knoche}, {Knox}, {Kunz}, {Kurki-Suonio}, {Lagache},
  {L{\"a}hteenm{\"a}ki}, {Lamarre}, {Lasenby}, {Lattanzi}, {Lawrence}, {Leahy},
  {Leonardi}, {Lesgourgues}, {Levrier}, {Lewis}, {Liguori}, {Lilje},
  {Linden-V{\o}rnle}, {L{\'o}pez-Caniego}, {Lubin}, {Mac{\'\i}as-P{\'e}rez},
  {Maggio}, {Maino}, {Mandolesi}, {Mangilli}, {Marchini}, {Maris}, {Martin},
  {Martinelli}, {Mart{\'\i}nez-Gonz{\'a}lez}, {Masi}, {Matarrese}, {McGehee},
  {Meinhold}, {Melchiorri}, {Melin}, {Mendes}, {Mennella}, {Migliaccio},
  {Millea}, {Mitra}, {Miville-Desch{\^e}nes}, {Moneti}, {Montier}, {Morgante},
  {Mortlock}, {Moss}, {Munshi}, {Murphy}, {Naselsky}, {Nati}, {Natoli},
  {Netterfield}, {N{\o}rgaard-Nielsen}, {Noviello}, {Novikov}, {Novikov},
  {Oxborrow}, {Paci}, {Pagano}, {Pajot}, {Paladini}, {Paoletti}, {Partridge},
  {Pasian}, {Patanchon}, {Pearson}, {Perdereau}, {Perotto}, {Perrotta},
  {Pettorino}, {Piacentini}, {Piat}, {Pierpaoli}, {Pietrobon}, {Plaszczynski},
  {Pointecouteau}, {Polenta}, {Popa}, {Pratt}, {Pr{\'e}zeau}, {Prunet},
  {Puget}, {Rachen}, {Reach}, {Rebolo}, {Reinecke}, {Remazeilles}, {Renault},
  {Renzi}, {Ristorcelli}, {Rocha}, {Rosset}, {Rossetti}, {Roudier},
  {Rouill{\'e} d'Orfeuil}, {Rowan-Robinson}, {Rubi{\~n}o-Mart{\'\i}n},
  {Rusholme}, {Said}, {Salvatelli}, {Salvati}, {Sandri}, {Santos},
  {Savelainen}, {Savini}, {Scott}, {Seiffert}, {Serra}, {Shellard}, {Spencer},
  {Spinelli}, {Stolyarov}, {Stompor}, {Sudiwala}, {Sunyaev}, {Sutton},
  {Suur-Uski}, {Sygnet}, {Tauber}, {Terenzi}, {Toffolatti}, {Tomasi},
  {Tristram}, {Trombetti}, {Tucci}, {Tuovinen}, {T{\"u}rler}, {Umana},
  {Valenziano}, {Valiviita}, {Van Tent}, {Vielva}, {Villa}, {Wade}, {Wandelt},
  {Wehus}, {White}, {White}, {Wilkinson}, {Yvon}, {Zacchei}, \&
  {Zonca}}]{2016A&A...594A..13P}
{Planck Collaboration}, {Ade}, P.~A.~R., {Aghanim}, N., {et~al.}
  2016{\natexlab{b}}, \aap, 594, A13

\bibitem[{{Pomakov} {et~al.}(2022){Pomakov}, {O'Sullivan}, {Br{\"u}ggen},
  {Vazza}, {Carretti}, {Heald}, {Horellou}, {Shimwell}, {Shulevski}, \&
  {Vernstrom}}]{2022MNRAS.515..256P}
{Pomakov}, V.~P., {O'Sullivan}, S.~P., {Br{\"u}ggen}, M., {et~al.} 2022,
  \mnras, 515, 256

\bibitem[{{Reinhardt}(1972)}]{1972A&A....19..104R}
{Reinhardt}, M. 1972, \aap, 19, 104

\bibitem[{{Reiprich} {et~al.}(2014){Reiprich}, {Basu}, {Ettori}, {Israel},
  {Lovisari}, {Molendi}, {Pointecouteau}, \&
  {Roncarelli}}]{2014efxu.conf..362R}
{Reiprich}, T.~H., {Basu}, K., {Ettori}, S., {et~al.} 2014, in Suzaku-MAXI
  2014: Expanding the Frontiers of the X-ray Universe, ed. M.~{Ishida},
  R.~{Petre}, \& K.~{Mitsuda}, 362

\bibitem[{{Riseley} {et~al.}(2020){Riseley}, {Galvin}, {Sobey}, {Vernstrom},
  {White}, {Zhang}, {Gaensler}, {Heald}, {Anderson}, {Franzen}, {Hancock},
  {Hurley-Walker}, {Lenc}, \& {Van Eck}}]{2020PASA...37...29R}
{Riseley}, C.~J., {Galvin}, T.~J., {Sobey}, C., {et~al.} 2020, \pasa, 37, e029

\bibitem[{{Roncarelli} {et~al.}(2006){Roncarelli}, {Ettori}, {Dolag},
  {Moscardini}, {Borgani}, \& {Murante}}]{2006MNRAS.373.1339R}
{Roncarelli}, M., {Ettori}, S., {Dolag}, K., {et~al.} 2006, \mnras, 373, 1339

\bibitem[{{Ryu} {et~al.}(2008){Ryu}, {Kang}, {Cho}, \&
  {Das}}]{2008Sci...320..909R}
{Ryu}, D., {Kang}, H., {Cho}, J., \& {Das}, S. 2008, Science, 320, 909

\bibitem[{{Schnitzeler} {et~al.}(2019){Schnitzeler}, {Carretti}, {Wieringa},
  {Gaensler}, {Haverkorn}, \& {Poppi}}]{2019MNRAS.485.1293S}
{Schnitzeler}, D.~H.~F.~M., {Carretti}, E., {Wieringa}, M.~H., {et~al.} 2019,
  \mnras, 485, 1293

\bibitem[{{Schroetter} {et~al.}(2019){Schroetter}, {Bouch{\'e}}, {Zabl},
  {Contini}, {Wendt}, {Schaye}, {Mitchell}, {Muzahid}, {Marino}, {Bacon},
  {Lilly}, {Richard}, \& {Wisotzki}}]{2019MNRAS.490.4368S}
{Schroetter}, I., {Bouch{\'e}}, N.~F., {Zabl}, J., {et~al.} 2019, \mnras, 490,
  4368

\bibitem[{{Shimwell} {et~al.}(2022){Shimwell}, {Hardcastle}, {Tasse}, {Best},
  {R{\"o}ttgering}, {Williams}, {Botteon}, {Drabent}, {Mechev}, {Shulevski},
  {van Weeren}, {Bester}, {Br{\"u}ggen}, {Brunetti}, {Callingham}, {Chy{\.z}y},
  {Conway}, {Dijkema}, {Duncan}, {de Gasperin}, {Hale}, {Haverkorn}, {Hugo},
  {Jackson}, {Mevius}, {Miley}, {Morabito}, {Morganti}, {Offringa}, {Oonk},
  {Rafferty}, {Sabater}, {Smith}, {Schwarz}, {Smirnov}, {O'Sullivan},
  {Vedantham}, {White}, {Albert}, {Alegre}, {Asabere}, {Bacon}, {Bonafede},
  {Bonnassieux}, {Brienza}, {Bilicki}, {Bonato}, {Calistro Rivera}, {Cassano},
  {Cochrane}, {Croston}, {Cuciti}, {Dallacasa}, {Danezi}, {Dettmar}, {Di
  Gennaro}, {Edler}, {En{\ss}lin}, {Emig}, {Franzen}, {Garc{\'\i}a-Vergara},
  {Grange}, {G{\"u}rkan}, {Hajduk}, {Heald}, {Heesen}, {Hoang}, {Hoeft},
  {Horellou}, {Iacobelli}, {Jamrozy}, {Jeli{\'c}}, {Kondapally}, {Kukreti},
  {Kunert-Bajraszewska}, {Magliocchetti}, {Mahatma}, {Ma{\l}ek}, {Mandal},
  {Massaro}, {Meyer-Zhao}, {Mingo}, {Mostert}, {Nair}, {Nakoneczny},
  {Nikiel-Wroczy{\'n}ski}, {Orr{\'u}}, {Pajdosz-{\'S}mierciak}, {Pasini},
  {Prandoni}, {van Piggelen}, {Rajpurohit}, {Retana-Montenegro}, {Riseley},
  {Rowlinson}, {Saxena}, {Schrijvers}, {Sweijen}, {Siewert}, {Timmerman},
  {Vaccari}, {Vink}, {West}, {Wo{\l}owska}, {Zhang}, \&
  {Zheng}}]{2022A&A...659A...1S}
{Shimwell}, T.~W., {Hardcastle}, M.~J., {Tasse}, C., {et~al.} 2022, \aap, 659,
  A1

\bibitem[{{Sofue} {et~al.}(1979){Sofue}, {Fujimoto}, \&
  {Kawabata}}]{1979PASJ...31..125S}
{Sofue}, Y., {Fujimoto}, M., \& {Kawabata}, K. 1979, \pasj, 31, 125

\bibitem[{{Subramanian}(2016)}]{2016RPPh...79g6901S}
{Subramanian}, K. 2016, Reports on Progress in Physics, 79, 076901

\bibitem[{{Thomson} \& {Nelson}(1982)}]{1982MNRAS.201..365T}
{Thomson}, R.~C. \& {Nelson}, A.~H. 1982, \mnras, 201, 365

\bibitem[{{Tjemsland} {et~al.}(2024){Tjemsland}, {Meyer}, \&
  {Vazza}}]{2024ApJ...963..135T}
{Tjemsland}, J., {Meyer}, M., \& {Vazza}, F. 2024, \apj, 963, 135

\bibitem[{{Turner} \& {Widrow}(1988)}]{1988PhRvD..37.2743T}
{Turner}, M.~S. \& {Widrow}, L.~M. 1988, \prd, 37, 2743

\bibitem[{{Vacca} {et~al.}(2010){Vacca}, {Murgia}, {Govoni}, {Feretti},
  {Giovannini}, {Orr{\`u}}, \& {Bonafede}}]{2010A&A...514A..71V}
{Vacca}, V., {Murgia}, M., {Govoni}, F., {et~al.} 2010, \aap, 514, A71

\bibitem[{{Vacca} {et~al.}(2018){Vacca}, {Murgia}, {Govoni}, {Loi}, {Vazza},
  {Finoguenov}, {Carretti}, {Feretti}, {Giovannini}, {Concu}, {Melis},
  {Gheller}, {Paladino}, {Poppi}, {Valente}, {Bernardi}, {Boschin}, {Brienza},
  {Clarke}, {Colafrancesco}, {En{\ss}lin}, {Ferrari}, {de Gasperin},
  {Gastaldello}, {Girardi}, {Gregorini}, {Johnston-Hollitt}, {Junklewitz},
  {Orr{\`u}}, {Parma}, {Perley}, \& {Taylor}}]{2018MNRAS.479..776V}
{Vacca}, V., {Murgia}, M., {Govoni}, F., {et~al.} 2018, \mnras, 479, 776

\bibitem[{{Vachaspati}(2021)}]{2021RPPh...84g4901V}
{Vachaspati}, T. 2021, Reports on Progress in Physics, 84, 074901

\bibitem[{{van Haarlem} {et~al.}(2013){van Haarlem}, {Wise}, {Gunst}, {Heald},
  {McKean}, {Hessels}, {de Bruyn}, {Nijboer}, {Swinbank}, {Fallows},
  {Brentjens}, {Nelles}, {Beck}, {Falcke}, {Fender}, {H{\"o}randel},
  {Koopmans}, {Mann}, {Miley}, {R{\"o}ttgering}, {Stappers}, {Wijers},
  {Zaroubi}, {van den Akker}, {Alexov}, {Anderson}, {Anderson}, {van Ardenne},
  {Arts}, {Asgekar}, {Avruch}, {Batejat}, {B{\"a}hren}, {Bell}, {Bell}, {van
  Bemmel}, {Bennema}, {Bentum}, {Bernardi}, {Best}, {B{\^\i}rzan}, {Bonafede},
  {Boonstra}, {Braun}, {Bregman}, {Breitling}, {van de Brink}, {Broderick},
  {Broekema}, {Brouw}, {Br{\"u}ggen}, {Butcher}, {van Cappellen}, {Ciardi},
  {Coenen}, {Conway}, {Coolen}, {Corstanje}, {Damstra}, {Davies}, {Deller},
  {Dettmar}, {van Diepen}, {Dijkstra}, {Donker}, {Doorduin}, {Dromer}, {Drost},
  {van Duin}, {Eisl{\"o}ffel}, {van Enst}, {Ferrari}, {Frieswijk}, {Gankema},
  {Garrett}, {de Gasperin}, {Gerbers}, {de Geus}, {Grie{\ss}meier}, {Grit},
  {Gruppen}, {Hamaker}, {Hassall}, {Hoeft}, {Holties}, {Horneffer}, {van der
  Horst}, {van Houwelingen}, {Huijgen}, {Iacobelli}, {Intema}, {Jackson},
  {Jelic}, {de Jong}, {Juette}, {Kant}, {Karastergiou}, {Koers}, {Kollen},
  {Kondratiev}, {Kooistra}, {Koopman}, {Koster}, {Kuniyoshi}, {Kramer},
  {Kuper}, {Lambropoulos}, {Law}, {van Leeuwen}, {Lemaitre}, {Loose}, {Maat},
  {Macario}, {Markoff}, {Masters}, {McFadden}, {McKay-Bukowski}, {Meijering},
  {Meulman}, {Mevius}, {Middelberg}, {Millenaar}, {Miller-Jones}, {Mohan},
  {Mol}, {Morawietz}, {Morganti}, {Mulcahy}, {Mulder}, {Munk}, {Nieuwenhuis},
  {van Nieuwpoort}, {Noordam}, {Norden}, {Noutsos}, {Offringa}, {Olofsson},
  {Omar}, {Orr{\'u}}, {Overeem}, {Paas}, {Pandey-Pommier}, {Pandey}, {Pizzo},
  {Polatidis}, {Rafferty}, {Rawlings}, {Reich}, {de Reijer}, {Reitsma},
  {Renting}, {Riemers}, {Rol}, {Romein}, {Roosjen}, {Ruiter}, {Scaife}, {van
  der Schaaf}, {Scheers}, {Schellart}, {Schoenmakers}, {Schoonderbeek},
  {Serylak}, {Shulevski}, {Sluman}, {Smirnov}, {Sobey}, {Spreeuw}, {Steinmetz},
  {Sterks}, {Stiepel}, {Stuurwold}, {Tagger}, {Tang}, {Tasse}, {Thomas},
  {Thoudam}, {Toribio}, {van der Tol}, {Usov}, {van Veelen}, {van der Veen},
  {ter Veen}, {Verbiest}, {Vermeulen}, {Vermaas}, {Vocks}, {Vogt}, {de Vos},
  {van der Wal}, {van Weeren}, {Weggemans}, {Weltevrede}, {White}, {Wijnholds},
  {Wilhelmsson}, {Wucknitz}, {Yatawatta}, {Zarka}, {Zensus}, \& {van
  Zwieten}}]{2013A&A...556A...2V}
{van Haarlem}, M.~P., {Wise}, M.~W., {Gunst}, A.~W., {et~al.} 2013, \aap, 556,
  A2

\bibitem[{{Vazza} {et~al.}(2017){Vazza}, {Br{\"u}ggen}, {Gheller}, {Hackstein},
  {Wittor}, \& {Hinz}}]{2017CQGra..34w4001V}
{Vazza}, F., {Br{\"u}ggen}, M., {Gheller}, C., {et~al.} 2017, Classical and
  Quantum Gravity, 34, 234001

\bibitem[{{Vazza} {et~al.}(2021){Vazza}, {Paoletti}, {Banfi}, {Finelli},
  {Gheller}, {O'Sullivan}, \& {Br{\"u}ggen}}]{2021MNRAS.500.5350V}
{Vazza}, F., {Paoletti}, D., {Banfi}, S., {et~al.} 2021, \mnras, 500, 5350

\bibitem[{{Vernstrom} {et~al.}(2017){Vernstrom}, {Gaensler}, {Brown}, {Lenc},
  \& {Norris}}]{2017MNRAS.467.4914V}
{Vernstrom}, T., {Gaensler}, B.~M., {Brown}, S., {Lenc}, E., \& {Norris}, R.~P.
  2017, \mnras, 467, 4914

\bibitem[{{Vernstrom} {et~al.}(2019){Vernstrom}, {Gaensler}, {Rudnick}, \&
  {Andernach}}]{2019ApJ...878...92V}
{Vernstrom}, T., {Gaensler}, B.~M., {Rudnick}, L., \& {Andernach}, H. 2019,
  \apj, 878, 92

\bibitem[{{Vernstrom} {et~al.}(2018){Vernstrom}, {Gaensler}, {Vacca}, {Farnes},
  {Haverkorn}, \& {O'Sullivan}}]{2018MNRAS.475.1736V}
{Vernstrom}, T., {Gaensler}, B.~M., {Vacca}, V., {et~al.} 2018, \mnras, 475,
  1736

\bibitem[{{Vernstrom} {et~al.}(2021){Vernstrom}, {Heald}, {Vazza}, {Galvin},
  {West}, {Locatelli}, {Fornengo}, \& {Pinetti}}]{2021MNRAS.505.4178V}
{Vernstrom}, T., {Heald}, G., {Vazza}, F., {et~al.} 2021, \mnras, 505, 4178

\bibitem[{{Vernstrom} {et~al.}(2023){Vernstrom}, {West}, {Vazza}, {Wittor},
  {Riseley}, \& {Heald}}]{2023SciA....9E7233V}
{Vernstrom}, T., {West}, J., {Vazza}, F., {et~al.} 2023, Science Advances, 9,
  eade7233

\bibitem[{{Vovk} {et~al.}(2023){Vovk}, {Korochkin}, {Neronov}, \&
  {Semikoz}}]{2023arXiv230607672V}
{Vovk}, I., {Korochkin}, A., {Neronov}, A., \& {Semikoz}, D. 2023, arXiv
  e-prints, arXiv:2306.07672

\bibitem[{{Welter} {et~al.}(1984){Welter}, {Perry}, \&
  {Kronberg}}]{1984ApJ...279...19W}
{Welter}, G.~L., {Perry}, J.~J., \& {Kronberg}, P.~P. 1984, \apj, 279, 19

\bibitem[{{Wen} \& {Han}(2015)}]{2015ApJ...807..178W}
{Wen}, Z.~L. \& {Han}, J.~L. 2015, \apj, 807, 178

\bibitem[{{You} {et~al.}(2003){You}, {Han}, \& {Chen}}]{2003AcASn..44S.155Y}
{You}, X.~P., {Han}, J.~L., \& {Chen}, Y. 2003, Acta Astronomica Sinica, 44,
  155

\bibitem[{{Zonca} {et~al.}(2019){Zonca}, {Singer}, {Lenz}, {Reinecke},
  {Rosset}, {Hivon}, \& {Gorski}}]{2019JOSS....4.1298Z}
{Zonca}, A., {Singer}, L., {Lenz}, D., {et~al.} 2019, The Journal of Open
  Source Software, 4, 1298

\bibitem[{{Zou} {et~al.}(2019){Zou}, {Gao}, {Zhou}, \&
  {Kong}}]{2019ApJS..242....8Z}
{Zou}, H., {Gao}, J., {Zhou}, X., \& {Kong}, X. 2019, \apjs, 242, 8

\end{thebibliography}

\begin{appendix}
   
\section{Gas  density contrast versus distance from centre  in a galaxy cluster}
\label{app:deltag_R200}

 Using galaxy cluster simulations, \citet[][]{2006MNRAS.373.1339R} find the gas density contrast ($\delta_g = \rho_g/(\Omega_b\rho_c)$ profile in galaxy clusters as a function of the distance from the centre in $R_{200}$ units ($x$), where $\rho_g$ is the gas density, $\Omega_b$ is the baryonic density parameter, and $\rho_c$ is the critical density of the Universe. The relation is a broken power law: 
\begin{equation}
      \delta_g(x)  =
      \begin{cases}
           a x^{-b_1}    &  {\rm if}\,\,\,\,x \leq R_b/R_{200} \\ 
            a \left(\frac{R_b}{R_{200}}\right)^{-(b_1-b_2)} x^{-b_2}    &  {\rm if}\,\,\,\,x > R_b/R_{200}
            \label{eq:deltag_R200}
      \end{cases}
\end{equation}
where $b_1=2.46\pm 0.03$, $b_2=3.38\pm 0.29$, and $R_b/R_{200}=1.14 \pm 0.20$. The normalisation can be estimated from their Fig. 5 as $a\approx 100$. That  relation is estimated using  clusters of virial mass in the range $1.5\times 10^{14}$--$3.5\times 10^{15} \, M_\odot$ and in the distance range of $0.3$--3~$R_{200}$. We found that our  observed source closest to a cluster is at a projected separation of 0.27 $R_{200}$ that is close to the lower limit of that range. Also, Eq. (\ref{eq:deltag_R200}) returns $\delta_g(3\,R_{200})= 2.75$ that is lower than the typical density contrast of 10 in filaments and it does not contribute much to the RRMs we measure. Thus, when we draw the separations to clusters from their distribution we restrict them to the range above.  

\section{Fitting with field strength dependent on the gas density}
\label{app:fitting}

Results of the fitting of Eq. (\ref{eq:fitting}) using $B_f$ depending on $\rho_g$ as for Eq. (\ref{eq:Bdens}) (Table~\ref{tab:rrmf_fit_Bdens}).

\begin{table*}
	\centering
	\caption{As for Table \ref{tab:rrmf_fit} except the magnetic field is dependent on the gas density as $B_f \propto \rho_g^{2/3}$. Columns are the same except  $B_{f,0}^{10}$ is the filament field strength at $z=0$ and gas density contrast  $\delta_g =10$. A local component term as $A_{rrm}/(1+z)^2$ is also used.  } 
	\label{tab:rrmf_fit_Bdens}
	\begin{tabular}{lcccc}
	  \hline 
        model & $\alpha $ &  $B_{f,0}^{10}$ &  $A_{rrm}$  &   $\beta$  \\ 
           &   & [nG]  & [rad m$^{-2}$]  &   \\ 
        \hline    
       stochastic $\alpha_s$ = -1.0  & $2.7 \pm 0.4$ & $4.0 \pm 1.0$ & $1.30 \pm 0.08$ & $0.7 \pm 0.4$ \\ 
       stochastic $\alpha_s$ = 0.0 & $2.5 \pm 0.7$ & $4.5 \pm 2.3$ & $1.30 \pm 0.10$ & $0.5 \pm 0.7$ \\ 
       stochastic $\alpha_s$ = 1.0 & $2.6 \pm 0.5$ & $4.2 \pm 1.2$ & $1.28 \pm 0.09$ & $0.6 \pm 0.5$ \\ 
       stochastic $\alpha_s$ = 2.0 & $2.3 \pm 0.8$ & $5.1 \pm 2.4$ & $1.28 \pm 0.12$ & $0.3 \pm 0.8$ \\ 
       astroph & $2.1 \pm 0.5$ & $7.9 \pm 2.2$ & $1.15 \pm 0.11$ & $0.1 \pm 0.5$ \\ 
       astroph 2 & $2.0 \pm 0.6$ & $9.7 \pm 3.2$ & $1.11 \pm 0.12$ & $0.0 \pm 0.6$ \\ 
       astroph 2 + stochastic $\alpha_s$ = -1.0 & $2.2 \pm 0.6$ & $7.5 \pm 2.4$ & $1.23 \pm 0.11$ & $0.2 \pm 0.6$ \\ 
       uniform & $2.2 \pm 1.0$ & $5.3 \pm 4.1$ & $1.26 \pm 0.13$ & $0.2 \pm 1.0$ \\ 

       \hline 
	\end{tabular}
\end{table*}


\end{appendix}

\end{document}